\documentclass[11pt]{article}
\usepackage[utf8]{inputenc}
\usepackage{cmap}
\usepackage[T1]{fontenc}
\usepackage{amsmath,amssymb}
\usepackage{graphicx}
\usepackage{float}
\graphicspath{{figures/}{./}}
\usepackage{booktabs,array,tabularx}
\usepackage{xltabular}
\usepackage[margin=1in]{geometry}
\usepackage{xcolor}
\definecolor{boxbg}{RGB}{244,248,251}
\definecolor{boxln}{RGB}{150,170,190}
\usepackage{mdframed}
\usepackage[hidelinks]{hyperref}
\usepackage{caption}
\usepackage{parskip}
\title{One Inference, Four Failure Modes: Formal Models of Why Pain Location Fails}
\author{Adam Y. Shavit\\ \small Hunter College and the Graduate Center, CUNY \\ \small ORCID: \href{https://orcid.org/0009-0008-1235-0995}{0009-0008-1235-0995}}
\date{}
\begin{document}
\maketitle
\begin{abstract}Patient-reported pain location is diagnostically decisive for some presentations and nearly uninformative for others. The companion paper argues that this is not one gradient of diagnostic utility but three distinct failures of localization. This paper gives those three failures their mathematics, and shows that they are one object seen from three sides: a single Bayesian generative model failing at different \textbf{nodes} — different points in one inference, namely the likelihood, the model class, and group- or context-dependence in that same likelihood. The companion's three failures and this paper's four failure modes are not a disagreement about the count: the first three are failures of the inference that produces a \textbf{felt} location, and the fourth, added here, is a downstream failure of \textbf{reporting} it.\par Each failure gets the lens that fits it. \emph{Anatomical multiplexing}, where many structures share one felt location, is a \textbf{non-identifiable inverse problem}: under this paper's continuous-mixture source model, source contrasts that survive even after each measured direction is rescaled by its own noise are contrasts that location cannot recover. The qualifier is load-bearing rather than decorative, and §4 supplies the counterexample that makes it so: for a single categorical diagnosis, a rank-deficient matrix can still send every source to a distinct observation. Rank deficiency alone is therefore not the general criterion; what it implies depends on what the source is assumed to be (§4). \emph{Delocalized amplification}, clinically central sensitization or nociplastic pain, is a \textbf{change of generative model}, and its dynamics are a neural-field bifurcation in which spatial extent is the order parameter — the quantity that changes state at the transition — while spatial instability is the marginal-stability signature, the sign that the leading restoring mode is losing what stability it had (§7). \emph{Referred and atypical displacement} is \textbf{group- or context-dependence in the presentation likelihood} $p(\ell \mid s, m, g, z)$ — a probability, not a loss. Its clinical consequence is a separate object one layer downstream: an optimal decision threshold that differs by group prevalence and cost (§6). Repeated observation cannot reduce recoverable information about a fixed inferential \textbf{target} — a trajectory counts, so biological drift is not the exception; changing the estimand is. The chain rule gives that as an exact identity, not as the data-processing inequality and not as an empirical finding, and it has a decision-theoretic twin: optional extra data cannot raise the optimal Bayes risk either. Practical value can still fall, because acquisition burden is not part of the information term, and a misspecified rule's actual risk is a different object from the optimum (§5).\par The paper's core is a fourth node: \textbf{the report itself}. The first three concern where pain \emph{is}. The reporting node concerns where pain is \emph{said to be}. A spatial Bayesian model factorizes mislocalization into a peripheral referral blur, a fixed anatomical projection bias, and a precision-weighted cognitive override, and it represents phantom-limb, mirror-box and central-post-stroke reports as regimes of one equation rather than as separate phenomena — a mapping into the model, not a claim to have identified their biological causes (§8). The model is \textbf{estimable}, under conditions the paper states rather than assumes, and the conditions are exact. The anatomical bias and the cognitive prior are identifiable from trials at known report weights \textbf{if and only if} the weight contrasts $W_j - W_1$ have full rank $d$ — the report-weight contrasts must span the whole report space, which with two trials means $W_2 - W_1$ is nonsingular. A single trial leaves the pair jointly degenerate, and in more than one dimension merely \emph{distinct} weights are not enough. A zero-expectation control trial is one sufficient and especially transparent second trial, and it is constructive: it reads the bias off directly and then returns the prior in closed form. The mean report geometry does \textbf{not} identify absolute cue precision. The mean weights fix only relative fusion structure, and the family of precision pairs consistent with an observed weight is $d$-dimensional, not a single common rescaling, so absolute precisions need variability or reliability information beyond the mean report. Recovering the referral operator and the bias additionally needs provocations that load enough independent source directions to make the augmented design full rank, and separately anchored cue locations (§8, §8.1).\par Two sensitivity sweeps show that the headline numerical patterns persist across the specified kernel, noise and prior ranges rather than appearing only at the illustrative parameter point. What they establish is robustness over the families swept, which §9 is careful to distinguish from structural invariance and, in particular, from robustness of the clinical migration prediction. One internal control is reported against the paper's own interest: the transport statistic first used to measure spatial migration scores a stationary, spreading profile as though it had travelled. Migration is therefore retained as a falsifiable hypothesis, not as a demonstrated result (§7.2, §9).\end{abstract}

\par\medskip

\emph{Companion to "Three Failures of Pain Location: Why Its Diagnostic Utility Is Three Quantities, Not One" (Shavit, 2026, arXiv:2607.26297), which carries the clinical argument and the evidence. This paper carries the models. It was previously the mathematical appendix of that article and is developed here as a standalone treatment. Model code and the numerical results are archived in the public supplement (Open Science Framework, \url{https://doi.org/10.17605/OSF.IO/ZFHKQ}).}

\textbf{Scope.} The models here are illustrative existence proofs with chosen parameters, not fitted instruments. §10 states, claim by claim, which results are proved, which hold numerically under a stated model, and which remain open empirical hypotheses. Nothing here is medical advice.

\section{Introduction}

A patient points to where it hurts. Sometimes that gesture is close to decisive — it localizes the problem well enough to act on. Sometimes it is nearly uninformative, and the same gesture from a different patient, or the same patient a month later, would point somewhere else. The standard reading treats this as one gradient: some body regions are anatomically complicated, so location tells you less there.

The companion paper argues that the gradient hides three different failures, and that they call for different instruments and have different consequences (Shavit, 2026). This paper is the formal half of that argument. It asks what each failure \emph{is}, mathematically, and whether the three are related or merely adjacent.

They are related. Write the clinical encounter as one generative model — hidden cause, felt map, report, decision — and each failure is that single model breaking at a different point. One qualification keeps the claim honest, and §3's factorization is what forces it: the \emph{full} model carries the model class $m$ and the group and context variables $g, z$ as variables of its own, so nothing in it breaks when a patient's class or group changes. What breaks is the \textbf{reduced} diagnostic model a clinician actually runs — the one that fixes the class as peripheral, or assumes the presentation likelihood is the same across groups. Each failure below is therefore an assumption of that reduced model failing, at the factor named.

\begin{itemize}\setlength{\itemsep}{2pt}\item \textbf{Anatomical multiplexing} (mechanism a) breaks the \textbf{likelihood}. Many structures project to one felt location, so the map from cause to report is many-to-one and cannot be inverted from location alone.\item \textbf{Delocalized amplification} (mechanism b) breaks the \textbf{model class}. A centrally driven pattern replaces the peripheral generator, so the inference is not mis-estimating a parameter; it is fitting the wrong family.\item \textbf{Referred and atypical displacement} (mechanism c) breaks the \textbf{group-conditional presentation likelihood}. The distribution of felt location given the diagnosis shifts with a group variable. A threshold calibrated on one group is then wrong for another — but that threshold lives one layer downstream, in the decision rule, and is a consequence of mechanism (c) rather than its definition. The distinction is the discipline §3 sets up, and is worth the extra clause: a probability and a loss are different objects.\end{itemize}

Naming the shared object does two things: it says which mathematics each failure needs, and it makes the failures comparable rather than a list. §3 sets out the generative model, and also what is \emph{outside} it — the decision rule and the treatment step, which are a separate layer and not additional probability nodes. Conflating them lets a treatment disparity be read as a perceptual one.

\textbf{The fourth node, and this paper's main contribution.} The three mechanisms all concern where pain \emph{is}. A report is a further step, and it has its own machinery: what a patient says is the outcome of a second inference, in which a noisy percept is weighted against an expectation. §8 develops a spatial Bayesian model of that step. Prior computational pain models do represent location: Boström et al. (2014) put nociceptive receptors at explicit two-dimensional skin sites feeding a somatotopic map. That model has no perceiver and computes no posterior, and the Bayesian prior art for \emph{inferring} a location is in innocuous touch (Goldreich, 2007). What §8 claims as new is narrower than either — an explicitly Bayesian model of the location a patient \emph{reports} — and it joins the two lineages. It factorizes mislocalization into three named parts — referral blur, a fixed anatomical offset, and a precision-weighted cognitive override — and phantom-limb pain, the mirror box and central post-stroke pain are then representable as regimes of one equation. Each such mapping is a mechanistic hypothesis about how the model renders a phenomenon, not evidence about what causes it.

\textbf{What the models here establish, and what they do not.} The models are illustrative: parameters are chosen to demonstrate that a mechanism \emph{can} produce a pattern, not fitted to data. That is a real limit and it is stated per claim in §10, which sorts every result into proved, numerical under a stated model, or open empirical hypothesis. Two results are reported against the paper's own interest. The transport statistic first adopted to measure spatial migration turns out to score a stationary, spreading profile as though it had moved (§7.2), and the no-repulsion property of cue fusion holds only in the scalar case the figures use, with a checked counterexample in general (§8). Readers trust a paper of illustrative models that reports the controls that came out against it.

\section{Notation}

Notation: $H$ = hidden cause; $X$ = observed location/report; $\Phi $ = standard-normal CDF; $I(\cdot ;\cdot )$ = mutual information (bits); $S$ = the diagnosis-relevant cause; $m$ = the latent generative-model class (peripheral vs. central); $\ell $ = the felt percept location; $R$ (roman, in §4/§8, not to be confused with report $r$) = the referral/lead-field operator; $G$ = a scientifically justified group variable; $Z$ = clinical/measurement context; $A$ = clinical action; $T$ = treatment; $Y$ = outcome; $D$ = the diagnosis indicator, $D = 1$ when the target condition is present. Symbols are scoped to the section that introduces them, and six carry a second, local meaning: $M$ is the number of body-surface mesh cells in §4; $s$ is the disinhibition parameter in §7 and §9 and the cause vector everywhere else; $T$ is the number of repeated observations in §5; $C$ is the conditioning context in §5 and the expected cost in §6; $I$ is the localized input in §7 and the identity matrix in §8; and $W$ is the lateral-connection kernel in §7 and the reporting weight in §8. Code for each model is in the supplement.

\section{Generative model, decision layer, and care pathway}

Diagnosis inverts a probabilistic generative model of cause, percept, and report — probabilities only: $p(s, m, \ell , r \mid g, z) = p(s, m \mid g, z) \cdot  p(\ell \mid s, m, g, z) \cdot  p(r \mid \ell , g, z)$, In words: group and context set the prior over cause and model class; those generate a felt location; and the felt location is then converted into a report. Here, with cause $s$, generative-model class $m$ (peripheral vs. central), a scientifically justified group variable $g$, clinical/measurement context $z$, percept location $\ell $, and report $r$. The diagnostic posterior integrates out the latent percept: $p(s, m \mid r, g, z) \propto  p(s, m \mid g, z) \int  p(\ell \mid s, m, g, z) p(r \mid \ell , g, z) d\ell $. Clinical action is \emph{not} a fourth factor inside this posterior — a loss function is not a probability node. It is chosen under an explicit decision rule, outside the factorization, with $L_{\mathrm{dec}}$ the decision loss — what a wrong action costs: $a^{*}(r, g, z) = \operatorname{arg\,min}_a E[L_{\mathrm{dec}}(a, S, g, z) | r, g, z]$. In words, the action taken is whichever choice carries the smallest posterior expected cost, so a change in prevalence, or in the cost of a miss against a false alarm, can change the action even when the likelihood has not moved at all. Treatment and outcome follow the action, in their own layer: $p(t, y \mid a, s, g, z) = p(t \mid a, s, g, z) \cdot  p(y \mid t, a, s, g, z)$. Each mechanism is now a failure at a distinct, correctly-typed object: the likelihood $p(\ell \mid s,m,g,z)$ (a, overlap or non-injectivity), the model class $m$ and its dynamics (b), group- or context-dependence in that same $p(\ell \mid s,m,g,z)$ (c — a probability, not a loss), the observation model $p(r \mid \ell ,g,z)$ (d, formalized in §8), and — downstream of all four, in the separate action/ treatment layer rather than in the perceptual model — unequal treatment uptake $p(t \mid a,s,g,z)$. This separation does three things a single mixed graph could not: it stops a loss function from being read as a probability, it stops a treatment disparity from being read as a perceptual displacement (the error corrected in mechanism (c) above), and it lets a group effect enter at whichever stage the data actually support, rather than forcing it into one node by construction. The decomposition is a diagnostic tool for locating a failure, not a claim that every patient occupies exactly one box; mechanisms can coexist. Nor is it claimed \textbf{exhaustive}. A pain location can also be uninformative for reasons that live outside this factorization altogether — how the map was acquired, how it was coded, what is missing from it, how a clinician read it, or how a downstream model was calibrated. What the paper claims is narrower and testable: these four failure types are formally distinct, they are confusable with one another in a single static report, and different measurements can help distinguish them, each targeting a different failure mode. \emph{Help} is the accurate verb: §5 shows that added measurements raise recoverable information without necessarily eliminating ambiguity, and §8's identifiability results say which ambiguities survive which measurements. Figure 1 is a one-page visual primer to the kinds of mathematics this paper uses — probability, information, linear algebra, decision theory, causal inference, and dynamical systems — one intuitive picture each. Figure 2 draws the confusability itself and the way out of it: two models that differ in which node failed produce the identical single report, and come apart the moment reports are repeated.

\begin{figure}[H]\centering
\includegraphics[width=\linewidth]{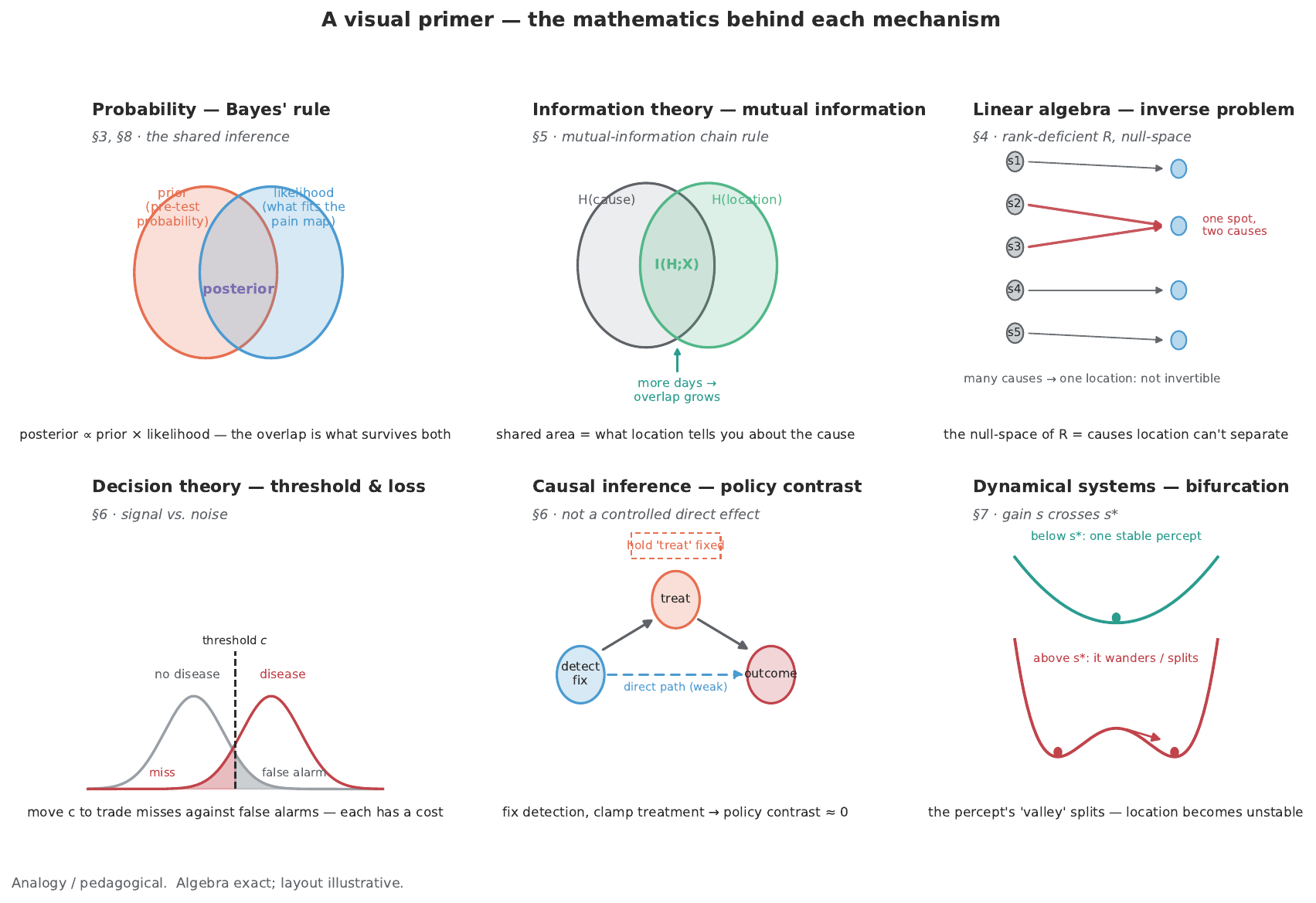}
\caption{Figure 1. A visual primer — one intuitive picture per type of mathematics used in this paper, so the formalism can be read alongside its gist. Probability (§3, §8): Bayes' rule as the overlap of prior and likelihood — the posterior is what survives both. Information theory (§5): mutual information I(H;X) as the shared area of two entropy circles, which cannot decrease as observation days accumulate (the repeated-observation chain rule). Linear algebra (§4): a rank-deficient referral matrix R sends two same-segment causes to one location — under the continuous-mixture source model of §4, the null-space is what location cannot separate. (The qualifier matters: §4 shows a rank-deficient matrix can still separate every source of a single categorical diagnosis.) Decision theory (§6): a threshold on overlapping signal and noise distributions, trading misses against false alarms, each with its own cost. Causal inference (§6): the standardized policy contrast — holding treatment fixed while improving detection — is near zero under current uptake; a controlled direct effect is a distinct potential-outcomes estimand, zero only under the stated causal graph's fixed-mediator assumptions. Dynamical systems (§7): a potential 'valley' that splits as central gain crosses $s^{*}$, so the stable percept becomes unstable. [Analogy / pedagogical — algebra exact, layout illustrative.]}
\end{figure}

\begin{figure}[H]\centering
\includegraphics[width=\linewidth]{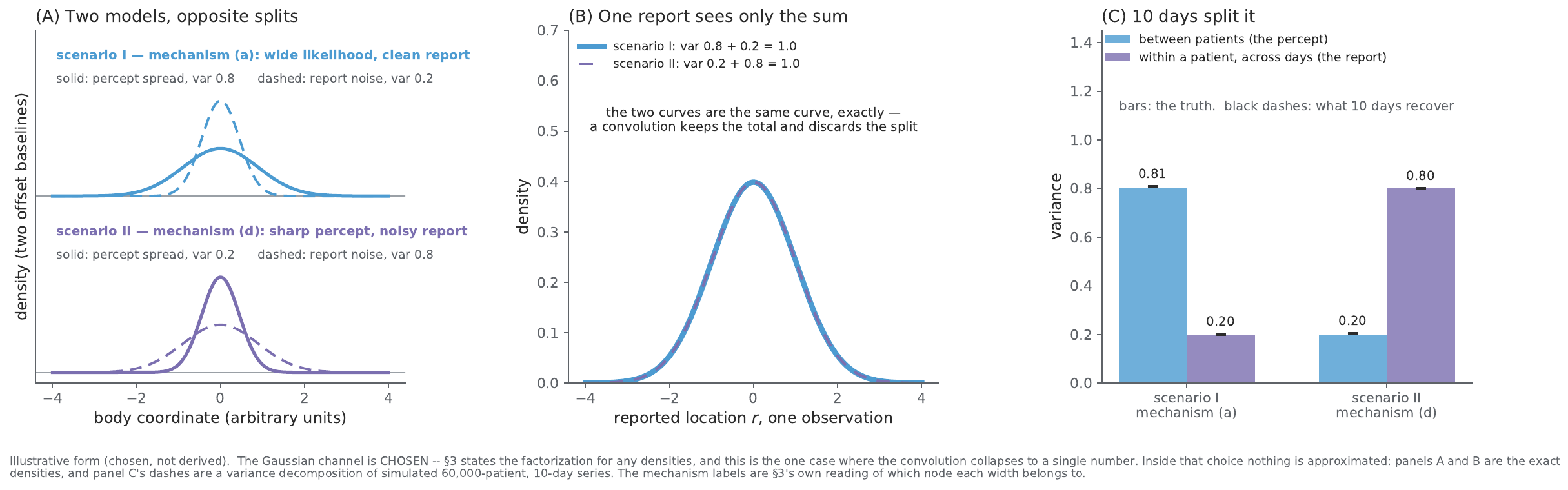}
\caption{Figure 2. Why one static report cannot say which node failed, and what ends the tie. §3's diagnostic posterior integrates the felt location out, because nobody observes it, and that integral is a convolution — an operation that keeps the total and discards the split. (A) Two models of the same patient with opposite splits: a wide presentation likelihood with a clean report, which is mechanism (a), and a sharp percept with a noisy reporting node, which is mechanism (d). Each model has its own baseline; the solid curve is the percept spread and the dashed one the report noise, and the two rows are each other's mirror. (B) The distribution of a single report under the two models. They are not similar curves, they are the same curve — the two variances sum to 1.0 either way, and one observation sees the sum and nothing else. That is what "confusable in a single static report" means, drawn exactly rather than worried about. (C) Repeated observation splits the same total, because the percept is drawn once per patient and the report afresh each day: ten days recover 0.81 and 0.20 against a true 0.80 and 0.20, and 0.20 and 0.80 against a true 0.20 and 0.80. A two-sample test that cannot tell the single reports apart (p = 0.63) separates the day-to-day differences decisively, so the check fails in both directions and neither direction alone would mean anything. Read across fields: one photograph of a blurred edge cannot say whether the subject moved or the lens was soft, and a burst of photographs can. §5 and §8 are explicit that added measurements raise recoverable information without necessarily eliminating it; this figure draws a case where they happen to separate it completely, which is not the general rule. [Illustrative form — the Gaussian channel is chosen, because it is the case where the convolution collapses to a single number; inside that choice every quantity is exact, and panel C is a variance decomposition of 60,000 simulated ten-day series.]}
\end{figure}

\section{Anatomical multiplexing as a linear inverse problem}

Model the pain map on $M$ body-surface mesh cells (not as a bare 2-D coordinate — a two-dimensional surface still supports an $M$-dimensional field, and rank deficiency must come from similar referral columns, not from the surface's intrinsic dimension): $y = R s + \varepsilon $ — in words, the observed body-surface map is the superposition of each active source's referral pattern, plus measurement noise — with $y \in  \mathbb{R}^M$, referral matrix $R \in  \mathbb{R}^{M\times K}$ (column $k$ = structure $k$'s referral pattern), and measurement noise $\varepsilon  \sim  N(0, \Sigma _y)$. If an instrument retains only a 2-D centroid $c$, that is a lossy summary $c = B(y)$, and any ambiguity $B$ adds is measurement loss, not evidence that anatomy itself converges to two dimensions.

The identifiability statement this paper uses depends on what $s$ is assumed to be, and the paper's own illustrative design — a continuous posterior over source strength, $\Sigma _{\mathrm{post}}$ below and in §8 — assumes an \textbf{unconstrained continuous mixture} $s \in  \mathbb{R}^K$. Under that assumption, every source direction is likelihood-identified iff the noise-whitened operator $\tilde{R} = \Sigma _y^{-0.5} R$ has full column rank; the identifiability \emph{floor} is the dimension of $\tilde{R}$'s near-null-space, the structure combinations $v$ with $\|\tilde{R} v\| \approx  0$. This full-rank condition is \textbf{not} the general identifiability criterion, and would be the wrong one for a single-cause categorical diagnosis $S \in  \{1,\dots ,K\}$: a rank-deficient $R = [[1,0,1],[0,1,1]]$ still sends its three one-hot sources to three \textbf{distinct} noiseless means $(1,0)$, $(0,1)$, $(1,1)$, so the categorical model is identifiable even where the continuous one is not — a checked counterexample (\texttt{math\_checks.py / categorical\_rank\_check}). Distinct means are what identifiability asks for and are not the same as perfect classification: with nonzero measurement noise the three categories still overlap, and a classification error rate remains. For a sparse or nonnegative mixture, identifiability depends on the restricted null space — only those null vectors that can themselves be written as the difference of two admissible sources matter — and cannot be read off rank alone. The manuscript's own referral-matrix illustrations below therefore describe the continuous-mixture case; a categorical or sparse re-analysis would use the corresponding restricted criterion instead. Figure 3 draws both readings: the two continuous mixtures that collide on one body map, the three categorical sources that land apart, and the error rate that survives at nonzero noise.

\begin{figure}[H]\centering
\includegraphics[width=\linewidth]{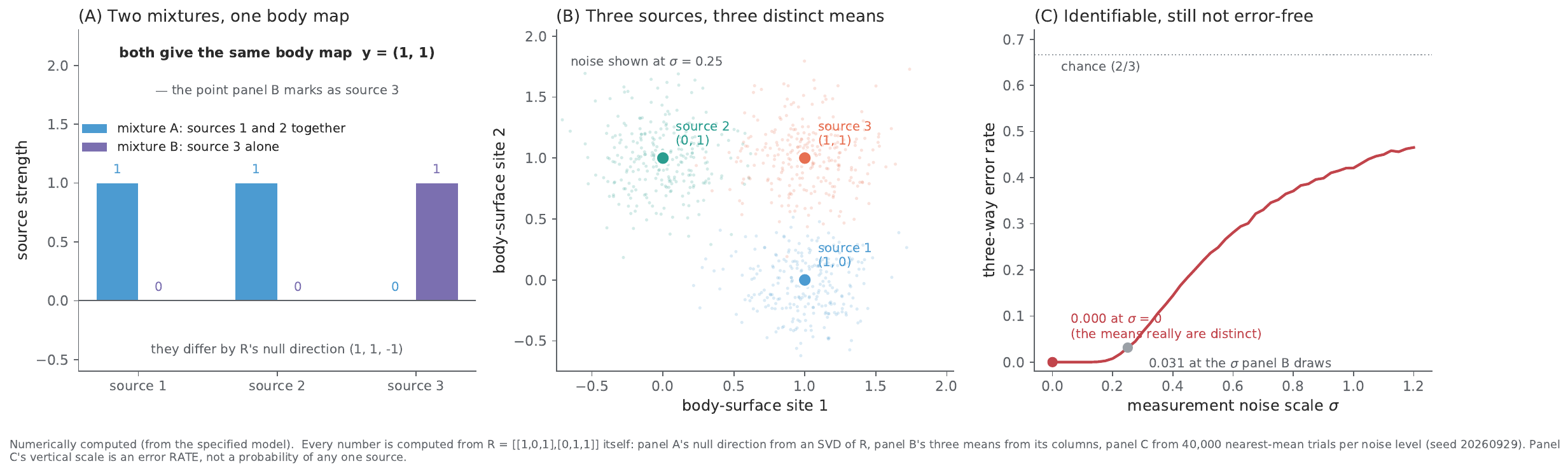}
\caption{Figure 3. A rank-deficient referral matrix can still separate a single-cause diagnosis, which is why §4's source-ontology qualifier is load-bearing rather than decorative. All three panels are computed from $R = [[1,0,1],[0,1,1]]$ itself. (A) Under a continuous mixture, sources 1 and 2 acting together and source 3 acting alone produce the identical body map y = (1,1); the two differ by $R$'s null direction (1,1,$-$1), taken here from a singular value decomposition rather than written in by hand. That is the identifiability floor §4 defines, and no measurement of location resolves it at any precision. (B) Under a single-cause categorical model the admissible sources are only the three one-hot vectors, and they land on three distinct points — (1,0), (0,1) and (1,1) — so the categorical model is identifiable exactly where the continuous one is not. (C) The sentence §4 adds immediately afterwards, drawn: distinct means are not the same as perfect classification. The three-way error of the optimal nearest-mean rule is exactly 0.000 at zero noise, 0.031 at the noise panel B draws, and climbs toward but not to the two-thirds chance line. Read across fields: two ingredients mixed in a bowl and a third that tastes like their mixture are one problem if any amounts are allowed, and a different problem if exactly one ingredient may be used. [Numerically computed — the null direction from R's own singular value decomposition, panel B's means from its columns, panel C from 40,000 nearest-mean trials per noise level. Control: the same collision search on a full-rank operator finds nothing, so the collision is not an artefact of the search.]}
\end{figure}

\textbf{Sensitivity to the source ontology.} Treating the source as a continuous mixture is a modeling commitment, so it was tested against the alternative rather than asserted. Modeling $s$ as a \textbf{sparse nonnegative} mixture — few structures active at once, none contributing negatively — recovers all five single sources instead of two, on the same referral matrix and noise (\texttt{source\_ontology\_comparison.py}). That gain is \emph{not} the provocation gain reported below, and the matching counts are a coincidence worth naming before either is read: provocation adds measurements, whereas sparsity adds an assumption. Sparsity also restores the floor only for sources acting alone. Where two floor structures are active together, sparse recovery still returns the wrong support in 2 of the 10 pairs. The two weakest members are recovered marginally rather than comfortably: forbidding the true column costs a fit penalty of 0.25 for muscle and 0.18 for ligament, against 0.90 for the disc and 1.63 for the sacroiliac joint. The penalty is the rise in the nonnegative-least-squares residual norm when the true column is forbidden, on referral columns scaled to unit activation against a per-site measurement noise standard deviation of 0.02 over $M = 10$ body sites (\texttt{source\_ontology\_comparison.py}).

\textbf{What those penalties should be compared against is the statistic's own null, not the noise standard deviation.} The two are in different units — the penalty is a difference of two Euclidean residual norms taken over all ten sites, while 0.02 is a standard deviation at one site — so dividing one by the other gives a number with no clean reading. Nor is the obvious dimensional repair right: because the statistic is a \emph{difference} of two nested residual norms fitted to the same data, it scales like the noise in the single design direction the restricted model cannot reach, which is of order one standard deviation and not $\sigma \sqrt{M}$. The null's own 95th percentile is 0.0011, which is 0.056 of a per-site standard deviation rather than the $\sqrt{M} = 3.16$ of them that a norm-level correction assumes, so scoring the penalty against $\sigma \sqrt{M} = 0.063$ would overstate the noise floor about fifty-sixfold and understate the finding. The comparison that needs no multiplier is the statistic against itself: over 20,000 maps built from pure noise on this referral matrix, the largest penalty the statistic ever produced was \textbf{0.011}, and every one of the four penalties above exceeds it — muscle, the weakest, by more than twentyfold (\texttt{p37\_noise\_norm\_units\_2026-09-29.py}). One caution on reading that: the null depends on the geometry of the referral matrix, so the ratio of a penalty to its null is not comparable across designs and must not be used to rank them. Within this design it says the gaps are not noise.

Location data alone do not establish sparsity as the correct source ontology. This paper therefore keeps the continuous mixture, which leaves the facet, muscle and ligament floor unresolved and reports it, rather than narrowing that floor with an assumption the data cannot check.

Define each structure's unique signal as $u_k = \|r_k - R_{-k} \hat{\beta }_k\| / \|r_k\|$ (residual of column $k$ regressed on the others); $u_k$ small $\Rightarrow $ structure $k$ is in the floor. In words, $u_k$ is the fraction of structure $k$'s referral pattern that the other structures cannot synthesize between them, so a value near zero means location carries almost nothing unique about that source. Provocation tests append rows $P$ (movement $\times $ structure loadings) to form $R_{\mathrm{aug}} = [R; P]$. \emph{Illustrative result:} location alone, 2 of 5 structures clear the practical unique-signal criterion; with provocation, 5 of 5. The word \emph{identifiable} is deliberately not used for this count, because it is reserved above for the structural condition — injectivity, full column rank, a restricted null space — and a threshold on a residual is a practical resolution criterion, not that. The count is the number of columns whose unique signal $u_k$ clears a 0.30 residual bar; the bar sets the exact count, but the \emph{direction} — provocation adds identifying information — is invariant to it across the range we checked: over thresholds from 0.01 to 1.00 the provocation-augmented count never falls below the location-only count, and it is strictly larger throughout $u_0 \in  [0.148, 0.886]$. The control that makes this a test rather than an arithmetic identity is a provocation block drawn from $R$'s own row space, adding no new direction: it raises the count at no threshold at all.

Provocation is guaranteed to add Fisher information — the local precision with which the likelihood tells nearby source vectors apart, so that larger in the positive-semidefinite order means no source contrast is measured less precisely than before. With an independent provocation measurement $q = P s + \eta $, the information matrix changes from $J_0 = R^{\top}\Sigma _y^{-1}R$ to $J_1 = J_0 + P^{\top}\Sigma _p^{-1}P$, and the increment is positive semidefinite, so information cannot decrease, posterior covariance cannot increase, and rank cannot fall. \textbf{Provocation is not guaranteed to improve the condition number.} A numerical counterexample: starting from $R = I_2$ (condition number 1) and appending one strongly informative but one-sided row $P = (100, 0)$ gives $J_1 = \operatorname{diag}(10001, 1)$, condition number 10001 — information rose sharply in one direction while the \emph{balance} of the information worsened (\texttt{math\_checks.py / condition\_number\_check}). The earlier claim that “provocation lowers the condition number” is withdrawn as a general statement; a good provocation design instead maximizes an explicit criterion such as the minimum eigenvalue of $J_1$, its log determinant, or expected reduction in diagnostic loss, not a smaller condition number. The appended rows $P$ are also not perfectly selective — biomechanically, loading a facet joint also stresses the neighbouring muscle, ligament, and nerve — so the model \emph{quantifies} the surviving uncertainty (as the posterior width $\Sigma _{\mathrm{post}}$ over the still-correlated columns, §8) rather than claiming to eliminate it.

\section{Repeated observation and the information chain rule}

Diagnostic information $U = I(H; X | C) = \mathbf{H}(H | C) - \mathbf{H}(H | X, C)$, where $H$ is the hidden cause, $X$ the observed location or report, $C$ the conditioning context (the covariates held fixed — the clinical setting and patient group), and $\mathbf{H}(\cdot )$ the entropy operator, written in bold to keep it distinct from the hidden cause $H$. Both terms are conditional entropies in the standard sense, each already averaged over everything it conditions on, so no further expectation is written around the second. In words, diagnostic information is the expected reduction in uncertainty about the target once the location or report data are in hand. Naming it \emph{diagnostic information} rather than \emph{utility} matters: clinical value is defined separately, through the decision loss of §3/§6, and the two meet only at the action layer (§6) — a threshold-specific summary such as a likelihood ratio does not by itself read out mutual information without prevalence and the full joint distribution.

For any fixed target $H$, context $C$, and any sequence of observations — no conditional independence or Markov assumption required — the exact mutual-information \textbf{chain rule} (Cover \& Thomas, 2006) gives $I(H; X_{1:T} | C) = I(H; X_{1:T-1} | C) + I(H; X_T | X_{1:T-1}, C) \ge  I(H; X_{1:T-1} | C)$. This is \emph{not} the data-processing inequality (which needs a Markov emission structure); it is a simpler and more general identity, and it is exact, not an empirical finding: retaining another observation cannot reduce recoverable information about a fixed target. The increase is strict only when the new observation carries positive conditional information.

\textbf{What has to be fixed is the inferential target, not the biology.} The hypothesis the chain rule needs is that the same random variable $H$ appears on both sides of the inequality. It may perfectly well be an entire trajectory $H_{1:T}$: retaining one more observation still cannot reduce information about that trajectory-valued target, however much the underlying generator moves meanwhile. Drift is therefore not a mathematical exception to the identity, and requiring the hidden cause to be biologically static asks for strictly more than the theorem uses. What does remove the ordering is comparing two \emph{different} targets, such as $I(H_{T-1}; X_{1:T-1})$ against $I(H_T; X_{1:T})$; nothing orders those, and the gap can be the whole quantity. A constructed pair shows how large: under one estimand the first observation carries a full bit, $I(H_1; X_1) = 1.000$, while under a changed estimand the longer record carries $I(H_2; X_{1:2}) = 0.000$ bits — all of the information, not a fraction of it (\texttt{p37\_reporting\_node\_identifiability\_2026-09-29.py}, control NC-BR-2). Clinically the change of estimand is the thing that actually happens, and it is exactly the onset of mechanism (b): as acute pain persists the generator can move from a localized tissue source toward a centralized loop, so the target of interest becomes a current state rather than a single $H$. What longitudinal tracking should then estimate is a switching, non-stationary state-space model with \emph{change-point} detection, scored by filtering error or detection delay — a hidden state allowed to change over time, judged by how accurately and how quickly the tracker catches those changes, not by monotone accumulation toward a static answer.

\textbf{A decision-theoretic companion, and where misspecification actually lives.} The identity has a risk-side twin. Write $R^{*}(X)$ for the smallest expected decision loss reachable when the action may depend on $X$, holding the target, the action space and the loss fixed. Then

$R^{*}(X_{1:T}) \le  R^{*}(X_{1:T-1})$.

In words: optional extra data cannot raise the optimal Bayes risk. The proof is one sentence. A rule that uses only the shorter record is still available once the extra observation arrives, because it can ignore it, so the richer problem minimizes over a superset of rules. That is Blackwell dominance in its elementary form; a 3,000-problem finite-alphabet enumeration reproduces it with a worst violation of $1.1 \times  10^{-16}$, which is evidence that the encoding is right and not a substitute for the proof.

That inequality then fixes how practical value has to be written. Setting

$V_T = R^{*}(X_{1:T-1}) - R^{*}(X_{1:T}) - \lambda \cdot cost(X_T)$,

the first difference is \textbf{never negative}, so $V_T$ turns negative only through the acquisition and burden term $\lambda \cdot cost(X_T)$. Misspecification cannot enter through $R^{*}$ and should not be attributed to it: $R^{*}$ is by definition the optimum, and a misspecified rule is by definition not the optimum. A forced or misspecified estimator $\delta _T$ therefore needs its own \textbf{actual} risk $R(\delta _T)$, stated separately, and that quantity can and does worsen with more data — forcing a misspecified rule on the longer record raised its actual risk to 0.582 against an optimal 0.526, an excess of 0.056 in the same loss units, in the same enumeration (control NC-BR-1). Three distinct things can therefore make an extra observation worth less than it costs: acquisition and adherence burden, delay once it is priced into the loss, and the use of a rule that is not the optimal one. None of the three is a failure of the information identity, and keeping them apart is the point of writing the risk term this way. Figure 4 draws the pair: the information total that cannot fall, and the value that turns negative once the burden is priced in.

\begin{figure}[H]\centering
\includegraphics[width=\linewidth]{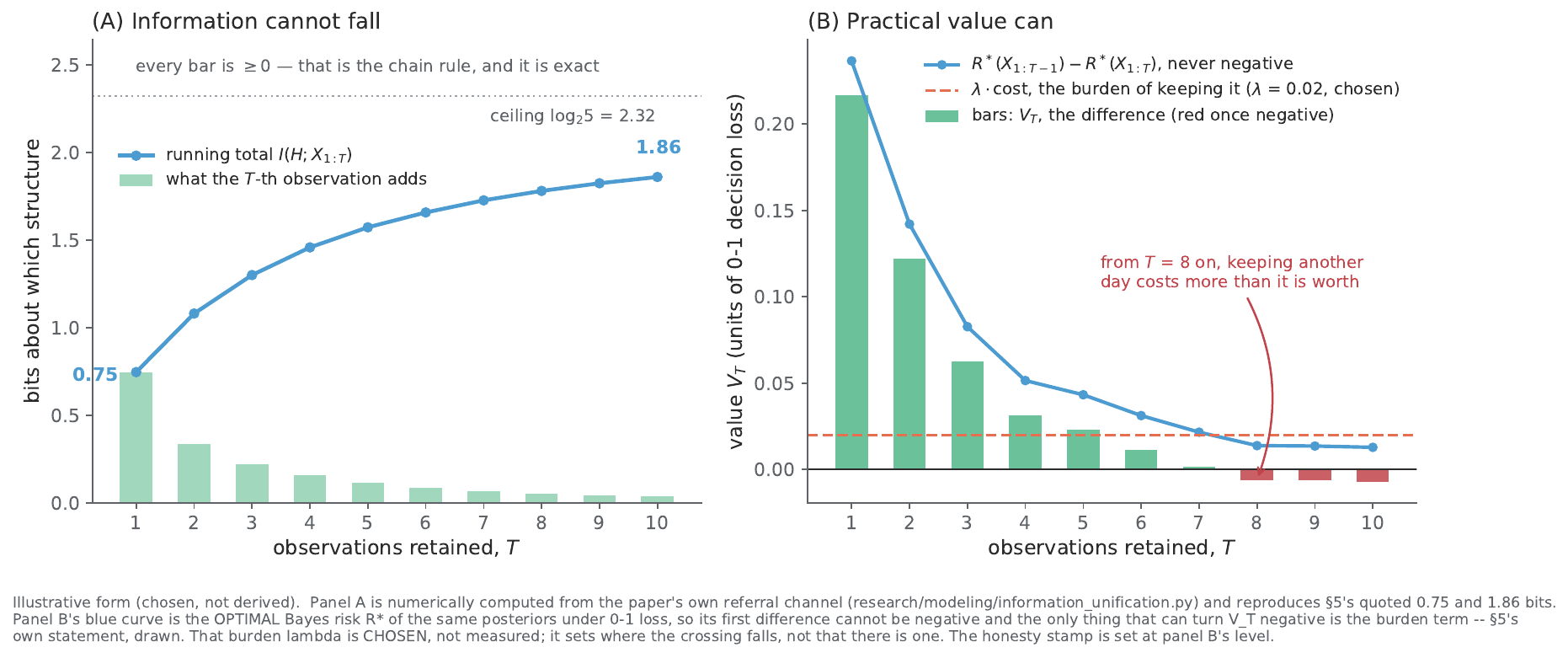}
\caption{Figure 4. Information cannot fall; practical value can, and only the burden term can push it down. (A) The many-to-one referral channel of §5, re-run: the running total $I(H; X_{1:T})$ climbs from 0.75 bits at one observation to 1.86 at ten against the $\log_{2}5 = 2.32$ ceiling — the numbers §5 quotes — and each bar is what the $T$-th observation adds. Every bar is non-negative, and that is the chain rule itself, exact rather than empirical. (B) The risk-side twin, on the same posteriors under 0-1 loss. The curve is the first difference $R^{*}(X_{1:T-1}) - R^{*}(X_{1:T})$, which cannot be negative because the richer problem minimizes over a superset of rules; the dashed line is the burden $\lambda \cdot cost$; the bars are $V_T$, their difference, and they turn red from $T$ = 8, where another day costs more than it removes. Bits and losses are different scales and get different panels rather than two axes on one plot. Read across fields: a longer recording never contains less of the signal and can still be not worth making. [Illustrative form — panel A is computed from the paper's own \texttt{information\_unification.py} and reproduces its published values to four decimals; panel B's burden $\lambda $ = 0.02 is chosen, not measured, and fixes where the crossing falls, not that there is one. Controls: a channel carrying no information returns exactly 0 bits, and halving the sample under a new seed moves the curve by 0.007 bits.]}
\end{figure}

\emph{Illustrative results (bits).} (a) $I(structure; location)$ rises 0.75 $\to $ 1.86 as $T$ grows 1 $\to $ 10 (ceiling $\log_{2}5 = 2.32$); (b) $I(H; snapshot) = 0.21$ $\to $ $I(H; 14-day series) = 0.999$ (ceiling 1). (c) The third example needs all five of its information quantities on the page, because three different questions are easy to conflate here and only one of them is the one the clinical reading wants. This model is one-dimensional and Gaussian, so each quantity is an exact integral rather than a histogram estimate; 4,000-node Gauss–Legendre quadrature gives, in bits (\texttt{p37\_conditional\_mi\_exact\_2026-09-29.py}):

\begin{itemize}\setlength{\itemsep}{2pt}\item $I(D; X) = 0.176$ — how diagnostic the signal is, group ignored.\item $I(D; G) = 0.000$ — how diagnostic group alone is.\item $I(D; X | G) = 0.204$ — how diagnostic the signal is \emph{within} a group.\item $I(D; G | X) = 0.028$ — what observing group adds \emph{after} the signal is in hand.\item $I(D; X, G) = 0.204$ — what the two together supply.\end{itemize}

\textbf{The premise that licenses reading that 0.028 as the information group adds, printed rather than assumed.} In general the stratification difference $I(D;X|G) - I(D;X)$ and the incremental information $I(D;G|X) = I(D;X,G) - I(D;X)$ are different objects, and the first can take either sign. Here they agree, and they agree for a reason specific to this parameterization rather than for a general one: group is drawn independently of disease and prevalence does not differ by group, so $I(D; G) = 0$ exactly, and the two chain-rule decompositions $I(D;X) + I(D;G|X)$ and $I(D;G) + I(D;X|G)$ collapse onto each other, each reproducing $I(D;X,G)$ to better than $3.3 \times  10^{-15}$ bits. Under that premise — and it is \textbf{sufficient}, not merely convenient — the 0.028-bit gap \emph{is} the incremental information from observing group. Without it the two quantities separate, and by an amount this paper's own §6 makes room for: letting prevalence differ by group, 0.25 against 0.55 at the same average of 0.40, raises $I(D; G)$ to 0.069 bits and pulls the stratification gap (0.043) and the incremental information (0.111) apart by exactly that much. That error would be 2.4 times the size of the effect reported here, and §6's threshold formula allows precisely that case, so the paper contains its own counterexample to the general reading. The equality above is a consequence of independence, not a fact about conditional mutual information.

In this illustrative parameterization, then, observing group adds little mutual information relative to the effect $G$ has on the optimal decision threshold (§6), so the principal modeled consequence is decision-theoretic rather than informational. That is what the numbers support. A small information gain under one chosen parameterization does not establish that the problem \emph{is} a decision rather than an information problem — the category is not settled by two numbers.

\section{Group-conditional location, threshold risk, and the care-pathway corollary}

The location-displacement hypothesis is a statement about the \textbf{reported} location distribution $p(r \mid s, g, z)$: that it varies with a scientifically justified group variable $g$, holding the diagnosis and clinical context fixed. Reported, not felt, because reports are what every dataset contains — §3's factorization puts a whole reporting stage between the two, and §8 gives that stage its own machinery. Direct evidence requires group-stratified location data under that fixed reference; it is marked \emph{open} in the companion paper's evidence table (Shavit, 2026) because no cited study provides it. Call this \textbf{group-conditional presentation shift}, not “covariate shift” in the usual machine-learning sense, since here it is the conditional presentation law — and consequently $p(g \mid \ell ,z)$ — that is allowed to vary, not merely a marginal input distribution.

\textbf{A group difference in reports has three rival explanations, not one, and naming them is a design requirement rather than a caveat.} The canonical factorization types mechanism (c) through $p(\ell \mid s, m, g, z)$, which carries the model class $m$. Dropping $m$ and writing $p(\ell \mid s, g, z)$ does not hold $m$ fixed, because $m$ is neither the diagnosis $s$ nor the context $z$. Marginalizing puts the omission in plain sight:

$p(\ell \mid s, g, z) = \Sigma _m p(\ell \mid s, m, g, z) \cdot  p(m \mid s, g, z)$.

In words: the observable law is a mixture, so a group difference on the left can come from the components or from the mixing weights. The two are not merely hard to tell apart — they are \textbf{not identified} from group-conditional location data at fixed $s$ and $z$. Two explicit five-cell models make that concrete: a pure mechanism-(b) model whose model-class weight differs by group, 0.75 against 0.30, with identical presentation likelihoods, and a pure mechanism-(c) model with weight 0.5 in both groups and the whole difference in the likelihoods. They agree on \textbf{every} observable to 1e-12, with all four component vectors valid probability distributions (\texttt{p37\_pathway\_and\_mechanism\_typing\_2026-09-29.py}, control NC-PD-1). Stated as a condition: mechanisms (b) and (c) are separately identifiable from these data \textbf{if and only if} $m$ is observed, or the family of presentation likelihoods is constrained enough to make the mixture decomposition unique. Neither holds in the proposed design, which tells that design what to add.

The third rival explanation is the reporting node itself, and §8's own equation is enough to show it. With $E[r] = W(R s + b) + (I - W) \mu _a$, a group difference in the reporting weight $W$ or the cognitive prior $\mu _a$ produces a group difference in \textbf{reports} with no difference in felt location at all: identical felt input $(1.0, 0.5)$ and identical $\mu _a$, with reporting weights 0.80 and 0.3333, give expected reports $(0.80, 0.40)$ against $(0.3333, 0.1667)$ — 0.5217 report units of group difference from the reporting stage alone (control NC-VI4). So the repair is not to choose between the felt location and the report. It is to state the hypothesis about the report, name all three routes — a group-dependent presentation likelihood (mechanism c), a group-dependent model-class mixture (mechanism b), and a group-dependent reporting weight or prior (the reporting node) — and note that §8's two-trial anchoring design identifies $W$ and $\mu _a$ and can therefore subtract the third. The rank condition of §8 is load-bearing for this section, not only for its own.

The threshold mathematics below is a separate, well-posed decision-theory result about \emph{detection}, useful whether or not the location hypothesis above holds. With $X | D=0 \sim  N(0,1)$ and $X | D=1, G=g \sim  N(\mu _g, 1)$, expected cost at threshold $c$ is $C(c, g) = \pi _g C_{\mathrm{FN}} \Phi (c - \mu _g) + (1-\pi _g) C_{\mathrm{FP}} (1 - \Phi (c))$, where $\pi _g$ is the group's prevalence and $C_{\mathrm{FN}}$ and $C_{\mathrm{FP}}$ are the costs of a missed case and of a false alarm. For $\mu _g > 0$ the group-specific optimum is $c_g^{*} = \mu _g/2 + (1/\mu _g) \log[(1-\pi _g)C_{\mathrm{FP}} / (\pi _g C_{\mathrm{FN}})]$. The cutoff therefore moves up when false alarms or a low prevalence weigh more, down when misses weigh more, and grows more sensitive to both as the signal separation $\mu _g$ shrinks. Different $\mu _g$, prevalence, or costs generally imply different group-specific optima; a pooled threshold need not be suboptimal for \emph{every} group, so the comparison should be calculated, not assumed (a checked illustrative example: equal costs and 10\% prevalence give $c^{*} = 2.70$ at $\mu  = 1.0$ and $c^{*} = 2.12$ at $\mu  = 1.8$, \texttt{math\_checks.py / threshold\_check}). The better-separated group therefore uses a threshold 0.58 lower, in units of the healthy distribution's standard deviation — the Bayes-optimal rule calls more of that group positive rather than holding one common cutoff. Figure 5 puts both cutoffs on the distributions themselves and then sweeps prevalence and cost, including the prevalence at which the two groups' optima cross.

\begin{figure}[H]\centering
\includegraphics[width=\linewidth]{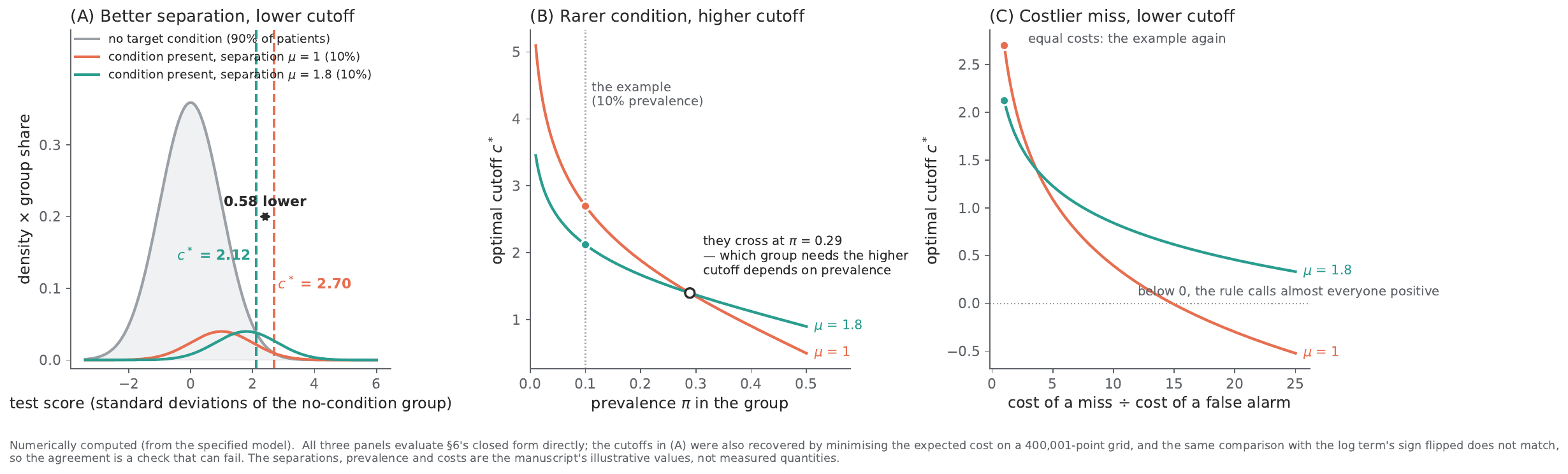}
\caption{Figure 5. The optimal cutoff is not a property of the test. All three panels evaluate §6's $c_g^{*} = \mu _g/2 + (1/\mu _g) \log[(1-\pi _g)C_{\mathrm{FP}} / (\pi _g C_{\mathrm{FN}})]$ directly. (A) §6's own checked example, on the distributions themselves: with equal costs and 10 per cent prevalence, the group separated by $\mu $ = 1.0 is best served by a cutoff at 2.70 and the group separated by $\mu $ = 1.8 by 2.12, so the better-separated group's cutoff sits 0.58 lower in units of the no-condition group's standard deviation — the Bayes-optimal rule calls more of that group positive rather than holding one common cutoff. (B) The same two cutoffs across prevalence. They cross, at $\pi $ = 0.29, solved for rather than read off the plot, which is §6's own warning drawn: which group needs the higher cutoff depends on prevalence, so a pooled threshold is not automatically wrong for everyone and the comparison has to be calculated. (C) The same two across the cost of a miss against a false alarm. Read across fields: the same eyesight test passes different people depending on whether it is for reading glasses or for a pilot's licence. [Numerically computed — the closed form, cross-checked against minimising the expected cost on a 400,001-point grid; with the log term's sign flipped that check fails, so its passing is evidence. The separations, prevalence and costs are §6's illustrative values, not measured quantities.]}
\end{figure}

For the care-pathway corollary, let $p_c(a,g)$ be detection probability under policy $a$, $q_g$ effective-treatment uptake among detected cases, $e_0$ baseline event risk, and $RRR$ relative risk reduction under effective treatment: $E[Y(a)|g] = e_0[1 - p_c(a,g)\cdot q_g\cdot RRR]$. The \textbf{standardized policy contrast under current treatment uptake}, holding $q_g$ at its currently observed group-specific value, is $\Delta _{\mathrm{current}}(g) = -e_0\cdot RRR\cdot q_g\cdot [p_c(a_1,g) - p_c(a_0,g)]$. In words, better detection reaches outcomes only through the fraction of detected patients who actually receive effective treatment: $q_g$ multiplies the whole benefit, so where uptake is small even a large detection gain moves outcomes very little.

\textbf{That equation is an identity, not an approximation — under premises worth naming, because one of them fails in the direction this paper cares about.} Given a common baseline event risk $e_0$ across the relevant cases, a common $RRR$, an uptake rate that does not depend on \emph{which} cases the policy detects, no path from detection policy to the outcome except through treatment, and false-positive harms kept outside $Y$, the displayed formula is exact to $1.1 \times  10^{-16}$ at three different detection probabilities (\texttt{p37\_pathway\_and\_mechanism\_typing\_2026-09-29.py}, control NC-PD-2). It is a decomposition with premises, not a general causal theorem.

The first two premises are homogeneity assumptions, and relaxing them has a computable cost with a known sign. Index cases by severity $v$. The exact quantity is $E[e(v)] - E[e(v)\cdot p(v)\cdot q(v)\cdot r(v)]$, so the displayed formula's error is \textbf{exactly} the gap between a product of averages and an average of products — a covariance across cases between baseline risk, detectability, uptake and treatment effect. Where detectability is independent of severity that covariance vanishes and the attribution is clean: the measured bias is $2.8 \times  10^{-17}$ events per case. But a more sensitive policy reaches \emph{milder} cases, which raises average detectability while the cases it adds carry below-average baseline risk, so multiplying the averages credits the policy with preventing events that were never going to happen. \textbf{The sign is the transportable part; the size is not.} In one constructed severity-indexed population — baseline risk rising linearly with severity, detection falling with it — a policy contrast of exactly that shape gives an exact effect of $-$0.002300 events per case against a displayed $-$0.003829, so the formula overstates the benefit by about two thirds of the true effect (control NC-PD-3). That number belongs to that severity distribution and to no other; what carries across populations is the direction, and the direction is the unwelcome one for a paper arguing for more sensitive detection. A stratified or individual-level pathway model is what would remove it.

The contrast above is the correct name for what this paper's illustrative model computes; it is \emph{not} automatically a controlled direct effect in the potential-outcomes sense, which would instead require a fixed mediator level — here the treatment $t$ — and identification assumptions for $CDE(t) = E[Y(a_1,t) - Y(a_0,t)]$. If the graph gives detection policy no path to outcome except through treatment, that direct effect is zero by construction and is not an empirical discovery. Take the illustrative model, whose parameters are set from the trial below rather than fitted to it. Fixing detection alone prevents just \textbf{1} adverse event per 1,000 (240 $\to $ 239). Fixing detection \emph{and} treatment uptake prevents \textbf{15} (240 $\to $ 225) — a \emph{reading} of the observed null (Lee et al., 2019), not a quantity that trial itself estimated, and a statement about the care pathway downstream of diagnosis, not about the location hypothesis above. All three counts are \textbf{synthetic}: they are an illustration parameterized from that trial's observed null, not a re-analysis of it and not the output of any fitting or calibration procedure — no such procedure is documented here, and the word \emph{calibrated} would claim one. They come from the mediation-events routine in \texttt{decision\_fairness.py}, and companion Figure 4, panel (c) — the care-pathway panel of the prediction atlas — plots them. Lee et al. supply the pattern the model's parameters were set to reproduce, the rise in detection with no matching rise in treatment or outcome; they do not supply these counts. One consequence of the identity above is worth stating against this model's own interest, and it is about external validity rather than arithmetic. The routine holds baseline risk, treatment effect and uptake at single scalar values for every case, so its population is homogeneous by construction and the displayed formula is \textbf{exact} for it — these three counts carry no composition error at all. What they carry instead is the homogeneity premise. If the real population is heterogeneous in the way the covariance above describes, with a more sensitive rule reaching milder cases, then these counts overstate what the policy would achieve, and by a margin the model as written cannot estimate.

\section{Delocalized amplification as a neural-field bifurcation}

An Amari (1977) neural field with additive noise $\tau  \partial u/\partial t = -u + (W_s * f(u)) + I + \eta $ — in words, activity decays locally, is rebuilt or suppressed by recurrent spatial interactions, receives a localized external drive, and is continually perturbed by noise — with membrane time constant $\tau $, sigmoid $f$, localized input $I$, noise $\eta $, and a Mexican-hat kernel $W_s = A_e K_e - A_i(s) K_i$ whose inhibitory amplitude falls as the disinhibition parameter $s$ rises. Writing the kernel with its $s$ rather than leaving $s$ out of the displayed equation matters, because the parameter enters \emph{inside} the kernel and not as a multiplier in front of it: the simulations turn $A_i(s)$ down linearly from 0.95 at $s = 0$ to 0.20 at $s = 1$ (\texttt{neural\_field.py}), so $s$ scales inhibition down rather than scaling the whole kernel up. §9 below sweeps a second, kernel-normalized parameterization in which $s$ multiplies the whole kernel; the two agree on where the transition sits qualitatively and disagree on its numeric value, and §9 says so where it uses it. Here $s$ is the \textbf{control} parameter — the quantity being turned, analogous to temperature in a phase transition — and spatial extent is the \textbf{order} parameter, the quantity that changes character at the critical value. (Dynamical systems calls this transition a bifurcation and statistical physics calls the analogous change of macroscopic character a phase transition. The vocabularies run parallel and the terms are not interchangeable: what is modelled here is a bifurcation, and the order-parameter language is borrowed from the physics.) As $s$ crosses a critical value ($s^{*} \approx  0.82$ in the illustrative kernel), the stable localized bump loses stability: spatial extent jumps and temporal instability peaks at the transition (marginal stability), then the field saturates. Thus spatial \emph{extent} is the order parameter, while spatial \emph{instability} is the marginal-stability signature that peaks at the crossing. The two are distinct quantities and only instability is non-monotone in $s$.

\begin{figure}[H]\centering
\includegraphics[width=\linewidth]{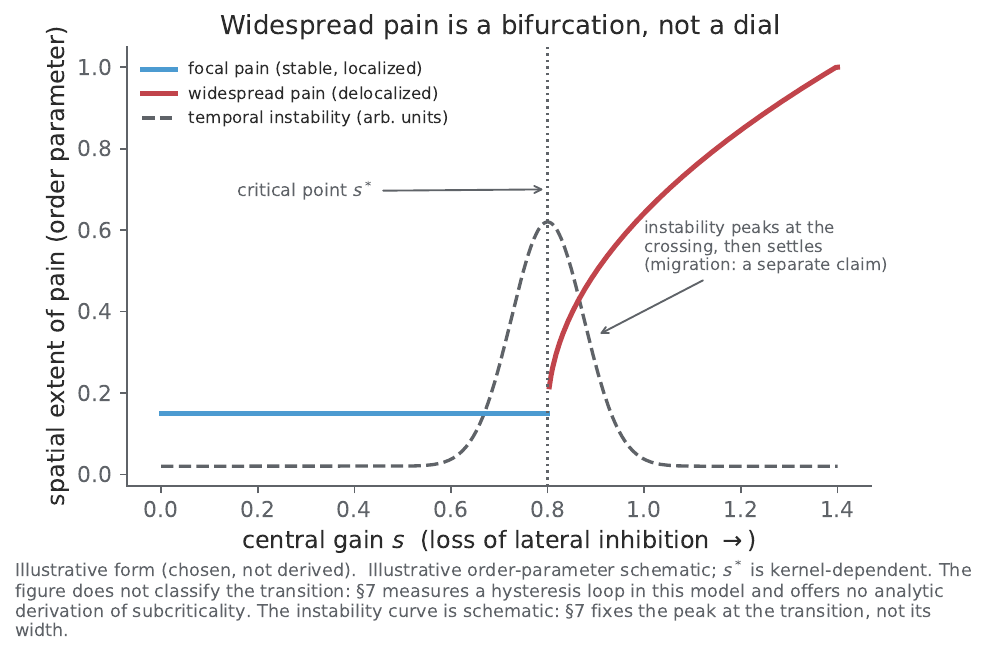}
\caption{Figure 6. Widespread pain as a bifurcation. Spatial extent, the order parameter, jumps at a critical central gain $s^{*}$ rather than rising smoothly with it. The dashed curve is a distinct quantity: a critical-slowing instability proxy that peaks at the crossing and settles as the field saturates. That the proxy peaks there is a derived, near-transition property of the linearized dynamics. It is not a migration claim: whether day-to-day spatial \emph{migration} tracks it needs the further assumption that the destabilizing mode is a displacement mode (§7.1), and the prediction that follows from that assumption is a \emph{decline} on the high-gain side, not the intermediate peak an earlier draft of this paper expected. Read across fields: a tipping point, not a dial — “critical” names the threshold, not clinical severity. One thing this schematic deliberately does not settle: $s^{*}$ here is the marginal-stability point of a branch, read off the linearization, and §7 measures the realized transition in the same field as \emph{hysteretic} — path-dependent, so which state the field occupies at a given gain depends on the direction it arrived from. The single crossing drawn here is the local stability picture, not a claim that the transition is supercritical; no analytic derivation of subcriticality is offered either way (§7). [Illustrative form — a schematic local stability picture, not the field's measured transition; $s^{*}$ is kernel-dependent, and the instability curve is schematic in width though not in where it peaks.]}
\end{figure}

\textbf{What this linearization does and does not establish.} Three claims carry different evidential weight and should not be read as one theorem — and the first of them has to be split again, because one part needs nothing but the linearization while two parts need the noise model as well. \emph{Derived from the linearized dynamics alone}: the leading mode's recovery time is $1/|\lambda _{1}(s)|$, so a perturbation takes longer and longer to decay as the eigenvalue $\lambda _{1}(s)$ approaches zero. Nothing stochastic enters there. \emph{Derived from the linearization together with the additive white noise the field equation already carries}: the leading mode is then an Ornstein–Uhlenbeck process, and its two early-warning statistics have closed forms — a stationary variance $\sigma ^{2}/(2|\lambda _{1}|)$ and a lag-$\Delta $ autocorrelation $\exp(-|\lambda _{1}|\Delta )$, both growing without bound as the eigenvalue approaches zero. Together the three are critical slowing, with no migration claim attached.

The split is not bookkeeping. A noise-free linearization has stationary variance \textbf{zero at every eigenvalue}, so "variance rises near the transition" is as much a statement about the driving noise as about the eigenvalue; and the lag autocorrelation exists as a statistic at all only because the mode is a stationary random process, even though its functional form turns out to be exactly the deterministic relaxation kernel. The split also tells a study which statistic to prefer. The variance scales with the \textbf{square} of the noise intensity — double $\sigma $ and it quadruples — while the autocorrelation does not contain $\sigma $ at all. So a cohort whose measurement noise varies between patients, or drifts across the follow-up, can still be read through the autocorrelation and cannot be read through the variance. \emph{Numerically illustrated}, under the stated kernel, noise, and metric: the chosen simulation produces a focal-to-widespread transition with an interior instability peak. \emph{Hypothesized}, and not derived from the stated bifurcation analysis alone: that this instability corresponds to day-to-day spatial \emph{migration} specifically. The further guess that migration \emph{peaks} at intermediate sensitization is the one §7.1 goes on to examine directly and then withdraw — the reduction that supplies a migration prediction licenses a \textbf{decline} on the high-gain side and no practically detectable intermediate rise, so the peak is not this paper's claim and is not retained anywhere below. The susceptibility formula below (§9) peaks at $s^{*}$ \textbf{by construction}: its denominator is built to vanish there. So sweeping kernels and finding $s^{*}$ robustly interior (§9, Figure 15A) shows that the \emph{critical point's location} is robust — not that migration peaks there. That would need the destabilizing eigenmode itself to be a spatial translation or deformation mode rather than a pure amplitude/spread mode, which the simulation should report explicitly and which this paper does not independently verify. What the paper carries forward as its falsifiable migration claim is therefore §7.1's, not a non-monotone one: \textbf{conditional on the stated sufficient conditions, finite-window migration declines on the high-barrier side, and the model does not predict a practically detectable intermediate rise.} The companion paper's evidence table (Shavit, 2026) records the same downgrade in its status column.

One measurement choice here is worth stating rather than leaving in the code, because it is the place where this prediction was most exposed and it did not survive its first test. The most direct measure of migration is the variability of the activity centroid's angle. That quantity is only defined while the activity has a direction: once the field saturates, the ring is uniform, the resultant vector length — the magnitude of the circular mean vector, and so a measure of how well any direction is defined at all — goes to zero, and the centroid angle is the phase of numerical noise, not a location. Unguarded, it returned its largest value at maximum gain — precisely where the intermediate-peak prediction is most vulnerable, and it would have read as a refutation. It was not migration. We therefore report the centroid measure only where the resultant length clears a floor, report that resultant length beside it so a reader can see where the question is well posed, and leave the entry undefined at saturation rather than deleting it. The instability measure, which carries the prediction, is defined throughout and is unaffected. Diagnostic and controls: \texttt{neural\_field\_metric\_check.py}.

The resultant-length guard reports only that the centroid angle is undefined at saturation; it does not measure migration there. To ask a well-posed question at saturation, the analysis has to replace the directional centroid with a statistic defined on the whole activity profile. Migration is mass moving across the body map, so we measure it as the \textbf{1-Wasserstein (earth-mover) distance between successive normalized activity profiles}. On a ring it has the closed form $W_{1} = \min_c \Sigma _x |F(x) - G(x) - c|$, \textbf{summed over all $N$ ring sites}, where $F$ and $G$ are the two profiles' \textbf{cumulative} mass around the ring and the minimising $c$ is the median of $F - G$. Operationally, the statistic compares the two cumulative profiles after removing whichever circular offset makes their $L^{1}$ discrepancy smallest. The sum rather than an average is what puts the answer in ring sites: on a unit-spaced ring the sum is the grid-scaled integral, and averaging over the $N = 64$ sites instead would report every distance in this section sixty-four times smaller. That is equation (11) of Rabin, Delon and Gousseau (2011), who write it as $\|F - G - \mu \|_{1}$ with $\mu $ the median of the values $F[i] - G[i]$, and who credit the formulation itself to earlier work on the circle rather than claiming it. It is a standard construction. Its \emph{use} here is not standard, and a targeted search found no precedent for it: optimal transport enters neuroscience to compare activity between conditions, subjects or sessions, and we found no source using a per-timestep or window-net transport distance as a drift statistic for a spatially distributed pattern, nor circular transport applied to a periodic neural variable's evolution. For exactly this question the field uses centroid tracking and population-vector shift — the first of which is the measure just shown to be undefined where the field saturates. We state that as an absence found by one search, not as a claim of priority. It is defined wherever the profile carries mass, so a uniform ring is a distribution rather than a degenerate direction; it reads in ring sites, so the number is a distance on the body map; and it was adopted in the belief that it separates spreading from moving, which the centroid angle confounds and extent cannot see. \textbf{That belief is wrong, and the control that shows it is below.} The controls it does pass: zero against itself, zero for uniform against uniform — the configuration where the angle statistic returned its column maximum — and exactly $k$ for a bump rigidly shifted $k$ sites. That last control is an identity rather than a tolerance the code was tuned to meet. Under a rigid shift the cumulative difference $F - G$ at site $x$ is just the mass lying in the $k$ sites ending at $x$, so each unit of mass is counted for exactly $k$ values of $x$ and the sum is $k$. The identity is exact under a condition worth stating, because it is necessary and not decorative: the profile's support plus the shift must stay inside half the ring, which is also what makes the median offset zero. Past the point where the profile overlaps itself, transport is correctly \emph{cheaper} than rigid translation because mass splits both ways.

Under that metric a large signal appears at the transition, in a quantity the earlier one could not name: the \textbf{window-net transport score}, which aggregates over the observation window rather than reporting step-to-step variability. It is called a transport score and never a displacement or a movement, here and everywhere below, because the widening control further on shows it cannot tell transport from spread. It is 0.05–0.10 sites at almost every sensitization and rises to \textbf{4.9 sites at $s = 0.81$}, a fiftyfold jump confined to the transition; every one of the five seeds peaks at $s = 0.81$ or $s = 0.82$, adjacent points on the 0.01 grid, three at the first and two at the second (Figure 7A). Two grids agree on the shape and disagree only in resolution: a coarse sweep stepping $s$ by 0.1 straddles the band and reads its edge as 2.3 sites at $s = 0.80$, while the fine sweep that draws Figure 7A steps by 0.01 and resolves the 4.9. By $s = 1.0$ the transport score is exactly zero. The collapse is nearly, not quite, complete one step earlier: at $s = 0.9$ the median realization scores 0.00 sites while 29 per cent still score above 0.25, so the signal \textbf{collapses across} $s = 0.9$ rather than having already gone at it. Per-step transport runs the other way, peaking at \emph{low} sensitization and falling to exactly zero at saturation — the field jittering while it is small, which is not migration. The window is a band rather than a knife edge: averaged over the five seeds, the transport score exceeds one ring site from $s = 0.80$ to $s = 0.85$ and nowhere else on the swept axis, about six per cent of it. Outside that band the seed-averaged score never reaches a third of a site, though single realizations at $s$ = 0.71, 0.76, 0.78 and 0.79 do touch half a site, so the band's edges are soft rather than sharp. \textbf{Status: numerically observed in the simulation} at the model's own parameters, on five seeds, with controls that are exact — zero against itself, zero for uniform against uniform, and exactly $k$ sites for a bump shifted $k$ sites, the last derived above rather than fitted. The metric is exact; the peak's location is a property of this parameterisation and is not a physical constant.

\begin{figure}[H]\centering
\includegraphics[width=\linewidth]{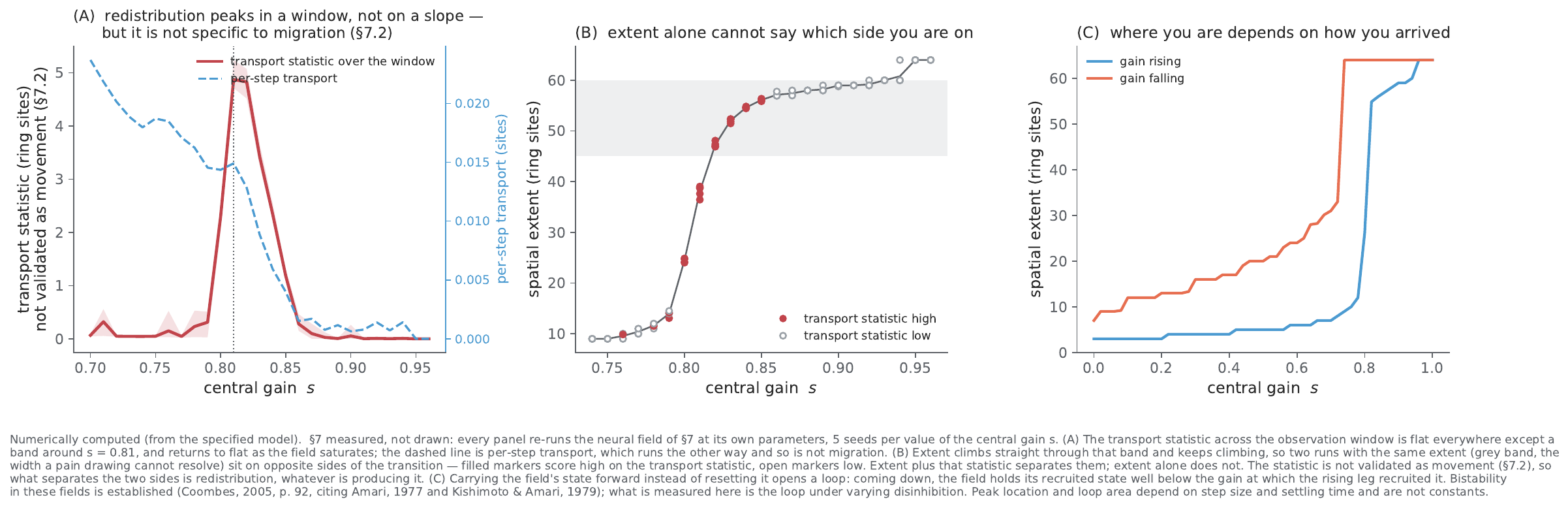}
\caption{Figure 7. The three §7 results, measured rather than described. Each panel re-runs the same neural field at the same parameters, five seeds per value of the central gain s. (A) The window-net transport/redistribution statistic is flat everywhere except in a band around s = 0.81, and flat again once the field saturates — a redistribution signal confined to a window, not a slope that rises with severity; the body's own control shows this statistic tracks width change and cannot by itself establish that the pain map moved. The dashed line is per-step transport, which runs the opposite way and is the field jittering while it is small, not a measure of relocation. (B) Spatial extent climbs straight through that window and keeps climbing, so two runs with the same extent (grey band, the width a pain drawing cannot resolve further) can fall on opposite sides of the transition: filled markers cross the redistribution band, open markers do not, and no value of s appears in both groups. Extent together with the redistribution statistic separates them; extent alone does not — though the redistribution statistic alone does not establish that the pain map moved (§7). (C) Carrying the field's state forward instead of resetting it opens a loop — coming down, the field holds its recruited state well below the gain at which the rising leg recruited it, so the same measurement means different things depending on which direction the patient arrived from. [Numerically computed — the §7 field at its own parameters; the peak's location and the loop's area depend on step size and settling time and are not constants.]}
\end{figure}

\textbf{A control refutes the reading, and it is the reason the last paragraph says “a large signal” rather than “migration.”} The measure was adopted because a bump that widens in place should transport mass only a short way, while a bump that translates should transport all of its mass the whole way. Test that directly: take a flat-topped profile, widen it from 21 to 30 ring sites symmetrically about the same site so that the centre of mass does not move at all — displacement exactly $0.00$ — and score it on the same circular transport distance. \textbf{It scores 2.24 ring sites}, which is thirty times the 0.05–0.10 off-transition baseline, and more than a rigid two-site shift scores (2.00). A Gaussian profile widened the same way scores 2.74. The exact number depends on the profile's shape; that a stationary peak scores like a travelling one does not.

Worse, this is not hypothetical for the runs above. Printing the centre's travel alongside the transport score at $s = 0.8$, seed by seed, gives 0.45, \textbf{0.03}, 0.06, 0.95 and 1.01 ring sites of centre movement against transport scores of 2.32, \textbf{2.03}, 2.53, 2.05 and 2.56 — while the width grows from about 20 sites to about 29 in every seed. \textbf{The score tracks the width, not the centre.} One seed scores 2.03 on a centre that travelled three hundredths of a site.

\textbf{Status: numerically observed in the simulation, and it observes the wrong thing.} The transport peak is a real and reproducible feature of the field at the transition, and it is \emph{not} evidence that the pain map moved, because this statistic cannot tell movement from spread at the widths the field actually reaches. That does not show the migration prediction is false. It shows this measure cannot establish it, which leaves the caution §7 already stated — that the destabilizing mode must be a translation mode rather than a pure amplitude or spread mode — exactly where it was. Control: \texttt{neural\_field\_widening\_control.py}, which asserts its own result and fails loudly if it stops holding. Scripts: \texttt{neural\_field\_transport.py} and \texttt{neural\_field\_moved\_test.py} for the coarse sweep and the detection rates; \texttt{make\_a5\_figure.py} for the fine sweep over five seeds that resolves the peak, which calls the same $transport_{\mathrm{run}}$ entry point; and \texttt{p37\_transport\_normalization\_check\_2026-09-29.py}, which derives the rigid-shift identity from the implementation as written and checks the closed form against an independent optimal-transport solver that shares none of its algebra.

The clinical form of that result is a two-variable claim, and it is the one worth testing, because \textbf{extent rises monotonically through the migration window and keeps rising past it}. Hold extent fixed at a band a drawing cannot resolve further, 45–60 sites in the model's units — a bin width chosen for illustration, which a real study would instead set from the drawing instrument's own test–retest and spatial-resolution error. Realizations with a nonzero transport score fall at $s$ = 0.82–0.85 and those with a zero one at 0.86–0.94, with no value of $s$ in both groups; widening the band past 60 sites admits exactly one, $s = 0.94$, where two realizations in twelve score above zero despite saturation. So extent alone cannot separate the transition from saturation — two patients with equally widespread pain fall on opposite sides of it — while extent together with the transport score can (Figure 7B). Both variables come from two pain drawings and a region count, and no new instrument is required. \textbf{Status: numerically observed in the simulation, and inheriting the caveat above} — the score separating the two groups is the same transport statistic, so the separation may be reading a change in width rather than a change in place. The two-variable measurement — spatial extent plus longitudinal profile change — survives the doubt in a weaker form, because a drawing shows extent and a change of extent equally well; what does not survive is calling the second variable \emph{movement}. The completeness of the separation is a property of the model's noise level and is not a predicted sensitivity.

The interior placement of the critical point is not tied to the illustrative kernel. Two different checks say so, and the distinction between them is worth stating because one sentence of an earlier draft ran them together. The critical ridge $s^{*}(\rho ) = 1/\max_q \hat{W}(q)$ comes from the \textbf{analytic} linear-stability expression, and it is \textbf{evaluated numerically} across a grid of kernel widths (§9, Figure 15A) — an analytic formula sampled on a grid, which is evidence of robustness over the swept family rather than a proof for all kernels. A separate \textbf{stochastic} sweep then reproduces the same interior placement. The stochastic check was run over a two-dimensional family of excitation and inhibition widths. It keeps the peak interior in every cell for inhibition-to-excitation ratios from 1.9 to 4.1, and $s^{*}$ rises monotonically with the ratio, as the analytic ridge predicts. Above a ratio of about 5 the critical gain exceeds the maximum this parameterization allows, so the transition is never reached; there the prediction is untestable, not false. Nor is it tied to the noise model. The interior peak is, first, an analytic property of the deterministic linearization — the marginal-mode susceptibility $\chi (s)$ peaks at $s^{*}$ (§9) with no noise term at all — and, second, reproduced by the stochastic simulation, where the noise $\eta $ is additive, spatially uniform, and white. That simulated peak survives both reseeding and a state-scaled (multiplicative) noise model in which the variance falls as the field saturates (the robustness sweep in the supplement: the peak stays at $s^{*} = 0.82$ in every seed and under both noise models, at a height of 0.123 additive and 0.128 state-scaled — the location is unchanged and the height moves by about 4\%, so the noise model perturbs amplitude far less than it would have to in order to threaten the location claim). The peak is sharp, not a broad shoulder. It rises about two orders of magnitude above baseline and returns to baseline within 0.04 of $s$, so any grid coarser than about 0.02 straddles it and understates its height. The one alternative the ascending sweep cannot see is a subcritical, hysteretic transition. Descending the same grid with the field state carried forward rather than reset now shows one: the two legs open a loop in every seed tried (Figure 7C), and coming down the field holds its fully recruited state to $s = 0.74$, where the ascending leg is still focal. \textbf{The transition is hysteretic in this model.} Bistability and hysteresis in neural fields are established phenomena and are not claimed as new here; what is narrower and numerical is the hysteresis loop this parameterization produces when lateral inhibition is varied. Coombes' review states it directly: \emph{“Amari (1977) was able to find localised stable pulses that are bistable with a homogeneous steady state, assuming a Heaviside firing rate function and Mexican hat connectivity”} (Coombes, 2005, p. 92). The same passage notes that Kishimoto and Amari (1979) extended the result to a smooth sigmoidal firing rate in the high-gain limit, which is the case this model is in. Coexisting stable states are the structure a hysteresis loop expresses. What a targeted search did not find is the sweep run here: bistability in these fields is established in the literature for \textbf{gain, threshold and input amplitude}, and we found no source that varies \textbf{disinhibition} — the lateral-inhibitory weight itself — and exhibits the loop. The mechanism transfers, since inhibition enters the same kernel integral as gain does, so the contribution is the measurement in this parameterisation rather than the phenomenon. Search record: the literature-search log in the supplement, 2026-08-29.

The hysteresis changes what a cross-sectional measurement can conclude — the state at a given sensitization depends on which direction the patient arrived from, so a widespread and \emph{stationary} pain means saturation on the way up and can mean a much lower sensitization on the way down. That is a longitudinal claim, and it qualifies the two-variable reading above rather than overturning it. \textbf{Status: numerically observed in the simulation}, three seeds, and now checked directly against the alternative explanation. The sweep on its own could not separate genuine bistability from slow relaxation, because a loop whose area depends on settling time is exactly what a slow relaxation also draws. Run instead at \textbf{fixed} $s$ from two initial conditions — the focal state and the fully recruited one — the deterministic field settles to two \emph{different} steady states for $s$ from 0.567 to 0.937, every run converging to a residual below $10^{-14}$. Outside that interval the two initial conditions converge to the same state, which is the control that makes the split inside it mean something. And the sweep's loop area moves by only 7 per cent when the settling time per step is lengthened tenfold, so the loop it draws is almost all static and hardly any of it rate. The coexisting region is \textbf{wider} than the descending leg suggested: the sweep held its recruited state down to $s = 0.74$, while the true lower edge sits at 0.567, because a finite-rate sweep does not linger long enough to find where the far branch really ends. No analytic derivation of subcriticality is offered, and none is claimed: this is a numerical location of two coexisting fixed points of a 64-dimensional system, not a proof. Scripts: \texttt{neural\_field\_hysteresis.py} and \texttt{research/validation/p37\_cf3\_static\_bistability\_2026-09-29.py}.

\subsection{A conditional displacement-mode model, not a derived reduction of the simulated field}

The stability analysis above establishes critical slowing and stops there: it says the leading mode gets slower near the transition, not that the mode moves the pain map. This subsection asks what would follow \emph{if} it did. The derivation has an unexpected shape — the prediction it delivers is the opposite end of the curve from the one this paper originally wanted.

Suppose the destabilizing mode is a spatial \textbf{displacement} coordinate, so a single number $z$ says how far the felt location falls from the midline between two candidate sites. Near a symmetry-breaking bifurcation — a \emph{pitchfork}, where one stable location gives way to two mirror-image ones — a center-manifold reduction — approximating the high-dimensional dynamics near the bifurcation by the few slow modes whose stability is changing — puts that coordinate's dynamics in a standard form. Written as a stochastic differential equation, with $\alpha $ increasing as central gain carries the system past the transition, $\beta  > 0$ keeping excursions bounded, and $\sigma $ the noise amplitude:

$dz = [\alpha  z - \beta  z^{3}] dt + \sigma  dW$,  $\beta  > 0$.

In words: the map is pushed away from the midline in proportion to how far past the transition the system is, pulled back by a cubic restoring term, and jostled by noise. Below the transition ($\alpha  < 0$) there is one preferred location. Above it there are two, at $z\pm  = \pm \sqrt{\alpha /\beta }$, separated by a barrier of height $\alpha ^{2}/(4\beta )$ in the potential $V(z) = \beta  z^{4}/4 - \alpha  z^{2}/2$. \textbf{The barrier grows as the square of the distance past the transition}, and that single fact drives everything below.

Day-to-day migration, on this reading, is the map hopping between those two locations. When the barrier is high relative to the noise, the classical Kramers escape rate gives the hopping frequency — an asymptotic approximation in which rare noise-driven switches occur at a rate exponentially suppressed by the barrier height relative to the noise, and therefore least trustworthy exactly where switching is not yet rare; multiplying it by the squared distance between the sites gives a proxy for observed displacement:

$M(\alpha ) \propto  \alpha ^{2} \exp[-\alpha ^{2}/(2 \beta  \sigma ^{2})]$.

That expression rises and then falls, because two effects pull against each other. Past the transition the two locations move further apart, which increases displacement, while the barrier between them grows quadratically, which makes the hop exponentially rarer. Differentiating gives an interior maximum at $\alpha _{\mathrm{peak}} = \sqrt{2\beta } \sigma $. Every step to here is exact and was verified symbolically (\texttt{p37\_pitchfork\_A6\_verification.py}, checks P1–P6).

\textbf{The interior maximum is real in the approximation and negligible in the process.} The proxy above rests on a rare-switch asymptotic that is weakest exactly where the peak falls. The stochastic differential equation was simulated directly instead, by Euler–Maruyama over three noise levels and eight seeds each. The maximum survives, but it is negligible: on a fifteen-point grid refined near the peak, every run climbs to it by \textbf{under 5\%} above the low-gain value and falls past it by \textbf{over 25\%}. Those two thresholds, and not the observed extremes, are what the verification script asserts. They are deliberately coarser than what any one seed set delivers, because the smallest climb and the largest fall in a set of twenty-four runs are sample extremes, and a later reseed moves them; a threshold a reseed cannot move is what a claim in the text can rest on. In the phone-based design this paper envisages, a rise of a few per cent is much harder to detect than a decline of a quarter or more — how much harder is a question of measurement error, cadence and sample size, and §7.2 states that no power calculation is offered for it. The simulated maximum also falls well below $\sqrt{2\beta } \sigma $, by a factor that is itself unstable across seeds, so that location should not be quoted as a target value.

\textbf{What this licenses the paper to claim.} Under the sufficient conditions below, the model predicts that finite-window spatial migration \textbf{declines} as central gain carries the system onto the high-barrier side of the transition. The reduction does not license a detectable intermediate peak. The two are easy to conflate because one curve contains both, and the earlier framing took the peak — the part of that curve the mathematics supports least. The size of that decline, though not its direction, turns out to depend on the coordinate running on an open line; the subsection below closes the domain and finds the same decline displaced to higher central gain.

The conditions are \textbf{sufficient, not necessary}: other mechanisms could produce the same decline, and failing one of them does not establish that migration is flat.

\begin{enumerate}\setlength{\itemsep}{2pt}\item the reduced coordinate represents spatial location or deformation, not amplitude or extent;\item two alternative spatial attractors exist past the transition;\item transitions between them are noise-driven on the observation timescale;\item the observation window is finite;\item central gain moves $\alpha $ through the relevant range within the study period.\end{enumerate}

Condition 1 carries the empirical risk, and it is the same condition §7 already flags: whether the destabilizing eigenmode is a translation mode is an open question about the pain field, not something this normal form settles. Conditions 3 and 4 act together — at low noise the simulated displacement curve is flat across the whole low-gain region before it falls, so a study sampling too infrequently, or a patient whose map is too stable, would see the decline with no structure before it.

Two things this subsection deliberately does not do. It does not replace the susceptibility curve of §9, which is a statement about the deterministic linearization and peaks at $s^{*}$ by construction. And it does not establish that the pain field has a displacement mode at all. This subsection changes the status of the migration claim: from a hypothesis with no derivation behind it, to a directional prediction that follows from stated assumptions, carrying the honest note that the model's headline feature is its falling limb rather than its peak.

\subsection{Two measures of migration, and which one the claim rests on}

§7 measures migration in the full neural field, as mass moving across the body map. §7.1 measures it in a one-number reduction of that field, as the map hopping between two locations. They appeared to disagree, and the disagreement is now resolved — not by choosing between them, but because the control above disqualifies one of them as a measure of movement at all.

\textbf{They never disagreed about saturation, which is the claim the paper leans on.} The full-field transport score is exactly zero at $s = 1.0$ — the score, not a displacement, since the widening control below disqualifies that reading of it. The reduction's proxy falls by over a quarter past its maximum, on every run, because the barrier suppressing hops grows as the square of the distance past the transition. What that licenses is a claim about \emph{change}, not about a single visit: a decline in migration from a previously mobile state is the directional prediction both accounts share. A single stationary observation is not enough to place a patient on the gain axis, and we add two reasons of our own. At low noise the simulated displacement curve is flat across the whole low-gain region before it falls (§7.1, condition 3–4), so "not moving" is also what a low-gain map looks like. And the transition is hysteretic (§7), so a widespread stationary pain means saturation on the way up and can mean a much lower sensitization on the way down. State cannot be read cross-sectionally without history.

\textbf{They appeared to disagree about the rising limb, and that appearance was an artefact.} The reduction finds a climb of under 5 per cent into its maximum on every run, which is unlikely to be practically detectable at the phone-based measurement precision and sampling cadence this paper envisages. Detectability is a function of effect size, measurement error, cadence and sample size, so that is a statement about the envisaged design and not a theorem; no power calculation is offered for it. The field's transport statistic finds a fiftyfold jump. But the widening control shows that statistic scores a stationary, spreading peak as though it had travelled, and that in these very runs the centre moves as little as 0.03 sites where the statistic reads 2.03. The fiftyfold jump is the field \emph{recruiting}, which extent already measures, rather than the map moving.

\textbf{So the paper's migration claim is §7.1's.} Under the sufficient conditions listed above, finite-window spatial migration \textbf{declines} as central gain carries the system onto the high-barrier side of the transition. The model does not license a detectable intermediate peak. The falling limb is the testable feature; a climb of a few per cent is not. The subsection below qualifies how large that limb is and where it falls, without disturbing its direction.

§7's transport work therefore contributes something narrower than it first appeared, and still worth keeping. It establishes that the centroid angle is undefined where the field saturates, so the field's standard drift statistic cannot be used at the sensitizations that matter. It replaces \emph{extent} dynamics with a well-posed alternative. And it produces the hysteresis result below, which is measured on extent rather than on transport and is untouched by the control. The control below removed this paper's most striking number. Reporting it rather than burying it is why the numbers that remain can be trusted.

\subsection{The same model on a closed body surface, and where its prediction actually lives}

The normal form's coordinate can run away. Past the transition the two attractors sit at $z\pm  = \pm \sqrt{\alpha /\beta }$, a separation that grows without limit as central gain rises. A body does not allow that. The trunk is a cylinder, so two felt locations are at most half a circumference apart, and beyond that range the normal form is a local approximation being read far outside the neighbourhood where it is local. This subsection closes the domain and asks what changes.

Replace the coordinate with an angle around the body. The replacement potential has to do two things at once: look like the normal form's near the midline, and have exactly \textbf{two} resting places on the circle. Two is what a pitchfork delivers, and a mirror-image pair of felt locations is what those two mean on a body. The first requirement is easy to meet. The second is easy to \emph{believe} you have met, and the way it fails is worth a paragraph, because we spent a round of this analysis on the wrong side of it.

\textbf{The obvious completion has four wells, not two.} Substituting $\sin \theta $ for the normal form's coordinate gives $V(\theta ) = -(\alpha /2) \sin^{2}\theta  + (\beta /4) \sin^{4}\theta $. It matches the normal form near the midline, it stays bounded, and it looks like the same double well wrapped onto a circle. It is not. $\sin^{2}\theta $ repeats every half turn, so $\sin^{2}\theta  = \alpha /\beta $ has four solutions on a full circle — $\theta ^{*}$, $\pi  - \theta ^{*}$, $\pi  + \theta ^{*}$ and $2\pi  - \theta ^{*}$ — and the second derivative is positive at all four. That landscape has four attracting locations, and its outer barrier $(\alpha -\beta )^{2}/(4\beta )$, which reads naturally as a \emph{second route between the same two places}, is in fact the barrier into a \emph{third} place. Anything measured on it describes a four-attractor system. The failure is reproduced, symbolically and by brute-force minimum counting, as check W1 in \texttt{p37\_periodic\_domain\_A5\_3.py}, which passes by exhibiting the four wells rather than by hiding them.

\textbf{A completion that does have two.} Build the potential from $1 - \cos \theta $ instead:

$\widetilde{V}(\theta ) = -\alpha (1 - \cos \theta ) + \beta (1 - \cos \theta )^{2}$.

What makes this work is a property of the substitution rather than of the algebra. $1 - \cos \theta $ climbs monotonically from 0 at the front midline to 2 at the back, so it takes each value once on a half circle where $\sin^{2}\theta $ takes it twice. Written in $u = 1 - \cos \theta $ the potential is the ordinary quartic $-\alpha u + \beta u^{2}$, which has one interior minimum — and one interior minimum in $u$ is the mirror-image pair $\pm \theta ^{*}$ on the circle, and nothing else. Near the midline $\widetilde{V}$ reproduces the normal form's $-(\alpha /2)\theta ^{2}$ exactly and stiffens its quartic term to $\beta  + \alpha /6$, a quarter of the stiffening the four-well completion imposed. A closed surface does resist large excursions more than a line does, and this model still says so; it says it far more gently than the discarded one did.

\textbf{The wells separate until they meet at the back.} They sit where $\cos \theta ^{*} = 1 - \alpha /(2\beta )$, so they start together at the midline and move apart as central gain rises, reaching half a circumference at $\alpha  = 4\beta $. Below that point the barrier over the midline is $\alpha ^{2}/(4\beta )$ — identical to the normal form's, so nothing in the short-route physics changes by closing the domain.

\textbf{A second way out opens, between the same two places.} On a circle the pair is joined twice: straight across the front midline, or the long way round the back. Those barriers are $\alpha ^{2}/(4\beta )$ and $(\alpha  - 4\beta )^{2}/(4\beta )$. They are equal at $\alpha  = 2\beta $, and past that the far route is the cheaper one. At $\alpha  = 4\beta $ the far barrier is \textbf{exactly zero} while the midline barrier is $4\beta $. So a second escape becomes free precisely as the first becomes expensive — and, unlike the four-well completion, both routes run between the two locations the model is actually about.

\textbf{Past that point there is only one place left.} At $\alpha  = 4\beta $ the two wells do not merely stop separating; they meet at the back of the trunk and merge. Above it the curvature at the far side, $\alpha  - 4\beta $, is positive, so the circle carries a single well there and no second location to move to. The model therefore predicts that between-location migration at high central gain does not fade gradually — it stops. That is a sharper and more falsifiable claim than a four-well landscape can support, because there the map always has somewhere else to go.

Those are statements about the landscape. What a study would measure is how far the map actually moves, so both models were simulated with the same integrator, the same noise levels and the same lagged-displacement measure, with circular distance substituted where the domain is a circle:

{\footnotesize\begin{xltabular}{\textwidth}{>{\raggedright\arraybackslash}X>{\raggedright\arraybackslash}X>{\raggedright\arraybackslash}X}
\toprule
central gain runs to & unbounded line & closed surface \\
\midrule
\endhead
$\alpha  = \beta $ (the normal form's grid) & migration falls 31.7–50.9\% & migration falls 25.2–41.9\% \\
$\alpha  = 4\beta $ (the routes swap at $2\beta $, the wells merge at $4\beta $) & — & lagged displacement \textbf{rises}, by a factor of 1.29 to 1.72 — but the range \emph{ends on} the merge, so the rise is the far barrier flattening rather than more travel between two places \\
$\alpha  = 8\beta $ (past the merge) & — & lagged displacement falls to 26.7–29.4\% of the curve's own maximum; with one well left, this is within-well fluctuation and not migration \\
\bottomrule
\end{xltabular}}

The last two rows are deliberately not called migration. Past $\alpha  = 4\beta $ the landscape holds a single well — $\widetilde{V}''(\pi ) = \alpha  - 4\beta $ changes sign there and $\cos \theta ^{*} = 1 - \alpha /(2\beta )$ reads $-1$ — so whatever displacement the measure still records is fluctuation inside one location rather than travel between two. The distinction in fact dissolves a little \emph{before} the merge. The two wells are separated by a density trough of depth $\exp(-(\alpha  - 4\beta )^{2}/(2\beta \sigma ^{2}))$, which stops being even a factor of two deep at $\alpha  = 3.82\beta $ for the quietest of the three noise levels simulated and at $3.53\beta $ for the loudest. Over the last tenth or so of the range the table calls a rise, nothing a measurement could resolve separates the two places. The rise is real, and it is the far barrier collapsing; calling it migration would claim the opposite of what the landscape is doing.

Ranges are across three noise levels, eight seeds each, on the normal form's fifteen-point grid. The line model's figures clear the over-25-per-cent threshold reported above, which is the control that makes the closed model's figures worth reading.

Figure 8 draws all three facts together: the landscape with its two exits, the barriers crossing at $\alpha  = 2\beta $, and what each implies for the migration the study would actually measure.

\begin{figure}[H]\centering
\includegraphics[width=\linewidth]{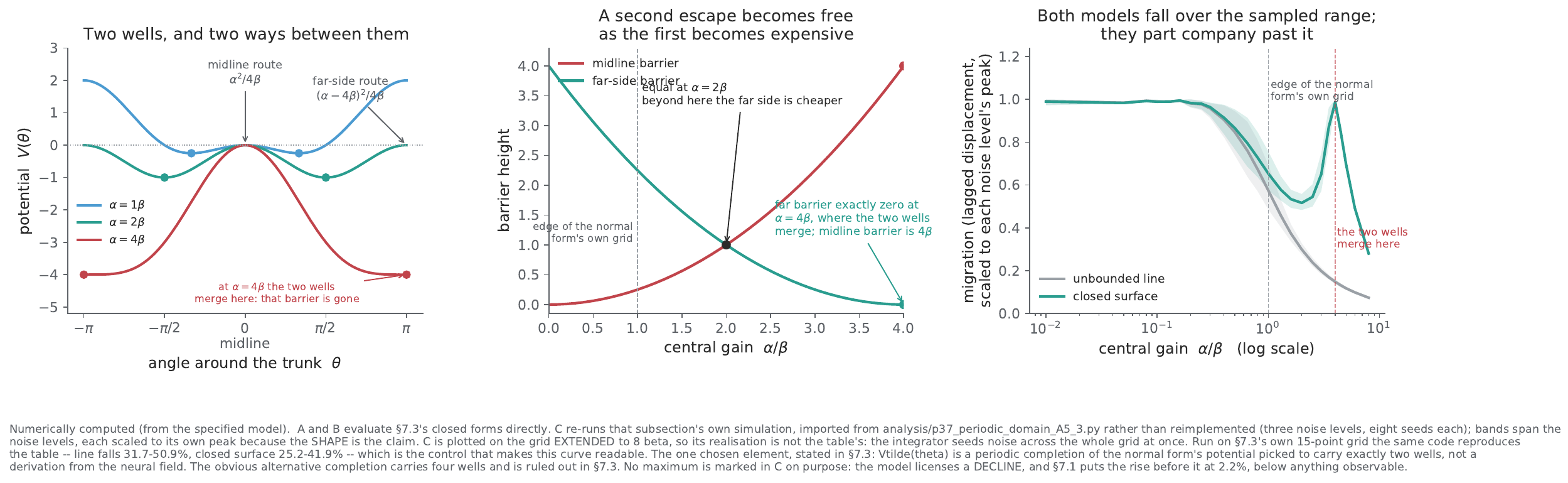}
\caption{Figure 8. Closing the body surface moves the prediction later and adds a stopping point. (A) The corrected periodic potential on a circle: two resting places, a mirror-image pair either side of the front midline, joined two ways — over the midline or round the back. (B) Those two barriers, $\alpha $$^{2}$/4$\beta $ and ($\alpha $$-$4$\beta $)$^{2}$/4$\beta $, cross at $\alpha $ = 2$\beta $; past it the far side is the cheaper escape, and at $\alpha $ = 4$\beta $ it costs nothing at all while the midline route costs 4$\beta $ — the gain at which the two wells also merge, leaving one location and nothing to migrate between. (C) The consequence for what the paper predicts: over the range the normal form sampled both models fall, the closed surface by somewhat less; between $\beta $ and 4$\beta $ the closed surface rises as the far route opens, and the decline arrives only past the merge. What to look at: the two curves track each other over the sampled range and part company only beyond it. [Numerically computed — panels A and B are §7.3 in closed form; panel C re-runs that subsection's own simulation. The potential drawn here is the two-well completion of §7.3, not the four-well one that subsection rules out.]}
\end{figure}

\textbf{How much the well count is worth.} Run on the four-well landscape, the same simulation gives 0.3 to 8.5 per cent for the first row and a two-thirds \emph{decline} for the second. Both are artifacts of the extra pair of wells, and the second has the wrong sign as well as the wrong size. $\alpha  = \beta $ to $\alpha  = 4\beta $ is exactly the range over which the far-side barrier $(\alpha  - 4\beta )^{2}/(4\beta )$ collapses to zero, so hopping gets cheaper across it rather than dearer. Read correctly, the closed surface has three phases. It tracks the line over the range the normal form sampled, so closing the domain costs some of the decline and not most of it. It parts company from the line as the far route opens. And at the merge it loses the prediction altogether: the two locations become one, and there is nothing left to move between.

\textbf{What this means for Study 1.} Suppose the closed surface barely moved over the sampled range. The design's power would then be conditional on recruiting patients whose central gain had carried them well past the transition, because a cohort near the transition would be measuring a flat curve. The two-well landscape does not put a flat curve there: it falls by a quarter to two fifths over exactly that range, and the roughly one-third decline the design is powered for sits inside that interval. The design no longer carries that assumption unstated. Its preregistration now measures where each patient sits, with a self-administered pressure-pain threshold as a \textbf{candidate proxy} for central gain, and reports the predicted decline conditional on it — together with a bridge-validation substudy of sixty participants that checks the self-measured threshold against a clinician-administered one before the covariate is trusted. The $n = 60$ is provisional: it carries no precision target, and its adequacy should be set by the confidence interval wanted for clinician–self agreement and by the attenuation tolerated in the gain-stratified analysis. That substudy validates the \emph{measurement method} and not the \emph{construct}: agreement between two ways of taking a pressure threshold says nothing about whether a pressure threshold tracks the neural field's control parameter $s$, which remains an assumption the design carries rather than tests. A phase-specific prediction sorted by a noisy or mis-scaled proxy degrades toward the pooled prediction, which is the null this subsection exists to avoid, so the proxy's error is a sensitivity analysis the study owes and not a detail. Stratifying by a gain-related covariate is what stops phases with opposite predicted slopes from being pooled: this subsection predicts a fall, then a rise as the far route opens, then nothing left to migrate between, so pooling patients across all three phases averages a fall against a rise and can return zero while every phase-specific prediction holds. A cohort that cannot be sorted by gain cannot report that honestly either way. A range assumption is still worth stating: the size varies by about a factor of one and a half across noise levels, and the model inherits every one of the normal form's conditions. But it is a weaker assumption than a four-well reading would demand.

\textbf{And two predictions the unbounded model cannot make.} Below $\alpha  = 2\beta $ the map hops across the midline, so migration should appear as displacement to the contralateral site; above it the map hops round the back. If the reduced coordinate really is a location on the trunk, rising central gain should change not merely how often pain moves but \emph{where it moves to}, from the mirror-image site to a position around the body. Past $\alpha  = 4\beta $ it should stop moving between sites at all. Nothing on a line has a far side, so the normal form could not have said either thing. Both are testable on the same body-map data Study 1 already collects, at no additional cost. Where they sit matters: the switch is at twice the transition, outside the gain range the normal form sampled, so a cohort inside that range would see the contralateral half of the prediction and not the far-side half. On the four-well landscape the switch appears at $\alpha  = \beta /2$, four times lower and well inside the sampled range. So the well count changes not only how big the effect is, but which half of this prediction a study could hope to see.

\textbf{What this subsection does not establish.} It inherits every one of the normal form's five sufficient conditions, and condition 1 — that the reduced coordinate is a location and not an amplitude or an extent — carries the same empirical risk here that it carries above. $\widetilde{V}$ is a periodic completion of the normal form's potential chosen to carry the right number of wells, not a derivation from the neural field itself, and a different closed-surface model could give different sizes. What the four-well case establishes is that the choice is not free: a periodic completion has to be checked for its topology before any number computed on it means anything, and this one is (checks W1 and S2). What closing the domain gives generically is a \textbf{second} route between the two wells: that is topology, and any periodic completion has it. That the second route becomes \textbf{free} — barrier exactly zero — at high gain is not topology but energetics, and it is a property of the particular completion $\widetilde{V}$ used here. The specific gains $2\beta $ and $4\beta $ at which the routes swap and the wells merge are likewise particular to it, and would move under a different completion. Verification, both the symbolic steps and the simulation, is in \texttt{p37\_periodic\_domain\_A5\_3.py}.

\subsection{What a measure blind to spreading can do, and the ceiling it cannot pass}

The widening control above disqualified a statistic, not a class, and the defect it exposed has a name. The transport distance charges for activity appearing where there was none, because it compares two profiles only after normalizing each to unit mass — and once the total is pinned at one, a plateau that widens has to get shorter, so the activity that left the top must be carried somewhere. Recruitment then reads as travel. The repair that control's own code names is a measure \emph{blind to symmetric spreading}. Stated as a property rather than a preference, that is \textbf{growth-blindness}: a functional $M(a, b)$ of two activity profiles reads zero whenever $b \ge  a$ at every site, so a profile that only gained activity is scored as not having moved.

The \textbf{vacated mass} has that property by construction. Write $V(a, b) = \sum _x \max(a_x - b_x, 0)$ — the activity that stopped being where it was, summed over ring sites and never normalized. On this paper's own widening control it scores \textbf{exactly zero} where the transport distance scores 2.24 ring sites, on the flat-topped profile and on the Gaussian alike. On a rigid two-site shift of the same profile it scores 2.00, so that zero is a discrimination and not a failure to respond. Normalizing the two profiles to unit mass first lifts the widening score from zero to 0.27, which locates the disagreement exactly: normalization is what manufactures the apparent movement, not the profiles. Both negative controls and the numbers above: \texttt{p37\_vacated\_mass\_widening\_control\_2026-09-22.py}.

That invites the question one statistic cannot answer — how far can \emph{any} measure of this kind be trusted to count movement? Write $R_k$ for rotation by $k$ sites on the ring and $D_a(k) = V(a, R_k a)$ for the activity a $k$-site rotation vacates, and call $M$ \textbf{exact at $k$} when it reads a $k$-site rotation back as $k$ steps, measured in the profile's own unit of single-step vacated mass $D_a(1)$. A profile's \textbf{geodesic radius} $\rho (a)$ is the largest $K$ for which $D_a(j) = j \cdot  D_a(1)$ at every $j \le  K$: how far it can be rotated before it begins to overlap itself, past which each further step vacates less than the rate the first step set.

The answer is a ceiling, and it is sharp. \textbf{No growth-blind functional that is 1-Lipschitz in the profile it compares against — 1-Lipschitz under $L^{1}$, meaning that changing that second profile by $\varepsilon $ of total activity can change the score by at most $\varepsilon $ — can count rotations, in the profile's own unit of single-step vacated mass, further than $V$ does; the ceiling is $\rho (a)$, and $V$ attains it.}

\textbf{The proof is a squeeze, and it assumes less than the statement looks like it needs.} A profile and its rotation carry the same total activity, so $D_a(k)$ is half the $L^{1}$ distance between them, and the triangle inequality caps it at $D_a(k) \le  k \cdot  D_a(1)$. Pushing the other way, evaluate $M$ at the pointwise maximum $\max(a, R_k a)$ — a profile that dominates $a$, where growth-blindness forces a reading of zero — and take one Lipschitz step from there back to $R_k a$; that caps whatever $M$ reports at $D_a(k)$. Exactness forces the two bounds to meet, so every rotation a competitor reads correctly is one $V$ already reads correctly.

\textbf{One step in that argument is a lemma rather than an observation, and it is what lets a single exact rotation be compared with the geodesic radius at all.} The squeeze delivers exactness at some rotation $k$ — that is, $D_a(k) = k \cdot  D_a(1)$. The definition of $\rho (a)$ asks for more: exactness at \emph{every} $j \le  k$. The two are the same statement, and the reason is subadditivity. Composing rotations, $a - R_{j+l}a = (a - R_j a) + R_j(a - R_l a)$; the positive part of a sum is at most the sum of the positive parts, $(x + y)_+ \le  x_+ + y_+$; and $R_j$ is a permutation of the ring, so it preserves total mass. Together these give

$D_a(j + l) \le  D_a(j) + D_a(l)$,

and in words, rotating in two stages never vacates less than rotating in one. Iterating it bounds $D_a(k) \le  k \cdot  D_a(1)$. Now suppose that bound is met at $k$. For any $j \le  k$,

$k \cdot  D_a(1) = D_a(k) \le  D_a(j) + D_a(k - j) \le  j \cdot  D_a(1) + (k - j) \cdot  D_a(1) = k \cdot  D_a(1)$,

so the chain closes on itself and every inequality in it is an equality. Since $D_a(j) \le  j\cdot D_a(1)$ and $D_a(k-j) \le  (k-j)\cdot D_a(1)$ separately, each must hold with equality. Exactness at $k$ therefore forces exactness at every $j \le  k$, the exactness set is an initial segment, and $k \le  \rho (a)$ follows.

\textbf{Subadditivity is doing the work there, not decoration.} A score that is \emph{not} subadditive can be exact on a set that is not an initial segment: a squared version of the same statistic has exactness set $\{1, 4\}$, with a subadditivity violation of +0.8346, so the prefix property genuinely fails for it (control NC-PE-2). A search of 849,341 triples found no violation for $D_a$ itself, the largest value being exactly zero. Two further things fall out of the same three lines: the containment is setwise rather than only over an initial run, and only the Lipschitz modulus in the second argument is used, leaving the first argument free. Full derivation, 21 checks in exact rationals with ten negative controls, every one of which fired, and a symbolic confirmation carried over two one-parameter families of non-arc profiles: \texttt{p37\_general\_profile\_exactness\_2026-09-22.py}.

\textbf{Three scope statements, because loose phrasing here would overstate the result.} Growth-blindness together with a second-argument Lipschitz modulus of at most one is \textbf{sufficient} for the ceiling, and nothing further about the profile is assumed — so this holds for every nonnegative profile on the ring that a single step actually disturbs, $D_a(1) > 0$, and not only for the indicator arcs on which it was first checked. That nondegeneracy condition is the one exception, and it is not a technicality: a profile constant around the ring is carried onto itself by every rotation, so it vacates nothing, and the unit the whole statement is measured in — the profile's own single-step vacated mass — does not exist for it. The profile's own unit is \textbf{necessary} rather than cosmetic: scored against a bare count of sites instead, the claim is false wherever a single-step rotation vacates more than one unit of activity, because the rescaled copy $V/D_a(1)$ is itself growth-blind and 1-Lipschitz and stays exact out to $\rho (a)$ on profiles where $V$ is exact at no rotation at all. And optimal is not unique: $\min(V, \rho (a)\cdot D_a(1))$ — the cap written in the same vacated-mass units as $V$, not in steps — is growth-blind, 1-Lipschitz and exact through $\rho (a)$ as well, so the result ranks $V$ first without singling it out, and no canonicity should be read into it. The unit in that cap is load-bearing rather than decorative: capping at the bare step count $\rho (a)$ instead breaks exactness on any profile whose single-step rotation vacates more than one unit of activity, which is the same regime where the previous sentence's rescaling argument applies. On a height-three arc of five sites, $D_a(1) = 3$ and $\rho (a) = 5$, and $\min(V, \rho (a))$ reads 5 at a two-step rotation where the profile's own unit calls for 6.

\textbf{Status: proved} — and it does not revive the migration claim. It says what the best statistic blind to spreading can resolve, which is movement out to the geodesic radius and nothing past it; it does not say the model's pain map moved. §7.1's declining-migration prediction remains the one this paper rests on. The reason to state the ceiling here is practical: a study that chooses a movement measure should know in advance where that measure stops being able to tell it anything.

\section{The spatial Bayesian mislocalization model: the reporting node}

The reporting node $p(r \mid \ell )$ flagged as \emph{proposed} in §3 is the model's novel core, and the novelty claimed for it is narrow: an explicitly Bayesian account of the pain location a patient \emph{reports}, in which a noisy percept is weighted against an expectation. Prior computational work already puts nociception in space, and the claim is narrowed against it. Boström et al. (2014) model phantom pain with tactile and nociceptive channels that carry receptive fields at explicit two-dimensional skin locations, feeding a self-organizing somatotopic map that encodes \emph{where} a signal arose and, by construction, not \emph{what} modality it was. That model is computational, nociceptive and spatial. It has no perceiver, though: in its authors' own words it "lacks a perceptual system", it computes no posterior over location, and it represents nothing a patient would say. The nearest prior art for the \emph{inferential} half is therefore in \emph{innocuous} touch, where Bayesian observer models of tactile spatial perception reproduce spatiotemporal mislocalization illusions such as the cutaneous rabbit (Goldreich, 2007). The model developed here joins the two: spatial nociceptive representation from the first, Bayesian read-out from the second, and the report as a further node on top of both. A series of targeted searches surfaced no prior Bayesian model of \emph{reported} pain location. That absence was found by targeted searching, which is weaker evidence than an absence established by a registered systematic search, and the same qualification applies to the optimal-transport and disinhibition-sweep statements below. Deep generators $s \in  \mathbb{R}^N$ project to a felt-location readout $\ell $ through a many-to-one referral (lead-field) matrix, $\ell  = R s + b + \varepsilon ,  \varepsilon  \sim  N(0, \Sigma _c)$. Here $\ell $ is a \textbf{retained low-dimensional summary} (a point or small region on the body surface) of a higher-dimensional felt map, not a claim that the body surface itself has only two effective dimensions (§4 represents the full map on $M$ mesh cells; Model E's $\ell $ is what a location report distills that map down to). Any extra ambiguity this summary introduces is measurement loss, not evidence of anatomical convergence — the convergence claim below rests on $R$'s column structure, which is checked independently in §9. $b$ is a \textbf{fixed central projection bias} — a spatial offset vector that does not vary with the person or the trial, and not the person-dependent decision bias of mechanism (c). A proximal-to-distal thalamocortical offset is the interpretation this paper has in mind for it, and the mathematics does not identify that one: $b$ is whatever systematic mapping offset the referral stage carries, and nothing below depends on which anatomy supplies it. $\Sigma _c$ is transmission noise. With prior $p(s) = N(\mu _s, \Sigma _s)$ the posterior is Gaussian with $\Sigma _{\mathrm{post}} = (R^{\top}\Sigma _c^{-1}R + \Sigma _s^{-1})^{-1}$. In words, posterior precision is prior precision plus whatever precision the referral measurement supplies, so a source direction $R$ barely sees gains almost no likelihood precision and stays close to its prior uncertainty. A word on what “identifiable” means here: because the prior precision $\Sigma _s^{-1}$ is positive-definite, $\Sigma _{\mathrm{post}}$ is \emph{always} proper, so the posterior is never ill-defined. This model treats $s$ as an \textbf{unconstrained continuous mixture} (§4's declared default), for which the source is \textbf{data-identifiable} (the likelihood pins it down) \emph{iff} $R$ has full column rank. What a rank-deficient $R$ then does to the posterior is subtler than it looks, and worth setting out as a small theorem, because the tempting version of that theorem is false. A rank-deficient $R$ leaves the likelihood flat along $R$'s null-space: no measurement, however precise, bears on that combination of generators directly. \textbf{What follows from that, always, is a statement about a conditional distribution} — the law of the null-space component \emph{given} the identified component is exactly the prior's, with the same spread and the same dependence on the identified part. \textbf{What does not follow is the marginal statement.} Take a unit vector $v$ in the null-space. The posterior along $v$ matches the prior in mean and in variance, for every observation, \textbf{if and only if} $R \Sigma _s v = 0$ — that is, if and only if the prior covariance carries the null-space back into itself. Read over the whole null-space, that condition says in words that the unidentified contrast is \emph{a priori independent} of everything the referral operator can see. It is a condition on the prior, and rank-deficiency of $R$ alone never supplies it. Where it fails, the posterior along $v$ is strictly \textbf{tighter} than the prior, by exactly $(R \Sigma _s v)^{\top} (R \Sigma _s R^{\top} + \Sigma _c)^{-1} (R \Sigma _s v)$: the data sharpen everything $R$ does see, and a correlated prior carries that sharpening across into a direction nothing measured. This posterior-variance reduction is one-way and bounded at both ends — never wider than the prior, and never tighter than the prior \emph{conditional} variance given the identified subspace, which it approaches as measurement noise vanishes and never crosses. Take a prior correlation of 0.9 between an unseen and a seen coordinate. A unit prior variance falls to 0.595 at unit noise and to 0.19 in the noiseless limit. That 0.19 is $1 - 0.9^{2}$, the prior conditional variance — not a return to the prior's 1.0. Model E's own prior $\operatorname{diag}(4, 4, 9)$ fails the condition along $R$'s exact null direction, so this paper's own model sits in the failure case rather than the safe one. The resulting posterior-variance change is 0.12 per cent, which moves no quantity reported below — but it moves in the direction the theorem predicts and not the one the tempting version does. Derivation, proof and numerical checks: \texttt{p37\_nullspace\_prior\_transport.py}. This full-rank criterion is specific to that continuous-mixture choice — a single categorical diagnosis over the same generators would not need it (§4) — and the manuscript states the choice here rather than leaving it implicit. Because dorsal-horn convergence is many-to-one, $R$ can be rank-deficient, and two generators sharing a spinal segment are the practical case: Model E's disc/facet pair below has near-collinear, not exactly collinear, columns: the 2$\times $2 submatrix holding those two columns alone has minimum singular value 0.25 and condition number 9.3, so the pair is \textbf{weakly}, not structurally, identifiable from bottom-up signal alone. Those two numbers describe that submatrix and nothing else. The full 2$\times $3 $R$ is a different object with different numbers — singular values 6.03 and 2.36, plus the exact null direction named above — and it is that null direction, not the disc/facet contrast, that §9's sweep measures. The distinction matters in principle — enough noise reduction would eventually separate the pair — but not in this model's practice: the posterior variance along the disc/facet contrast is within 1.1 per cent of the prior at realistic cutaneous precision and stays within it as precision improves, while along the exact null direction it never exceeds the prior-transport figure computed above, at any precision whatsoever — so for any noise level a provocation study could deliver, the pair is \emph{practically} — not structurally — unidentifiable from location alone, its posterior collapsing onto the prior along that axis (Model A's null-space, now as a posterior). That collapse is legitimate here for a reason the theorem above makes explicit rather than automatic: the disc/facet contrast is an eigenvector of Model E's prior $\operatorname{diag}(4, 4, 9)$, so it satisfies the independence condition. A contrast that failed that condition would not collapse, however unidentifiable it was — which is why the exact null direction, six degrees away from the disc/facet contrast and \emph{not} an eigenvector of that prior, lands at 0.9994 to 0.9998 of the prior rather than at exactly 1. The report adds a second Bayesian integration with a cognitive/attentional prior at $\mu _a$: $E[r] = W(R s + b) + (I - W) \mu _a,  W = (\Sigma _\ell ^{-1} + \Sigma _a^{-1})^{-1} \Sigma _\ell ^{-1}$, where $\Sigma _\ell $ is the covariance of the felt location $\ell $ (the bottom-up estimate carried forward from the perceptual stage), $\Sigma _a$ the covariance of the reporting prior, and $I$ the identity. This factorizes mislocalization into \textbf{R} (peripheral referral blur), \textbf{b} (anatomical projection error), and \textbf{W} (cognitive override): a strong reporting prior ($\Sigma _a \to  0$, $W \to  0$) collapses the report onto $\mu _a$ even with intact afferents — a criterion shift, formally distinct from the change of generative model that defines mechanism (b).

\begin{figure}[H]\centering
\includegraphics[width=\linewidth]{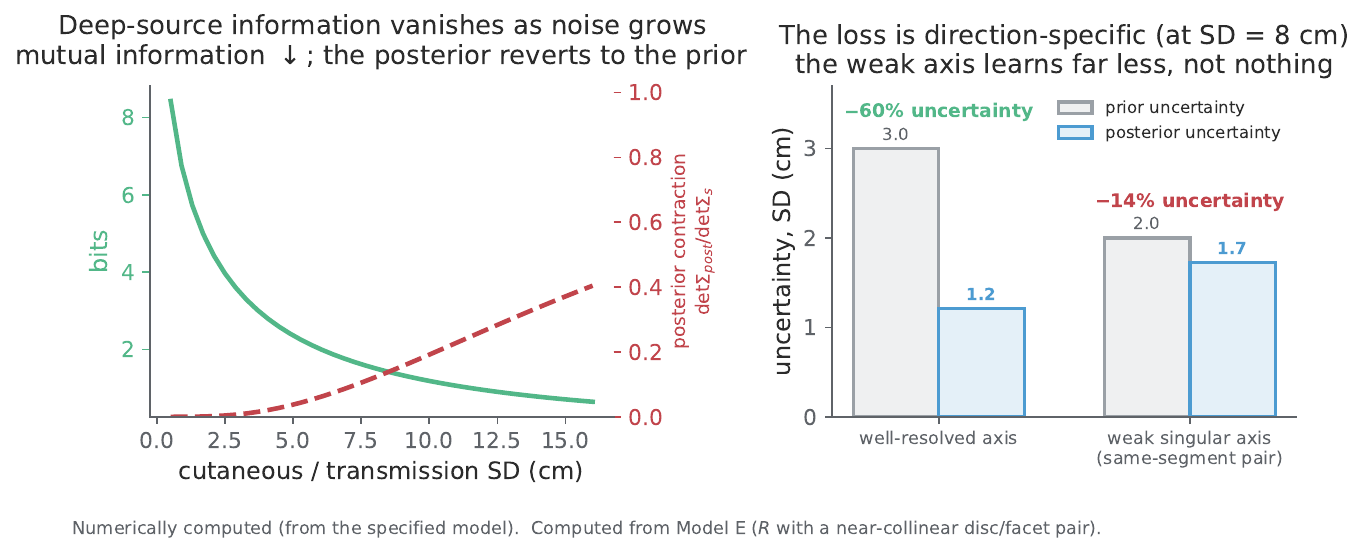}
\caption{Figure 9. Deep-source information collapses onto the prior as noise grows, and the loss is direction-specific, though not absolute. The bars compare the two measured singular axes of R: the well-resolved one loses 60 per cent of its prior uncertainty, the weak one — which lies in the same-segment disc/facet plane — only 14 per cent. Neither bar is the exact null direction, which loses nothing at any precision and is what Figure 15B draws; and the 0.25 minimum singular value quoted in §8 belongs to the 2$\times $2 disc/facet submatrix, not to either axis shown here. [Numerically computed — computed from Model E.]}
\end{figure}

\begin{figure}[H]\centering
\includegraphics[width=\linewidth]{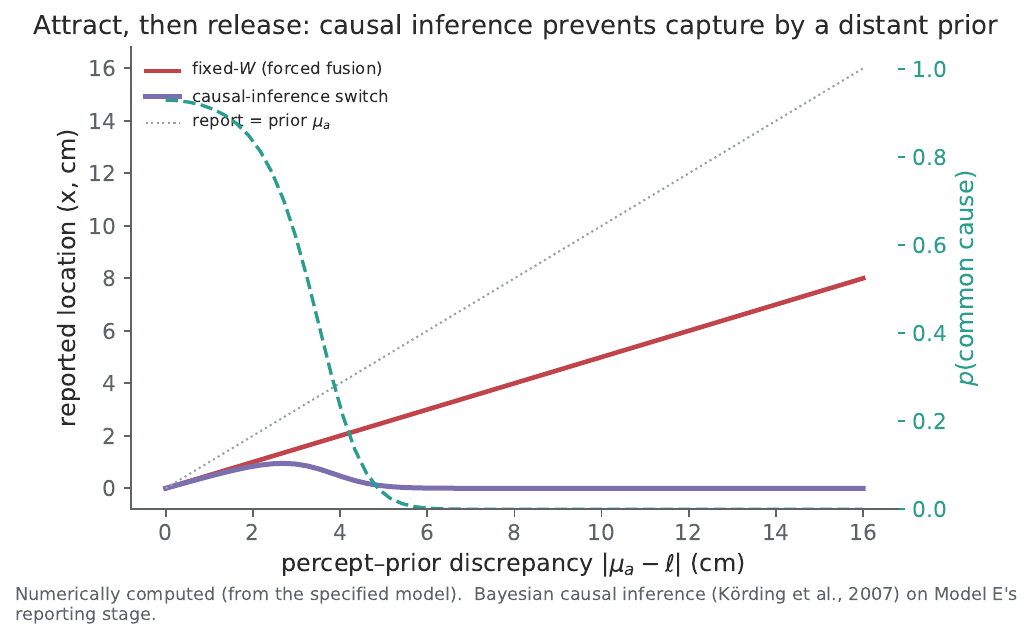}
\caption{Figure 10. The causal-inference switch. The report attracts to a nearby prior but releases back onto the percept once the discrepancy is large. Read across fields: “capture” = the report pulled toward where the pain is expected. [Numerically computed — Bayesian causal inference on Model E's reporting stage.]}
\end{figure}

\begin{figure}[H]\centering
\includegraphics[width=\linewidth]{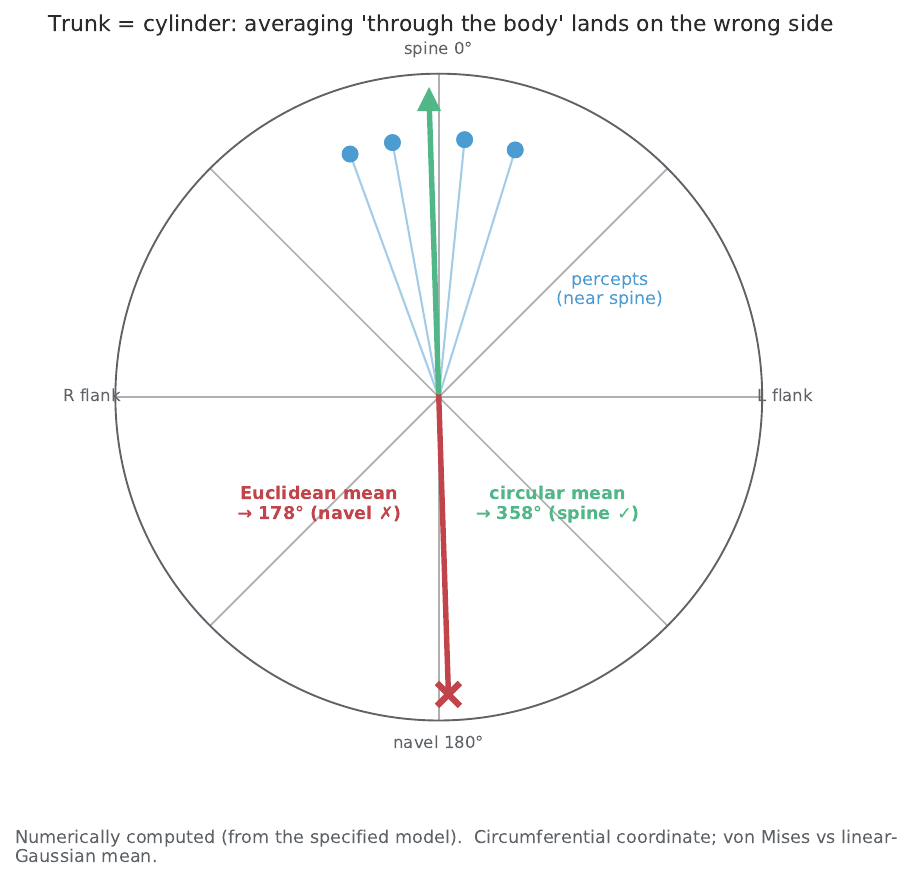}
\caption{Figure 11. On the body manifold (a cylinder), Euclidean averaging across the seam lands on the wrong side; the circular mean — the average that respects the body wrapping around — localizes correctly. [Numerically computed — von Mises vs linear-Gaussian mean on the circumferential coordinate.]}
\end{figure}

\begin{figure}[H]\centering
\includegraphics[width=\linewidth]{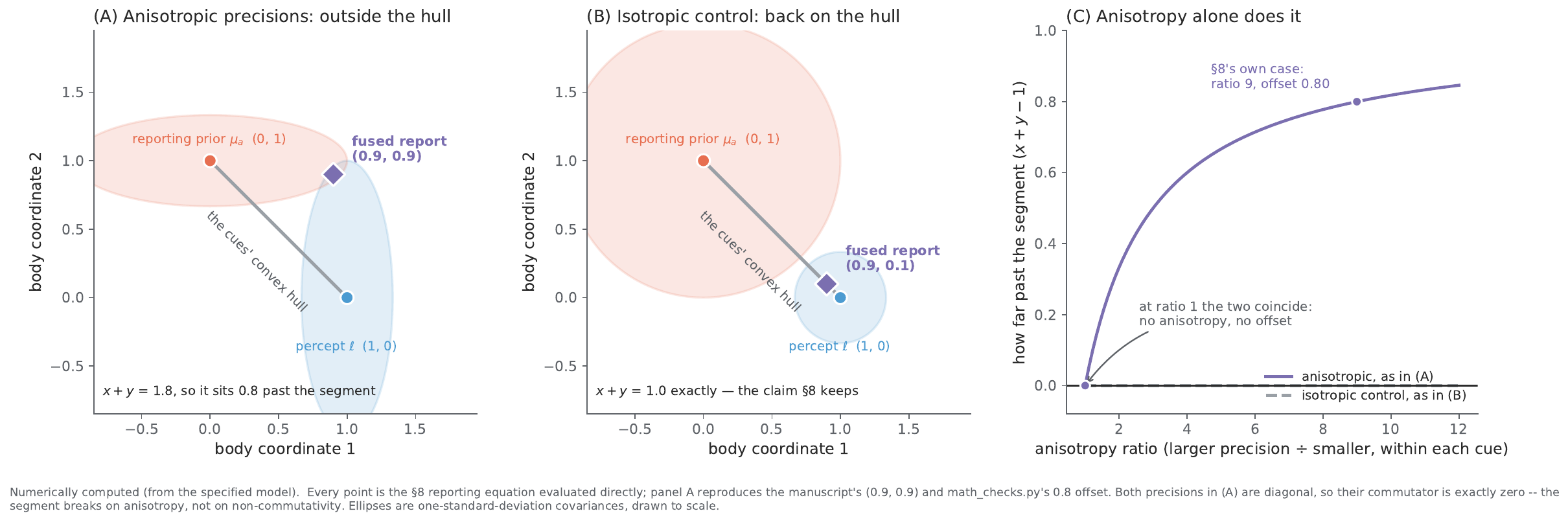}
\caption{Figure 12. A precision-weighted blend can land outside its own cues, and anisotropy alone is enough to do it. (A) §8's counterexample, with the covariance ellipses drawn to scale so the reason is visible: the percept at (1,0) is precise along the first body coordinate and vague along the second, the reporting prior at (0,1) is the mirror image, so the blend takes its first coordinate from the percept and its second from the prior and arrives at (0.9, 0.9) — 0.8 past the segment joining the two cues, which is the whole of their convex hull. Both precisions here are diagonal, so their commutator is exactly zero: what breaks the segment is anisotropy, not a failure of the precisions to commute. (B) The control, and it is the claim §8 keeps: the identical fusion with isotropic precisions lands on the segment, at x + y = 1.0 exactly. (C) The offset as the anisotropy ratio grows, starting at zero where the cues are isotropic. Read across fields: averaging two measurements normally gives something between them, and stops doing so once each instrument is trusted in a different direction. [Numerically computed — the §8 reporting equation evaluated directly; panel A reproduces the manuscript's (0.9, 0.9) and \texttt{math\_checks.py}'s asserted 0.8 offset. Ellipses are one-standard-deviation covariances.]}
\end{figure}

\begin{figure}[H]\centering
\includegraphics[width=\linewidth]{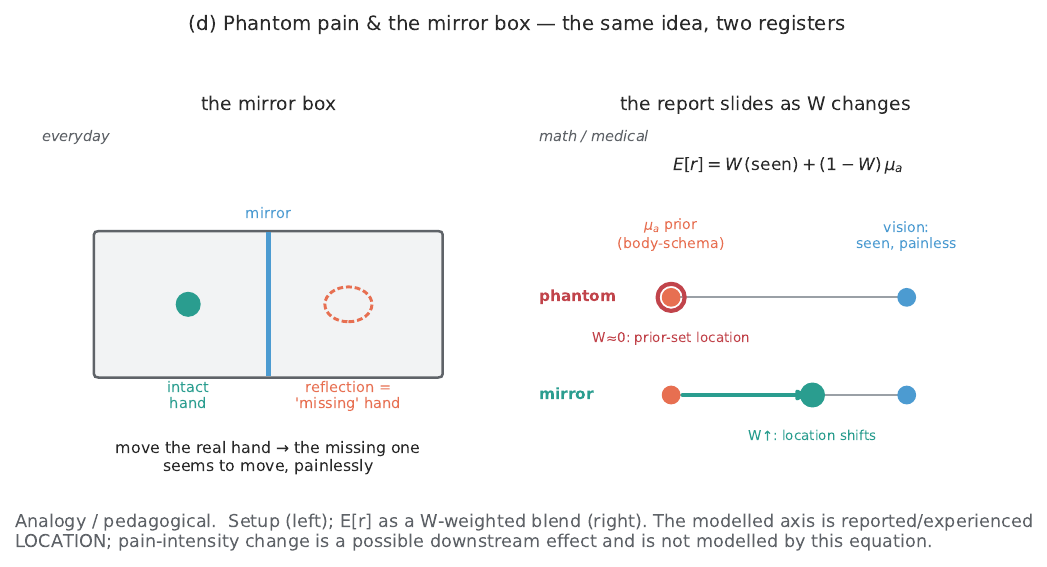}
\caption{Figure 13. Model E in one clinical picture: phantom limb pain and the mirror box. The reported location r is a precision-weighted blend of the bottom-up percept and the body-schema prior $\mu_{a}$, with the override weight W set by how precise the sensory evidence is: E[r] = W$\cdot $(percept) + (1$-$W)$\cdot $$\mu_{a}$. Amputation removes the afferent signal, so its precision collapses (W $\to $ 0) and the prior alone sets the report — a felt, often painful limb whose location is set without informative afferent input (the stump's ectopic firing is noisy, not location-informative; left ring). The mirror box injects a precise visual cue at that location — a limb seen moving painlessly — which enters as a high-precision sensory cue and shifts the integrated limb-location estimate off the prior toward the seen limb; repeated sessions could update $\mu_{a}$ itself in an extended learning model, which §5 does not derive. Read across fields: the same reweighting that provocation performs for mechanism (a) by adding rows to R, vision performs here by adding an informative channel to a prior-dominated report (formalized as the multisensory generalization in §8). What this equation models is LOCATION: the axis of the figure is where the limb is felt and reported, not how much it hurts. Pain-intensity change is a possible downstream clinical effect and is not modeled here (§8 states the same limit: "Model E governs location, not intensity"). [Analogy / pedagogical — the reporting equation is exact; the setup is illustrative.]}
\end{figure}

\begin{figure}[H]\centering
\includegraphics[width=\linewidth]{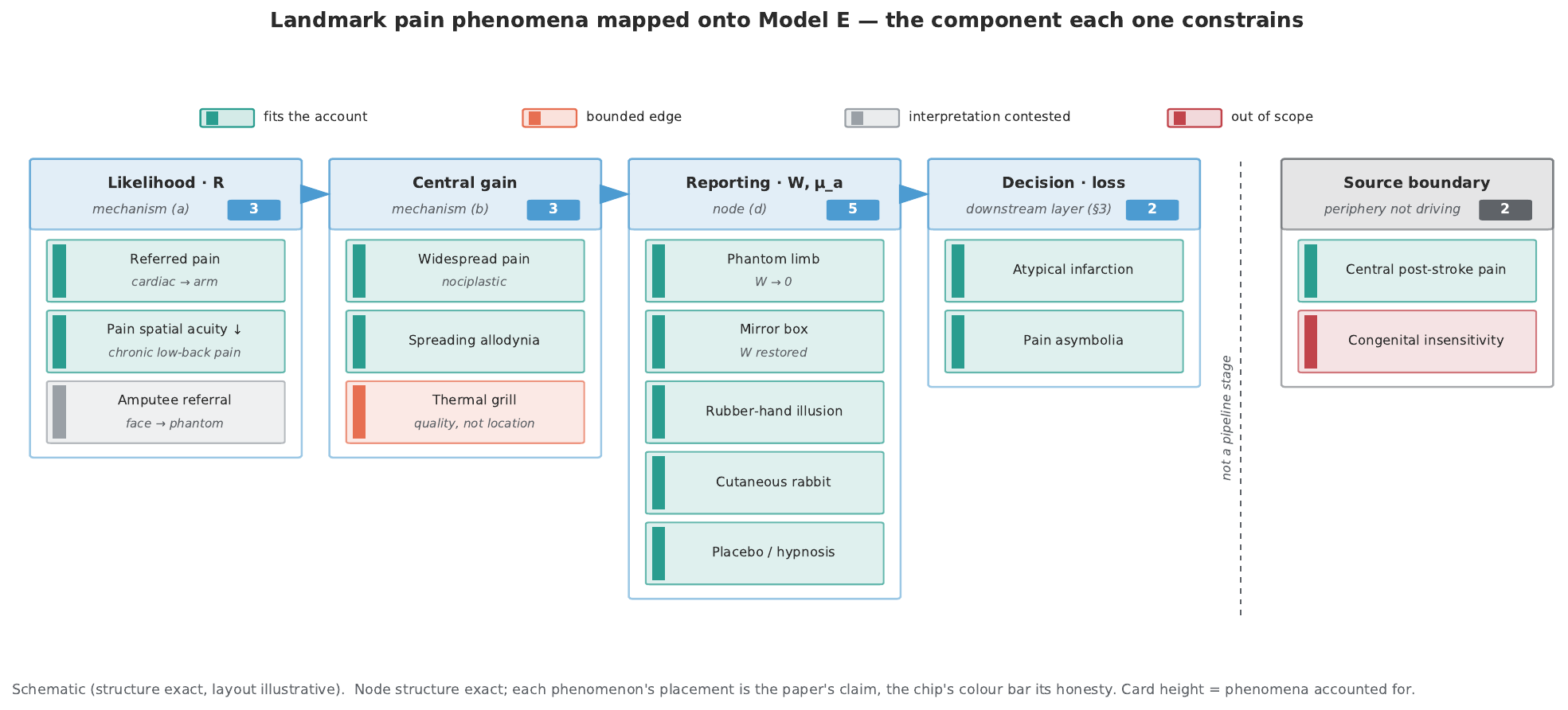}
\caption{Figure 14. Landmark pain phenomena mapped onto Model E — the component each one most directly constrains or illustrates, which is weaker than isolation and is the accurate verb here: most clinical phenomena are not clean interventions on one component, and §3 says mechanisms can coexist. The columns run in the order §3 sets out (likelihood R $\to $ central gain $\to $ reporting W, $\mu_{a}$ $\to $ decision/loss $\to $ source boundary), so the report precedes the decision that acts on it; each phenomenon is placed under the node it tests, and coloured by disposition: fits the account (referred pain, phantom limb, mirror box, rubber-hand illusion, pain asymbolia, central post-stroke pain, \dots), a bounded edge (the thermal grill — a quality, not a location, effect), a contested interpretation (the amputee face$\to $phantom referral, whose maladaptive-remapping reading is disputed), or out of scope (congenital insensitivity — no channel to localize). One caution the map itself makes: the decision column is not one of the four nodes. It sits in the downstream decision layer of §3, and it bundles two things — a \emph{loss} (what a detected pain is worth, the object pain asymbolia lesions — a genuine dissociation) and a \emph{threshold} (the decision rule, shifted by a demographic covariate, e.g. atypical infarction) — so those two chips test different facets, a biological deficit versus a statistical rule, not one node. Mechanism (c) itself is the group- or context-dependence of the presentation likelihood one stage earlier, and carries no chip here, because §6 marks the direct evidence for it open: the threshold shift is its downstream consequence, not the mechanism. The value is the coverage itself: the account engages the field's canonical demonstrations, and names its honest edges rather than hiding them. [Schematic — node structure exact; each placement is the paper's claim, chip colour its honesty.]}
\end{figure}

\textbf{The report cannot carry more diagnostic information than the percept it came from, and the factorization of §3 is what guarantees it.} That factorization writes the report stage as $p(r \mid \ell , g, z)$, conditioning on the felt location alone. So $S \to  L \to  R$ is a Markov chain given the group and context, and the data-processing inequality applies directly:

$I(S; R | G, Z) \le  I(S; L | G, Z)$,

with equality exactly when the report is a sufficient statistic for the cause. The hypothesis is load-bearing and not arithmetic: allowing the report to depend on the cause directly, $p(r \mid \ell , s)$, breaks the factorization and the inequality fails, by up to 0.3 bits over 300 draws (\texttt{p37\_round2\_new\_math\_2026-09-24.py}, control A2.C1). This is the tool §5 declines for the \emph{repeated-observation} chain, where no Markov emission structure is available and the chain rule is the right instrument. Here the structure is supplied by the model itself, so the paper is entitled to the inequality one section later, and the two statements do not conflict.

In the linear-Gaussian model the inequality has a closed form worth printing, because it names what the reporting node costs. Let the report carry its own noise, $r = W \ell  + (I - W) \mu _a + \nu $ with $\nu  \sim  N(0, \Sigma _\nu )$ and $W$ invertible, and write $\Sigma _L = R \Sigma _s R^{\top} + \Sigma _c$ for the marginal spread of the percept. Then

$I(S; R) = \tfrac{1}{2} \log [ \det(\Sigma _L + W^{-1} \Sigma _\nu  (W^{-1})^{\top}) / \det(\Sigma _c + W^{-1} \Sigma _\nu  (W^{-1})^{\top}) ]$,

and $I(S; L)$ is the $\Sigma _\nu  \to  0$ limit of the same expression. In words: the reporting node is \textbf{exactly} equivalent to inflating the transmission noise $\Sigma _c$ by $W^{-1} \Sigma _\nu  (W^{-1})^{\top}$, and nothing else. Two readings follow. With no report noise and an invertible weight the report is an invertible function of the percept, so it loses nothing at all — the override displaces the report without destroying information. All of the loss comes from $\Sigma _\nu $, and $W$ enters that term \textbf{inverted, twice}. So a strong reporting prior, the $W \to  0$ criterion shift above, does not merely pull the report toward $\mu _a$: it destroys diagnostic information at a rate set by the inverse square of the weight on the percept. That is a sharper statement of the reporting node's clinical cost than a displacement alone gives. A numerical implementation confirms three behaviors: the disc/facet posterior stays wide — SD 1.55 against 0.12 for a separated source, about thirteen times wider, in the model's own source/location units; as $\Sigma _c$ grows the mutual information $I(s;\ell )$ falls and the posterior approaches the prior as data precision vanishes (posterior variance \emph{expands} toward the prior's, not contracts — \textbf{the “blobs” identifiability risk}), though — a point owed to adversarial review — the collapse is \emph{anisotropic}, occurring only along $R$'s small singular values and not as uniform smearing (Figure 9), so identifiability must be read from $I(s;\ell )$ and $\det \Sigma _{\mathrm{post}}/\det \Sigma _s$ (both unchanged by any invertible linear re-expression of the source coordinates, unlike a raw matrix norm) and not a matrix-norm heuristic; and a fixated reporting prior drags the report 4.2 cm to $\mu _a$ while the percept is unchanged — 84\% of the 5 cm separating percept from prior in that illustration, against 20\% at a moderate prior and 2\% at a weak one (\texttt{spatial\_bayesian\_E.py}), so the fraction and not the centimetre is what carries the claim. This same-segment identifiability floor is not an artifact of the illustrative noise or prior: it persists across the whole $(\Sigma _c, \Sigma _s)$ sweep (§9, Figure 15B). Two estimability preconditions follow. First, the provocation experiment (saline injection at blinded lumbar levels + a sham-laser expectation manipulation + iPad-avatar EMA) must keep the mutual information high enough for stable estimation at the planned sample size. Second — because $b$ and $\mu _a$ are jointly degenerate from a single trial (a shift in the anatomical bias is exactly cancelled by a shift in the cognitive prior) — the design needs trials at more than one report weight $W$. Exactly how many, and how they must differ, is a rank condition rather than a qualitative requirement, and it is worth stating precisely because the qualitative version is wrong above one dimension.

\textbf{How many trials separate the anatomical bias from the cognitive prior.} Take the bottom-up term $x_j = R s_j$ known or separately anchored on trial $j$, and the report weight $W_j$ known. Subtract what the bottom-up term contributes and write what is left as

$z_j = E[r_j] - W_j x_j = W_j b + (I - W_j) \mu _a = \mu _a + W_j \delta ,   \delta  = b - \mu _a$.

In words: every trial measures the cognitive prior plus that trial's own weight applied to the gap between bias and prior, so trials differ only through how their weights act on that one gap vector. Stack the contrasts against the first trial into $D = [W_2 - W_1; \dots  ; W_J - W_1]$. Then the pair $(b, \mu _a)$ is identifiable \textbf{if and only if} $rank D = d$, the dimension of the report space — equivalently, if and only if the weight contrasts share no common null direction, so that the only vector every $W_j - W_1$ sends to zero is zero itself. The condition is necessary as well as sufficient, and it fails constructively rather than delicately: where a nonzero $v$ lies in every kernel, the substitution $\delta  \to  \delta  + t v$, $\mu _a \to  \mu _a - t W_1 v$ reproduces every trial's expected report for every real $t$, an indistinguishable one-parameter family rather than an unstable solve (\texttt{p37\_reporting\_node\_identifiability\_2026-09-29.py}, control NC-RK-1, agreeing to 1e-9 across $t$). Verified at $d = 2, 3, 4$ over 400 draws each.

With exactly two trials the condition collapses to a determinant: $W_2 - W_1$ must be \textbf{nonsingular}. This is where \emph{distinct} stops being enough. Two distinct weights whose difference is singular leave a whole direction unresolved, so the phrase the design needs is that \textbf{the report-weight contrasts must span the full report space}, not that the weights differ.

The zero-expectation control trial is still the transparent choice, and it is now a construction rather than an existence claim. Set $W_1 = I$ (an uninformative cue): the control trial reads $z_1 = b$ directly, anchoring $(R, b)$ before the prior is fit, and the second trial then gives $\mu _a = (I - W_2)^{-1}(z_2 - W_2 b)$ in closed form, since a nondegenerate Bayesian fusion weight makes $I - W_2$ invertible. Recovery error $3.5 \times  10^{-16}$ at $d = 2$ and $5.1 \times  10^{-16}$ at $d = 3$ — the estimator, not merely the possibility of one.

\textbf{Two scope conditions on that theorem, because it assumes $W$ known and §8.1 shows what a cue-line experiment can actually see.} §8.1 proves that a scalar cue-line probe recovers only the symmetric part of $W$. The two statements do not contradict each other — two trials can share a symmetric part exactly and still differ by a nonsingular skew matrix, and a worked pair does: $\|sym(W_2) - sym(W_1)\| = 6.3 \times  10^{-16}$, invisible to any cue-line probe, with $\det(W_2 - W_1) = 0.16$, from which both vectors are recovered exactly. But the design conclusion is sharp. \textbf{Retain the full vector report.} A cue-line summary discards exactly the part of $W$ doing the identifying, so a study that keeps only the scalar summary and then assumes $W$ known has assumed what its own measurement cannot supply; $W$ must instead come from full-vector displacements across spanning conflicts, from an isotropy restriction, or from independent calibration. And the skew escape is \textbf{even-dimensional}: every skew matrix of odd order is singular, confirmed at $d = 3$ and $d = 5$ (largest $|\det|$ over 200 draws, $1.8 \times  10^{-16}$ and $6.9 \times  10^{-16}$) against 2.78 at $d = 2$ and 12.4 at $d = 4$. So two trials differing only in skew never identify the pair in odd dimension, and this paper works in two dimensions and three in different places.

\textbf{What that separation does and does not buy.} Recovering $b$ and $\mu _a$ as distinct numbers is identifiability of two \emph{model components} under the stated observation model. It does not identify a unique biological substrate for either. A fixed additive $b$ can summarize several central mapping offsets at once, and a fitted $\mu _a$ can summarize expectation, body schema, response habit or any other top-down structure that pulls the report the same way. The phantom-limb, post-stroke, placebo and mirror-box mappings below are readings of the fitted components, and the algebra does not license reading them as measurements of the named mechanism. Three extensions sharpen the model. First, replacing the fixed reporting weight with a Bayesian causal-inference stage (Körding et al., 2007) — inferring whether the percept and the cognitive prior share a common cause — makes the report attract to $\mu _a$ only when the two are close and \emph{release} back onto the percept at large discrepancy (Figure 10), so a well-localized pain is not relabelled to a distant expectation. Second, because the body surface is a bounded manifold and not a plane, the linear-Gaussian form is a local approximation: on the trunk (a cylinder) a geodesic-distance, von Mises formulation is required, since Euclidean averaging across the midline can place the estimate on the wrong side of the body (Figure 11). Third, the mirror box and the rubber-hand illusion each introduce an \emph{independent visual estimate} of limb location that the two-cue blend above does not carry explicitly; the honest generalization is ordinary multisensory integration, $E[r] = \Sigma _k W_k x_k$ over the cues $\{R s + b$ (nociceptive), $v$ (visual), $\mu _a$ (prior)$\}$. For \textbf{independent} Gaussian cue errors with precisions $P_k = \Sigma _k^{-1}$ and total precision $P_{\mathrm{tot}} = \Sigma _k P_k$, the weights are exactly

$W_k = P_{\mathrm{tot}}^{-1} P_k,   \Sigma _k W_k = I$,

and this generalizes the two-cue $W$ used above rather than approximating it. Writing $W_k \propto  \Sigma _k^{-1}$ would be a scalar habit carried into a matrix setting, and this section's own subject is what happens when matrix weights fail to commute, so the ordering has to be written down: the reversed product $P_k P_{\mathrm{tot}}^{-1}$ also sums to the identity and is a \textbf{different} rule, giving a fused estimate 0.7223 report units away on a non-commuting three-cue problem, while for commuting precisions the two coincide exactly (\texttt{p37\_reporting\_node\_identifiability\_2026-09-29.py}, controls NC-MC-1 and NC-MC-2; the exact identity checked over 300 random three-cue problems at $d = 3$ to 1e-10). The normalization $\Sigma _k W_k = I$ is therefore not what pins the rule down — the ordering is. If the cue errors are correlated, even this formula must be replaced by the corresponding joint Gaussian estimator. The two-cue equation above is the special case with no visual input ($\Sigma _v^{-1} = 0$), so this leaves every earlier result unchanged.

\textbf{What the mean report geometry does not identify: absolute precisions.} The weight $W = (\Sigma _\ell ^{-1} + \Sigma _a^{-1})^{-1} \Sigma _\ell ^{-1}$ is a \textbf{relative} object, so no amount of mean-report data separates the two cue precisions in absolute terms. The obvious degeneracy is the common rescaling $(P_\ell , P_a) \to  (cP_\ell , cP_a)$, which leaves $W$ fixed to $4.9 \times  10^{-16}$, but the unidentified family is far larger than that one direction. Writing $S = P_\ell  + P_a$, the fusion weight fixes only $P_\ell  = S W$ and $P_a = S(I - W)$, so \textbf{every} symmetric $S$ for which both products are positive definite reproduces the same $W$. Requiring $S W$ to be symmetric imposes $d(d-1)/2$ equations on the $d(d+1)/2$-dimensional space of symmetric matrices, leaving a fibre of dimension exactly $d$ — measured as 2 at $d = 2$, 3 at $d = 3$ and 4 at $d = 4$. The common rescaling is one direction inside that fibre, not the whole of it: a fibre member exhibited by control NC-PR-1 has an entrywise ratio to the original spanning 1.0615 to 1.4344, so it is not a rescaling at all. Recovering absolute cue precisions therefore requires information the mean report does not carry — covariance or trial-to-trial variability data, an external reliability measurement, or an anchor on the total precision. \textbf{Scope of the no-repulsion claim.} In one dimension, or when the precisions are scalar/isotropic — the case used for the figures and worked examples throughout this section — the fused estimate always lies between the cues. With unrestricted multivariate covariance matrices, matrix-weighted fusion need not lie on the Euclidean line segment joining the cues (a checked counterexample: \texttt{math\_checks.py / cue\_fusion\_check}). The convex-hull/no-repulsion statement made in this section is therefore scoped to the scalar/isotropic model the figures use, not asserted for the general matrix-weighted case. What breaks the segment is anisotropy alone — precisions that differ by direction — and not a failure of the two precisions to commute. Two \emph{commuting} precisions show it: $\Sigma _\ell ^{-1} = \operatorname{diag}(9,1)$ and $\Sigma _a^{-1} = \operatorname{diag}(1,9)$, with the percept at $(1,0)$ and the reporting prior at $(0,1)$, fuse to $(0.9,0.9)$, which is outside the two cues' convex hull (\texttt{math\_checks.py / cue\_fusion\_commuting\_check}). Figure 12 draws that counterexample beside its isotropic control, which lands on the segment. Non-commuting precisions matter for a narrower failure, \emph{cue-line overshoot}, in which the fused weight along a fixed direction $d$ leaves $[0,1]$. That weight is the cue-line statistic $d^{\top}Wd/d^{\top}d$, and its range is the spectrum of the symmetric part of $W$. The symmetric part is all a cue-line probe can see, and what it discards has a name: the skew part of $W$ is $\tfrac{1}{2}S^{-1}[\Sigma _\ell ^{-1}, \Sigma _a^{-1}]S^{-1}$, the two precisions' failure to commute carried through the inverse of their sum $S = \Sigma _\ell ^{-1} + \Sigma _a^{-1}$ (\texttt{math\_checks.py / cue\_fusion\_skew\_commutator\_check}). That is the standard commutator-with-an-inverse identity, $[A^{-1}, B] = -A^{-1}[A, B]A^{-1}$ (Helton \& Howe, 1975, Eq. 2.4, p. 282), specialized to the fusion map, and no novelty is claimed for it. The consequence is about what an experiment can measure: scalar cue-conflict data cannot recover the skew part, so the direct signature of non-commuting precisions is not identifiable from them, however many conflict directions are probed. A targeted search of the cue-fusion, multisensory-integration, sensor-fusion and Kalman-filtering literatures found no prior use of this identity there (2026-09-06); that is an absence in searched sources, not a claim of priority. Non-commutativity is necessary for overshoot and not sufficient for it: the counterexample above has non-commuting precisions and no overshoot at all, its cue-line statistic staying inside $[0.099, 0.926]$. Leaving the Euclidean cue segment and overshooting in the one-dimensional cue-line coordinate are distinct failures, and the result above establishes neither. Anisotropy can produce the first even with commuting precisions; non-commutativity bears on the second without being sufficient for it. Phantom limb pain is then the corner where the nociceptive precision vanishes and the prior alone sets the report; the mirror box and the rubber-hand illusion add a high-precision visual cue that pulls the report toward the \emph{seen} limb — the same “append an informative channel” move as provocation in mechanism (a), now cross-modal. The added visual channel helps strictly \emph{somatically}, and the scope matters: a visual cue informs the report only where the source is visible on the body surface. For a deep, multiplexed source — the disc-versus-facet ambiguity of mechanism (a) — vision says nothing about which generator fired ($\Sigma _v^{-1} \to  0$ for that contrast), so the rank-deficiency of $R$ and its identifiability floor stand untouched. This extension adds a channel for limb and surface localization; it does \emph{not} resolve visceral inverse problems, and must not be read as if it did. The derivation, the adversarial review, the provenance trail, and the numerical results are in the supplement.

\begin{mdframed}[backgroundcolor=boxbg,linecolor=boxln,linewidth=0.4pt,roundcorner=3pt]\textbf{Intuition.} Locating a deep pain from the body-surface map is like locating an object from the \emph{shadow} it casts. Two things can go wrong. First, many objects cast the \emph{same} shadow — so the location is genuinely ambiguous (the matrix $R$ is not invertible). Second, what you \emph{expect} to see quietly bends where you think the shadow points (the prior $\mu _a$). The model's one job is to keep those two apart: the ambiguity of the shadow ($R$, $b$) versus the pull of expectation ($W$). It cannot do that from a still picture — hence the provocation experiment, which moves the object on purpose and adds a no-expectation trial so the two effects separate.\end{mdframed}

\begin{figure}[H]\centering
\includegraphics[width=\linewidth]{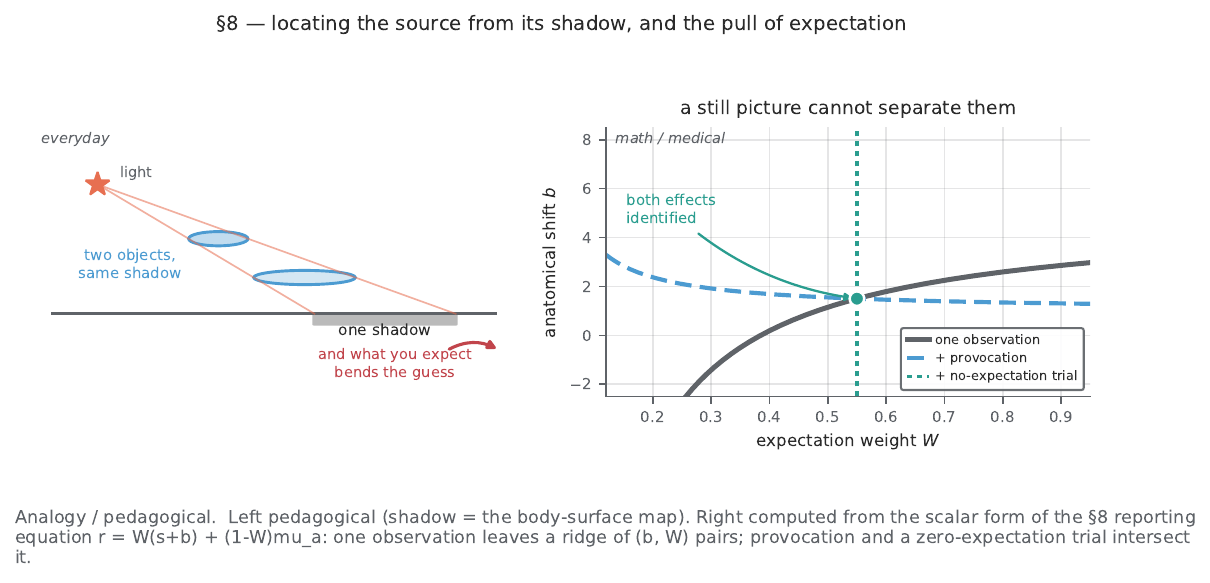}
\caption{The same idea in two registers. Everyday: an object located from the shadow it casts — two objects placed so each exactly fills the light cone throw the same shadow, and expectation quietly bends where you think it points. Math/medical: why a still picture cannot separate those two. One observed report is consistent with a whole ridge of (anatomical shift b, expectation weight W) pairs; a second observation at a different provocation level crosses that ridge, and the no-expectation control trial pins it. [Analogy on the left; the right panel is computed from the scalar form of the §8 reporting equation.]}
\end{figure}

\begin{mdframed}[backgroundcolor=boxbg,linecolor=boxln,linewidth=0.4pt,roundcorner=3pt]\textbf{Worked example: phantom limb pain and the mirror box (Figure 13).} The reporting equation makes a familiar clinical pair fall out as two settings of one knob. In phantom limb pain, amputation removes the limb's normal afferent input. The severed stump's ectopic and neuroma-driven firing is real, but it carries little reliable information about the limb's \emph{location}, so the location evidence loses precision and the override weight collapses ($W \to  0$). The report then rests on the body-schema prior ($E[r] \to  \mu _a$): a felt, often painful limb whose \emph{location} is set by the prior, not by an informative signal. Model E gives the clean, extreme case of the reporting node (\emph{d}): a location produced by the prior, not sensed (Model E idealizes the channel as uninformative, not literally silent).\end{mdframed}

The mirror box is the opposite move on the same knob. Placing the intact limb's reflection where the missing one would be injects a precise \emph{visual} cue: a limb seen moving, painlessly. That cue enters the report as a high-precision sensory term (the multisensory generalization below) and pulls it off the prior toward that painless percept. Repeated sessions could, in an extended learning model, update the body-schema prior $\mu _a$ itself — but that is a proposed extension and not a result this paper derives. §5's chain rule establishes monotone information accumulation about a \textbf{fixed} target; it supplies no update rule for $\mu _a$, and citing it for one would be exactly the conflation §5 warns against. An explicit state equation for $\mu _a(t)$ is what such a claim would need, and this paper does not write one. Structurally this is the same reweighting that provocation performs for mechanism (a): there by appending informative rows to $R$ to separate peripheral sources, here by adding an informative channel (vision) to a prior-dominated report. The congruence dependence follows from the causal-inference stage above (Figure 10): mirror feedback \emph{captures} the percept — pulls the felt location onto the seen one — only when vision and proprioception are inferred to share a common cause (congruent movement), and releases when they are not. That is a testable match to the clinical observation that the box helps when the reflection moves \emph{with} the intact hand. The same lens places neighbouring phenomena. The rubber-hand illusion is this same blend applied to touch and ownership, not to pain: synchronous seen-and-felt stroking captures the felt hand toward the visible one (Botvinick \& Cohen, 1998). Amputees' \emph{referred sensations} — a touch to the face felt in the phantom hand (Ramachandran et al., 1992; Ramachandran \& Hirstein, 1998) — are classically read as cortical remapping physically rewiring the referral matrix $R$. That \emph{causal} reading is now contested: the former hand's cortical representation appears preserved and not overwritten, and phantom pain tracks that preservation (Makin et al., 2013). Read through Model E, the newer finding is a fit, not a problem: with afferent input gone, the percept must be sustained by what remains — a \emph{maintained} representation of the hand — which the model abstracts as the stored prior $\mu _a$, the $W \to  0$ regime above. That abstraction is deliberate but lossy, and worth flagging: Makin's preserved \emph{cortical structure and function} is not the same object as a cognitive expectation, so folding it into $\mu _a$ blurs the line between a hardwired bottom-up map ($R$) and a top-down prior. The manuscript reads the phenomenon as a preserved prior and takes no position on the disputed remapping mechanism. Complex regional pain syndrome — where the affected limb can feel foreign or distorted, a body-perception disturbance (Lewis et al., 2007) — and spreading allodynia read as mechanism (b), the disinhibition bifurcation, where graded motor imagery and mirror work can reduce pain (Moseley, 2006). One possible extension is that such interventions lower the effective gain parameter and so move the system away from the delocalized regime — but nothing in the neural-field model derives a mapping from any treatment to $s$, so that is a hypothesis the model can express rather than a consequence it supplies. Two honest limits: Model E governs \emph{location}, not \emph{intensity}, so phantom-pain intensity dynamics still belong to (b); and mirror-box efficacy is empirically variable, so the model offers a mechanism by which it \emph{can} work — restore $W$, re-learn $\mu _a$ — not a claim that it always does.

\begin{mdframed}[backgroundcolor=boxbg,linecolor=boxln,linewidth=0.4pt,roundcorner=3pt]\textbf{Empirical handles on the nodes (Figure 14).} Three further phenomena each constrain a \emph{distinct} component most directly — one of them, asymbolia, by a genuine dissociation; the other two by compatibility rather than isolation, as noted below — and matter because they show the machinery is real and separable, not a single knob. \emph{Pain asymbolia} — after unilateral hemispheric lesions that in all six of the original patients damaged the insular cortex, patients still detect a noxious stimulus but lose the normal defensive and affective response (Berthier et al., 1988) — isolates the loss term of the \emph{decision layer} (§3, §6): the value the system places on a detected signal, here severed from an intact percept. This is a \emph{different object} from the one mechanism (c)'s care-pathway corollary exploits clinically — a shift in the decision \emph{threshold} under a scientifically justified group variable (the infarction missed more often in women). Decision theory carries both a loss and a threshold, and both now live in the downstream decision layer, not in the perceptual model; asymbolia is a lesion of the former, the group-conditional threshold shift a distortion of the latter, and reading them as one thing would conflate a biological deficit with a statistical rule. (The exact locus of the asymbolia deficit is itself debated; the detection-intact / response-absent pattern is not.) \emph{Hypnotic suggestion} selectively changes pain unpleasantness while intensity stays fixed, with matching anterior-cingulate activity (Rainville et al., 1997). Placebo effects move reported pain through expectation and conditioning, with the noxious stimulus itself unchanged (Colloca \& Benedetti, 2005). Both demonstrate expectation-dependent modulation of pain reports, and are qualitatively compatible with a reporting/prior mechanism. Neither identifies $W$, $\mu _a$, or any spatial localization effect: the findings concern unpleasantness and expectation, and this paper's contribution is \emph{location}. \emph{Central post-stroke pain}, generated by a lesion of the central somatosensory pathways with no peripheral source (Klit et al., 2009), is \textbf{representable} in the same prior-dominated corner as the phantom — a percept whose location the periphery no longer informs. That mapping is a mechanistic hypothesis about how the model would render the phenomenon, not evidence that prior domination is its biological cause. None of these requires new machinery; each exercises a node the model already posits — which is the point. Figure 14 maps these and the paper's other landmark cases onto the model component each one most directly constrains or illustrates, and marks the two the model bounds rather than explains (the thermal grill, congenital insensitivity). \emph{Constrains or illustrates} is the accurate verb across the whole map, and \emph{isolates} is not: most clinical phenomena are not clean interventions on one mathematical component, §3 says mechanisms can coexist, and two of the map's own chips are marked as edges the model does not explain at all.\end{mdframed}

\subsection{What cue-conflict data can and cannot identify}

The reporting weight $W$ is a matrix, but a cue-conflict experiment returns one number per trial: how far along the line between the two cues the report fell. This subsection asks what that number determines and what it cannot reach. The answer bounds the reporting node's estimability from the measurement side, where §8's earlier preconditions bound it from the design side. Every statement here is proved algebraically and checked numerically. The numbers quoted below are printed by \texttt{math\_checks.py / cue\_line\_identifiability\_check}; the fuller per-claim record is in \texttt{research/validation/deposit-verification-2026-09-06/}.

\textbf{The observable.} Put one cue at $x_1$ and the other at $x_2$, and write the conflict between them as $d = x_2 - x_1$. The fused report is $r = x_1 + Wd$. Its \textbf{cue-line coordinate} — the fraction of the conflict the report travels — is

$t(d) = (d^{\top}Wd)/(d^{\top}d)$.

A value of 0 puts the report on the first cue and 1 puts it on the second, so \emph{cue-line overshoot}, $t(d)$ outside $[0, 1]$, means the report passed a cue rather than landing between them. One caution that matters for reading any experiment: $t(d)$ is a coefficient along the conflict direction, not the whole report. For the worked pair used below, the direction with the largest off-line component gives $t = 0.513$ while the report sits 0.603 off the cue line relative to the length of $d$, so a cue-line statistic near the midpoint does not mean the report is near the midpoint.

\textbf{The cue-line statistic sees only the symmetric part of $W$.} Split $W$ into its symmetric part $A = (W + W^{\top})/2$ and its skew part $K = (W - W^{\top})/2$. Because $d^{\top}Kd = 0$ for any skew $K$, the coordinate reduces to $t(d) = (d^{\top}Ad)/(d^{\top}d)$ — a \textbf{Rayleigh quotient}, the standard ratio whose range over all directions is the interval between a symmetric matrix's smallest and largest eigenvalues. So $t$ ranges over $[\lambda _{\mathrm{min}}(A), \lambda _{\mathrm{max}}(A)]$ and nothing more. The skew part is invisible to it, which is the identifiability consequence stated above: since that skew part is the transported commutator of the two precisions, cue-line data cannot see the precisions failing to commute. The claim is about this statistic and not about the report. Holding $A$ fixed and flipping the sign of $K$ leaves every $t(d)$ unchanged to 3.3e-16 while moving the report itself by 0.378 — also relative to the length of $d$, so the two displacements are on one scale — so a second read-out taken off the cue line does recover what $t$ discards.

\textbf{What the full vector report recovers, exactly, and how many conflicts it takes.} Suppose the experiment keeps the whole displacement $y_j = W d_j$ rather than the scalar fraction. Collect the conflicts and the responses as the columns of $D = [d_1 \dots  d_m]$ and $Y = [y_1 \dots  y_m]$. Then $W$ — all of it, skew part included — is identifiable \textbf{if and only if} $D$ has rank $d$, meaning the conflict directions span the report space. The proof is two lines in each direction. If they span, $D$ has a right inverse and $W = Y D^{+}$; with exactly $m = d$ independent conflicts that is the square inverse, $W = Y D^{-1}$. If they do not span, pick any nonzero operator $N$ that annihilates every conflict direction — a rank-one $N$ built from a vector orthogonal to their span will do — and $W + N$ produces the same $Y$ on every trial, so nothing in the data separates them.

The design arithmetic that follows is the point. Recovering the whole operator from full-vector reports takes $d$ conflict directions; recovering only its symmetric part from scalar cue-line probes takes $d(d+1)/2$ of them. In two dimensions that is two probes against three, and in three dimensions three against six — fewer trials for strictly more of the matrix. Under noise the estimator is ordinary multivariate least squares and the conditioning of $D$ becomes the design question: spread the conflict directions across the space rather than clustering them. A study that collapses each two- or three-dimensional response to a fraction along the cue line is discarding exactly the part of $W$ that §8's identifiability theorem needs, and paying more trials for the privilege.

\textbf{How far the coordinate can travel, for a fixed conflict direction.} Hold the total precision $S = \Sigma _\ell ^{-1} + \Sigma _a^{-1}$ fixed and let the split between the two cues vary over everything admissible. Define

$r_d = \sqrt{(d^{\top}S^{-1}d)(d^{\top}Sd)} / (d^{\top}d)$,

which is at least 1 by Cauchy–Schwarz. $r_d$ is a direction-specific anisotropy factor: it measures how differently the total precision stretches the conflict direction in the primal and the inverse metric. The set of attainable cue-line coordinates is then the \textbf{open} interval $((1 - r_d)/2, (1 + r_d)/2)$. Open, not closed: reaching an endpoint would need a cue precision to lose rank, and a rank-deficient precision is not an admissible split. \emph{Which} precision has to go depends on the direction, and the two cases separate exactly along the line this section cares about. Writing $\Sigma _\ell ^{-1} = S^{1/2} M S^{1/2}$ makes the coordinate a linear function of the symmetric matrix $M$, whose supremum is carried by the positive eigenvalue of a rank-two matrix; the other eigenvalue is negative precisely when $r_d > 1$. So when the conflict direction \textbf{is} an eigenvector of the total precision — $r_d = 1$, the attainable interval is just $(0, 1)$ and no overshoot is available — the endpoint is approached by letting a \textbf{single} cue become uninformative while the other stays as well conditioned as $S$ itself. When the direction is \textbf{not} an eigenvector, so that $r_d > 1$ and overshoot is on the table, \textbf{both} cue precisions must degenerate together, each losing rank along a different direction. In a four-dimensional example, pushing the coordinate to within one part in a million of the endpoint takes the smallest eigenvalue of one cue precision to $2.0 \times  10^{-6}$ and of the other to $4.1 \times  10^{-6}$, against 4.07 for the total precision they sum to. Informative cues therefore stay strictly inside: over three thousand splits in which each cue retains at least a fifth of the total precision in every direction, none came within 0.23 of an endpoint (the numerical endpoint witness is in the supplement). Large overshoot is therefore not a prediction about well-conditioned cues; it is a statement about nearly degenerate ones, and an experiment using two informative cues should not expect to see it.

\textbf{When overshoot is possible at all.} $r_d$ equals 1 exactly when $d$ is an eigenvector of the total precision $S$, and then the attainable interval is $(0, 1)$, which contains no overshoot. So for a fixed conflict direction, \textbf{some} admissible split of a fixed total precision overshoots the cue line if and only if $d$ is not an eigenvector of $S$. The quantifier is the content here, and it is easy to misread. The claim is not that a given pair of cues overshoots along its non-eigenvector directions. The counterexample used above refutes that reading outright: its total precision has distinct eigenvalues 5.339 and 14.661, so almost every direction is a non-eigenvector, and yet its cue-line coordinate stays inside $[0.099, 0.926]$ — no direction overshoots at all. Nor does an eigenvector direction protect the report from leaving the segment between the cues — that is the separate failure scoped above, and anisotropy alone produces it.

\textbf{A bound in terms of how unequal the total precision is.} Let $\kappa $ be the condition number of the \textbf{total precision $S$} — the ratio of its largest to its smallest eigenvalue, a measure of how differently $S$ treats the best- and worst-determined directions. (This is a different matrix from §4's Fisher information, where a condition-number claim was withdrawn; the two should not be read together.) Kantorovich's inequality is the right tool here because it controls the product of a quadratic form and its inverse using nothing but the condition number, which turns total anisotropy directly into a worst-case statement. It bounds the spread above: $r_d \le  (\kappa  + 1)/(2\sqrt{\kappa })$. The bound is \textbf{sharp}, not merely an upper limit: equality holds when $d$ is the equal-weight combination of $S$'s extreme eigenvectors, and the two sides then agree to machine precision. A sweep over random directions never lands on that one and so reports a ratio just above 1, which is a fact about the sampling and not about the bound.

\textbf{Two dimensions, where the whole picture is one sinusoid.} Write $d$ at angle $\theta $. Then the cue-line coordinate is $t(\theta ) = \tau  + u \cos 2\theta  + v \sin 2\theta $, where $\tau  = \operatorname{tr}(A)/2$ is its mean over directions and $u = (A_{11} - A_{22})/2$ and $v = A_{12}$ are the two components of its variation. That is a single sinusoid in $2\theta $ with mean $\tau $ and half-swing $h = \sqrt{u^{2} + v^{2}}$ — the letter $h$ rather than the conventional $R$, because $R$ is this paper's referral operator and both appear in this section — offset by a phase that depends only on which direction is most extreme. Assume the eigenvalues of $W$ are real and satisfy $0 < \lambda _1 \le  \lambda _2 < 1$, which holds automatically here because $W$ is similar to $S^{-1/2} \Sigma _\ell ^{-1} S^{-1/2}$ — conjugate $W = S^{-1} \Sigma _\ell ^{-1}$ by $S^{1/2}$ — a symmetric matrix with spectrum strictly inside the unit interval. The \textbf{percept} precision is the one to name here: $S^{-1/2} \Sigma _a^{-1} S^{-1/2}$ is also symmetric with spectrum in $(0, 1)$, but it carries the spectrum of $I - W$, the weight the report puts on the cognitive prior rather than on the percept. The half-spread below is the same for both, so the decomposition that follows does not depend on which was named; the eigenvalues themselves are not. Write $\delta  = (\lambda _2 - \lambda _1)/2$ for the half-spread of those eigenvalues and $k = (W_{12} - W_{21})/2$ for the single number that carries the skew part in two dimensions. Then the half-swing decomposes as

$h^{2} = \delta ^{2} + k^{2}$,

so the swing of the cue-line coordinate across directions splits exactly into an eigenvalue-spread part and a non-commutativity part.

\textbf{Two overshoot criteria, not one.} Each side of the cue line has its own necessary and sufficient condition, and the two are different inequalities:

\begin{itemize}\setlength{\itemsep}{2pt}\item Some direction overshoots \textbf{above the second cue} if and only if $k^{2} > (1 - \lambda _1)(1 - \lambda _2)$.\item Some direction overshoots \textbf{below the first cue} if and only if $k^{2} > \lambda _1 \lambda _2$.\end{itemize}

Each right-hand side is a determinant, which is what makes the pair easy to read. $\lambda _1 \lambda _2$ is $\det W$, the product of the two weights the report places on the \textbf{percept}; $(1 - \lambda _1)(1 - \lambda _2)$ is $\det(I - W)$, the product of the two weights it places on the \textbf{cognitive prior}. In words: the report can run past the far cue only when the two precisions fail to commute by more than the report's total reliance on the prior, and it can fall back behind the near cue only when they fail to commute by more than its total reliance on the percept. Both equivalences were proved by showing that their negations are unsatisfiable over the model's own range $0 < \lambda _1 \le  \lambda _2 < 1$ (\texttt{p37\_overshoot\_symbolic.wls}, \texttt{p37\_overshoot\_iff.wls}; the algebra is two lines, since $(1 - \tau )^{2} - \delta ^{2} = (1 - \lambda _1)(1 - \lambda _2)$ and $\tau ^{2} - \delta ^{2} = \lambda _1 \lambda _2$).

\textbf{The two conditions coincide if and only if $\lambda _1 + \lambda _2 = 1$} — equivalently $\operatorname{tr} W = 1$, the two weights averaging to one half — and at no other pair of eigenvalues, since setting the two determinants equal cancels $\lambda _1 \lambda _2$ and leaves $1 - (\lambda _1 + \lambda _2) = 0$. Everywhere else one side can overshoot while the other does not, and the gap is not a curiosity of extreme parameters: over 20,000 fusion weights built from random positive-definite cue-precision pairs, 900 overshot above the second cue only, 819 below the first cue only, 599 both ways and 17,682 neither. Scoring the below-zero side with the above-one right-hand side therefore misclassifies 1,719 of those 20,000 cue pairs, 8.6 per cent of them (\texttt{p37\_overshoot\_criteria\_2026-09-29.py}).

A symmetry argument looks as though it should save the second statement from having to be printed, and it is worth saying why it does not: the symmetry moves the right-hand side as well. Relabelling which cue counts as the percept sends $W$ to $I - W$, and therefore each eigenvalue $\lambda $ to $1 - \lambda $, the skew number $k$ to $-k$, and the coordinate $t$ to $1 - t$. So “$W$ overshoots below the first cue” \emph{is} “$I - W$ overshoots above the second”, and feeding $I - W$ into the above-one criterion turns $(1 - \lambda _1)(1 - \lambda _2)$ into $\lambda _1 \lambda _2$. Holding one right-hand side fixed under a relabelling that changes it is the error the symmetry invites.

\textbf{Overshoot requires anisotropic fusion weights outright, and large overshoot requires ill-conditioned cues.} Equal eigenvalues force $k = 0$ exactly, because a matrix similar to $\lambda  I$ \emph{is} $\lambda  I$; over 4,000 equal-eigenvalue constructions the largest $|k|$ observed was $4.0 \times  10^{-16}$, which is zero to machine precision. With distinct eigenvalues every real $k$ is attainable, through $k = \delta  \cdot  \sin 2\theta  \cdot  (\rho  - 1/\rho )/2$ where $\rho $ is the eigenvalue ratio of the total precision $S$, so a large skew number demands a large $\rho $ — an ill-conditioned total precision. That is the quantitative form of this section's earlier statement that overshoot belongs to nearly degenerate cues rather than to well-conditioned ones. The below-zero witness shows the corrected criterion is not vacuous: $\lambda  = (0.015625, 0.031250)$, $k = -0.25$, both cue precisions positive definite, total-precision condition number 4,098, and a cue-line coordinate reaching \textbf{$-$0.227} — a report landing almost a quarter of the conflict \emph{behind} the nearer cue, on a pair the single printed criterion classifies as safe.

\textbf{The range assumption is doing the work in both statements, not the realness of the spectrum.} $W = \operatorname{diag}(1.5, 2.0)$ has a real spectrum, reaches $t = 2$, and satisfies neither criterion, because its eigenvalues sit outside $(0, 1)$. Inside the model both eigenvalues lie in $(0, 1)$, so both equivalences hold; stating the weaker hypothesis would not be enough.

\textbf{How many conflicts identify the symmetric part.} The symmetric part $A$ has $n(n + 1)/2$ free entries, and the \textbf{polarization identity} — the standard trick of recovering a symmetric form's cross terms from its values on sums, $2A_{\mathrm{ij}} = q(e_i + e_j) - q(e_i) - q(e_j)$ — recovers all of them from that many probes. Note the units, because they are the easy thing to get wrong. Polarization is an identity about the unnormalized form $q(d) = d^{\top}Ad$, while the cue-line coordinate divides that form by the probe's squared length. Reading $t$ rather than $q$ costs nothing, as long as the probe direction is recorded: $q(d) = \|d\|^{2} t(d)$ inverts the normalization exactly, so the requirement falls on the arithmetic and not on the instrument — which matters, because a cue-conflict experiment naturally reports a fraction of the conflict, and that number is perfectly usable. What fails is feeding $t$ into the polarization identity without multiplying it back first. In experimental terms: probe each coordinate direction and each pairwise-sum direction, and the pairwise probe contributes exactly the cross-term that the two single-coordinate probes cannot supply. The coordinate probes are also the ones that hide the mistake, since $\|e_i\|^{2} = 1$ makes $t$ and $q$ agree along them and the diagonal of $A$ comes out right either way; only the pairwise probes, with squared length 2, can expose it. When they are mishandled the damage has a closed form — polarizing $t$ directly returns $A_{\mathrm{ij}}/2 - (A_{\mathrm{ii}} + A_{\mathrm{jj}})/4$ in place of $A_{\mathrm{ij}}$ — so the error is set by the \textbf{diagonal} of $A$ and can dwarf the cross term being estimated. In a three-dimensional test it returned $-$2.907 for an off-diagonal entry whose true value was $-$0.092, thirty times too large and carrying the wrong sign. One probe fewer is never enough: $n(n + 1)/2 - 1$ directions always leave a one-parameter family of matrices fitting the data equally well. This counts what the conflict geometry determines about $A$ \textbf{given the cue locations}; it does not supply those locations, which is what the two-trial (e.g. zero-expectation) control design above is for, and it says nothing about the source-identifiability floor set by the rank of $R$.

\section{Parameter sensitivity and robustness over specified model families}

Because the models are illustrative rather than fitted, their parameters are chosen, not estimated; a fair objection is that the qualitative results might be artifacts of those choices. Two sensitivity sweeps test whether two headline numerical patterns persist over prespecified parameter families, and they do (Figure 15). That is parameter robustness over those families — the strongest thing two finite sweeps can establish — and not structural invariance of the models. \emph{(i) Neural-field bifurcation.} For a Mexican-hat kernel $\hat{W}(q) = A_e e^{-q^{2}/2} - A_i e^{-q^{2}\rho ^{2}/2}$ — $A_e$ and $A_i$ the excitatory and inhibitory amplitudes, $\rho $ the ratio of inhibitory to excitatory width, $q$ spatial frequency, so $\hat{W}(q)$ says how strongly the kernel amplifies a spatial pattern of that frequency, and the gain at which the most-amplified pattern first becomes self-sustaining is the transition — the marginal-stability gain is $s^{*} = 1/ \max_q \hat{W}(q)$ and the marginal-mode susceptibility, a critical-slowing proxy $\chi (s)$ whose denominator is constructed to vanish at $s^{*}$, so it peaks there by construction rather than by any independent evidence about migration: $\chi (s) = 1/(|s\cdot \max_q \hat{W} - 1| + \varepsilon )$. Sweeping the inhibition/excitation width ratio $\rho $ over $[2.0, 3.2]$ — a range spanning the Mexican-hat shapes for which the kernel has an interior spectral maximum at all, and reaching well past the illustrative point in both directions, so that persistence across it tests local tuning rather than claiming universality — the critical ridge $s^{*}(\rho )$ stays strictly interior — never at $s = 0$ nor at the saturation edge. This sweep shows the \textbf{critical point's location} is robust across the swept kernel family, not a tuned coincidence. "Across the swept family" is the honest strength: a finite sweep cannot establish a structural property, and this one does not. It demonstrates parameter robustness of \emph{where the transition falls}, not structural robustness of the clinical \emph{migration} prediction (§7), which remains a stated hypothesis this sweep does not independently test (though not analytically proven for all kernels). (The gain axis here uses the kernel-normalized convention $s^{*} = 1/\max_q \hat{W}(q)$, so its numeric values are not on the same scale as the illustrative $s^{*} \approx  0.82$ of Figure 6/§7; what transfers across figures is the \emph{qualitative} placement of the transition at an interior gain, not the number itself.) \emph{(ii) Model-E identifiability.} Across the transmission-noise range $\sqrt{\Sigma _c} \in  [0.5, 16]$ cm and the entire prior-width sweep $\kappa \Sigma _s, \kappa  \in  [0.5, 3]$, posterior contraction on the best-resolved singular axis falls with decreasing noise (the separable structure is learnable), while contraction along $R$'s exact null direction stays at 0.9994 to 0.9998 in every cell of the sweep — flat, and flat for the strongest available reason. That direction, a same-segment combination of the disc and facet coordinates sitting six degrees off the pure disc-vs-facet contrast, satisfies $R v = 0$ exactly, so \textbf{no} observation bears on it at \textbf{any} precision. The residual gap between the curve and 1 is not near-collinearity: it is the prior-transport effect of §8, which shifts the posterior variance along an unobserved direction by 0.12 per cent because Model E's prior is not aligned with that direction. The weaker sibling claim — that the disc/facet \emph{contrast} itself, whose 2$\times $2 submatrix has minimum singular value 0.25 and condition number 9.3, is identifiable in principle and unresolved in practice — needs the sweep re-run along that contrast, which gives 0.9947 to 0.9996 (\texttt{p37\_round2\_review\_checks\_2026-09-24.py}, check A17). So the identifiability floor of §8 survives every combination in the sweep, not just the illustrative parameters. The sweep code is in the supplement.

\begin{figure}[H]\centering
\includegraphics[width=\linewidth]{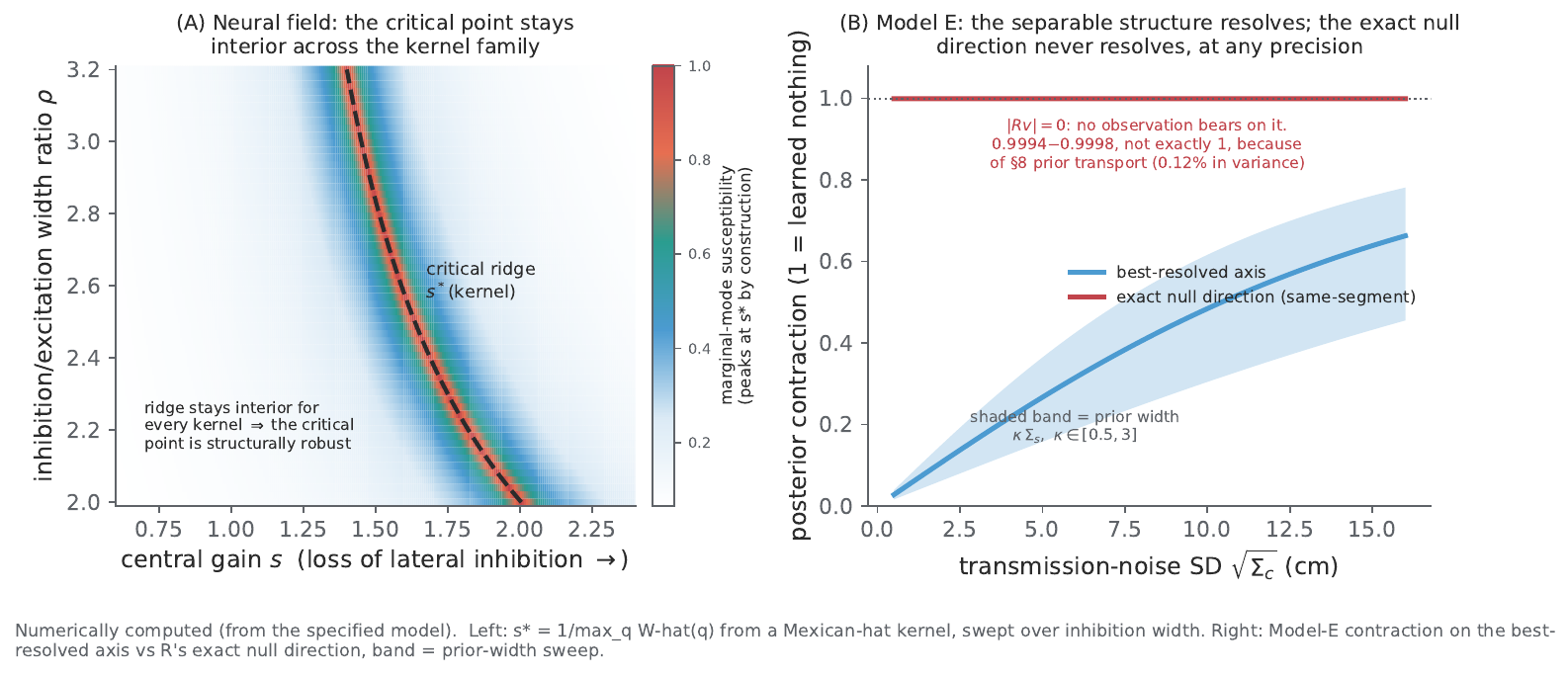}
\caption{Figure 15. Sensitivity analysis — persistence across the tested parameter families. (A) Across a whole family of Mexican-hat kernels, the critical gain $s^{*}$ stays strictly interior, so the critical-point location is robust over this kernel family. (B) Across the transmission-noise range and the entire prior-width sweep, the separable structure resolves while the red curve stays flat. The red curve is R's EXACT null direction: a same-segment combination of the disc and facet coordinates, six degrees off the pure disc-vs-facet contrast, satisfying R$\cdot $v = 0 to machine precision. It is flat because no observation bears on it at any precision whatsoever — unresolvable in principle, not merely unresolved in practice. It runs 0.9994 to 0.9998 rather than sitting at exactly 1 for a reason §8 derives and this figure illustrates: Model E's prior is not aligned with that direction, so prior transport moves the posterior variance along it by 0.12 per cent. The weaker claim — that the disc/facet contrast itself is identifiable in principle and unresolved in practice — is a separate computation, giving 0.9947 to 0.9996, and the 0.25 and 9.3 the text quotes describe the 2$\times $2 disc/facet submatrix rather than anything plotted here. [Numerically computed — computed sweeps of the neural-field kernel and Model E.]}
\end{figure}

\section{Status of the mathematical claims}

The mechanisms above mix results proved outright, results that follow given a stated model, purely numerical demonstrations, and open empirical hypotheses. Stating that mix explicitly, rather than letting the prose imply one register throughout, is the discipline this paper aims for.

{\footnotesize\begin{xltabular}{\textwidth}{>{\raggedright\arraybackslash}X>{\raggedright\arraybackslash}X}
\toprule
Claim & Status \\
\midrule
\endhead
Additional independent measurements add positive-semidefinite Fisher information (§4) & proved under the stated Gaussian model \\
Provocation adding information does not guarantee a lower condition number (§4) & proved by counterexample (\texttt{math\_checks.py}) \\
Full column rank is necessary for identifiability of an unconstrained continuous source; not necessary for a categorical source (§4) & proved algebraically / by counterexample, conditional on stated source semantics \\
Retaining another observation cannot reduce mutual information about a fixed inferential target (§5) & exact chain-rule identity; the target may be a trajectory, so a drifting generator is not an exception — a \textbf{changed estimand} is, and nothing orders the two mutual informations then \\
Optional extra data cannot raise the optimal Bayes risk, $R^{*}(X_{1:T}) \le  R^{*}(X_{1:T-1})$ (§5) & proved in one sentence (the shorter record's rule stays admissible); Blackwell dominance, elementary form. A 3,000-problem enumeration reproduces it to 1.1e-16, which is evidence the encoding is right, \textbf{not} the proof \\
Practical value of serial observation can be negative even as information is nondecreasing (§5) & model implication; untested for pain location. The negative term is acquisition burden — misspecification is \textbf{not} representable inside $R^{*}$ and needs the misspecified rule's own actual risk $R(\delta _T)$, which a control separates by 0.056 loss units \\
The 0.028-bit gap in §5's third example is the incremental information from observing group (§5) & exact under the stated independence premise, which is \textbf{sufficient}: group is drawn independently of disease here, so $I(D;G) = 0$ and the two chain rules collapse. All five quantities computed exactly by quadrature, both decompositions closing to 3.3e-15 bits. A control that lets prevalence differ by group separates the two readings by 0.069 bits, 2.4 times the reported effect \\
The leading mode's recovery time $1/|\lambda _{1}|$ grows as the leading eigenvalue approaches zero (§7) & linearized-model implication (derived); needs \textbf{no} noise model \\
Under the field equation's stated additive white noise, the leading mode's stationary variance $\sigma ^{2}/(2|\lambda _{1}|)$ and lag-$\Delta $ autocorrelation $\exp(-|\lambda _{1}|\Delta )$ grow the same way (§7) & derived from the linearization \textbf{and} the noise model, not from the linearization alone — a noise-free linearization has stationary variance zero at every eigenvalue. The variance scales with $\sigma ^{2}$; the autocorrelation does not depend on $\sigma $ at all, so it is the statistic to prefer when measurement noise is not constant across a cohort \\
The chosen neural field shows a focal-to-widespread transition with an interior instability peak & numerical, under the stated kernel/noise/metric; not proved for all kernels \\
The transition is hysteretic: two steady states coexist over $s \in  [0.567, 0.937]$ (§7) & \textbf{numerical evidence}, not proved — numerically located fixed points of a 64-dimensional system, with no analytic subcriticality argument. Checked by dual initialization at \textbf{fixed} $s$, which is what separates static bistability from slow relaxation; every run settled to a residual below 1e-14, and outside that interval both initial conditions converge to the same state, which is the control. The finite-rate sweep that first showed the loop understated the region, holding its recruited state only to $s = 0.74$ \\
Finite-window migration \emph{declines} on the high-barrier/high-gain side of the transition (§7.1) & conditional prediction — derived from the displacement normal form under five stated sufficient conditions, and simulated; \textbf{the model does not license a practically detectable intermediate peak}, so the earlier non-monotone version of this row is withdrawn rather than downgraded. Condition 1, that the destabilizing mode is a displacement mode, is the open empirical premise \\
The periodic completion $\widetilde{V}(\theta ) = -\alpha (1-\cos \theta ) + \beta (1-\cos \theta )^{2}$ has exactly two wells for $0 < \alpha  < 4\beta $, joined by barriers $\alpha ^{2}/(4\beta )$ and $(\alpha -4\beta )^{2}/(4\beta )$ that cross at $\alpha  = 2\beta $ (§7.3) & proved symbolically in free $\alpha , \beta $; independently confirmed by brute-force minimum counting \\
The obvious alternative completion, $-(\alpha /2)\sin^{2}\theta  + (\beta /4)\sin^{4}\theta $, has four wells and is therefore not a two-location model (§7.3) & proved; disproves that completion, and with it any quantity measured on it \\
Past $\alpha  = 4\beta $ the two wells merge into one at the far side, so between-location migration stops rather than declining (§7.3) & proved symbolically ($\widetilde{V}''(\pi ) = \alpha  - 4\beta $); the simulated collapse that follows is numerical \\
The sizes of the closed-surface migration decline, and its rise between $\alpha  = \beta $ and $\alpha  = 4\beta $ (§7.3) & numerical, three noise levels and eight seeds; the line model's reproduction of §7.1's published bounds is the control \\
No growth-blind functional that is 1-Lipschitz in the profile it compares against counts rotations, in the profile's own unit of single-step vacated mass, further than $V$ does; the ceiling is $\rho (a)$ and $V$ attains it (§7.4) & proved for every nonnegative profile on the ring that a single step disturbs ($D_a(1) > 0$; a constant profile has no single-step unit to be measured in), by a triangle-inequality and growth-cone squeeze; checked in exact rationals (21 checks, ten negative controls, all fired) and symbolically over two families of non-arc profiles. Sufficiency only — the statement is false against a bare site count rather than the profile's own unit, and $\min(V, \rho (a)\cdot D_a(1))$ attains the same optimum, so $V$ is optimal and not canonical \\
Exactness at a rotation $k$ forces $D_a(j) = j\cdot D_a(1)$ at every $j \le  k$, so the exactness set is an initial segment and $k \le  \rho (a)$ (§7.4) & proved in four lines from subadditivity, which follows from $(x+y)_+ \le  x_+ + y_+$ and $R_j$ being a permutation. The hypothesis is load-bearing rather than technical: a non-subadditive squared score has exactness set $\{1, 4\}$, which is not an initial segment (control NC-PE-2, subadditivity violation +0.8346). 849,341 triples searched for a violation of subadditivity itself, largest value exactly zero \\
Group-conditional felt/reported location distributions differ (c, §6) & open empirical hypothesis — no direct evidence yet \\
The preregistered decision rule recovers which of four explanations generated the data, at $\approx $ 1250 participants & \textbf{established in the companion paper (Shavit, 2026), not here} — specifically its Appendix A, \emph{Closed-form power (Study 1)}, which rests on the simulation in the preregistration §7b whose output ships with the code supplement. This manuscript develops no section behind the row, and it is listed only so the two papers' claim tables can be read against each other. Numerical there, under the stated generators and measurement-error model; the rule as first written did \textbf{not} recover them, and was repaired \\
Fixing detection alone yields little outcome benefit under low treatment uptake (care-pathway corollary, §6) & illustrated here on one setting (Lee et al., 2019), with empirical parallels reviewed in the companion; an algebraic policy decomposition, not a fitted controlled direct effect \\
Within a group, the pathway equation $E[Y(a)] = e_0[1 - p_c(a,g)\cdot q_g\cdot RRR]$ is \textbf{exact}, not approximate, under homogeneous baseline risk and treatment effect, uptake independent of which cases are detected, no policy-to-outcome path except through treatment, and false-positive harms outside $Y$ (§6) & proved as an identity with premises; exact to 1.1e-16 at three detection probabilities. Under heterogeneity the error is \textbf{exactly} a covariance across cases, zero to 2.8e-17 when detectability is independent of severity. Its \textbf{sign} is transportable and its \textbf{size} is not: in one constructed severity-indexed population a milder-reaching policy overstates its benefit by about two thirds ($-$0.003829 against an exact $-$0.002300). §6's own three counts come from a routine holding risk, effect and uptake at scalar values, so its population is homogeneous and the formula is exact \textbf{for it} — what those counts carry is the homogeneity premise, not a measured bias \\
Mechanisms (b) and (c) are separately identifiable from group-conditional location data \textbf{iff} the model class $m$ is observed or the likelihood family is constrained (§6) & proved; necessity exhibited by two five-cell models, one pure (b) and one pure (c), agreeing on \textbf{every} observable to 1e-12 with all four components valid distributions. A group difference in reports has a third rival explanation as well — a group-dependent reporting weight or prior, worth 0.5217 report units on its own — which §8's two-trial design can subtract \\
Threshold optimum $c_g^{*}$ differs by group prevalence/cost (§6) & proved algebraically; checked numerically \\
$(b, \mu _a)$ is identifiable from trials at known report weights \textbf{iff} the stacked contrast matrix $[W_2-W_1; \dots  ; W_J-W_1]$ has rank $d$; with two trials, \textbf{iff} $W_2-W_1$ is nonsingular (§8) & proved, necessary \textbf{and} sufficient. Necessity is exhibited rather than inferred: a rank-deficient contrast gives an explicit one-parameter family of indistinguishable $(b, \mu _a)$ agreeing to 1e-9. Verified at $d = 2, 3, 4$, 400 draws each \\
The zero-expectation control trial $W_1 = I$ is \textbf{constructive}, not merely sufficient: it reads $b$ directly and returns $\mu _a = (I-W_2)^{-1}(z_2 - W_2 b)$ (§8) & proved; recovery error 3.5e-16 at $d = 2$ and 5.1e-16 at $d = 3$ \\
Two trials can share their symmetric parts exactly and still identify $(b, \mu _a)$, so the design must retain the \textbf{full vector report} (§8, §8.1) & proved; a worked pair has $\|sym(W_2)-sym(W_1)\| = 6.3 \times  10^{-16}$ and $\det(W_2-W_1) = 0.16$. The escape is \textbf{even-dimensional} — odd-order skew matrices are always singular — so the two-trial criterion is dimension-dependent and this paper works in both 2D and 3D \\
The mean report geometry does \textbf{not} identify absolute cue precisions; the unidentified family has dimension $d$ (§8) & proved. $W(cP_\ell , cP_a) = W(P_\ell , P_a)$ to 4.9e-16 is only one direction of it: every symmetric $S$ with $SW$ symmetric and $SW$, $S(I-W)$ positive definite gives the same $W$, leaving $d(d+1)/2 - d(d-1)/2 = d$ dimensions. Measured as 2, 3, 4 at $d = 2, 3, 4$; an exhibited member has entrywise ratio 1.0615–1.4344, so it is not a rescaling \\
For independent Gaussian cues the multicue rule is exactly $W_k = P_{\mathrm{tot}}^{-1}P_k$ (§8) & proved; checked over 300 three-cue problems at $d = 3$ to 1e-10. $\Sigma _k W_k = I$ does \textbf{not} pin it down — the reversed product also sums to $I$ and fuses 0.7223 report units away when the precisions do not commute, and coincides exactly when they do \\
Matrix-weighted cue fusion lies on the segment between cues (§8) & true only in the scalar/isotropic case used by the figures; false in general, and anisotropy alone is enough to break it (two counterexamples in \texttt{math\_checks.py}, one with commuting precisions) \\
The cue-line statistic sees only the symmetric part of $W$, and its range is that part's spectrum (§8) & proved algebraically; elementary \\
Full-vector conflict reports identify \textbf{all} of $W$ \textbf{iff} the conflict directions span the report space, with estimator $W = Y D^{-1}$ at $m = d$ conflicts (§8.1) & proved, necessary and sufficient, two lines each way (right inverse one way; an annihilating operator $N$ giving $W + N$ the same data the other). The design count is $d$ conflicts for the whole operator against $d(d+1)/2$ scalar probes for its symmetric part alone \\
The report cannot carry more information about the cause than the percept, $I(S;R|G,Z) \le  I(S;L|G,Z)$ (§8) & proved; the data-processing inequality applied to §3's own factorization, which conditions the report on $\ell $ alone. Control A2.C1 fires: allowing $p(r \mid \ell ,s)$ breaks the factorization and the inequality fails by up to 0.3 bits. Distinct from §5, where no Markov emission structure is available and the chain rule is the right tool \\
The reporting node is \textbf{exactly} an inflation of the transmission noise $\Sigma _c$ by $W^{-1}\Sigma _\nu (W^{-1})^{\top}$ (§8) & proved in closed form for the linear-Gaussian model; re-derived by hand. With no report noise and invertible $W$ the report loses nothing, and all loss rides on $\Sigma _\nu $ amplified by the inverse square of the weight on the percept \\
The skew part discarded by that statistic is the transported precision commutator (§8) & proved; a specialization of the commutator-with-an-inverse identity, not a new result \\
For a fixed conflict direction and fixed total precision, the attainable cue-line coordinates form the open interval $((1-r_d)/2, (1+r_d)/2)$ (§8) & proved; an endpoint is approached only as a cue precision degenerates — \textbf{both} together whenever the conflict direction is not an eigenvector of the total precision, and \textbf{one alone} when it is. The dichotomy is exact, not a hedge: it is whether the rank-two matrix carrying the supremum has a strictly negative second eigenvalue \\
Some admissible split overshoots the cue line iff the conflict direction is not an eigenvector of $S$ (§8) & proved; existential over splits, not a claim about any given cue pair \\
$r_d \le  (\kappa +1)/(2\sqrt{\kappa })$ for $\kappa $ the condition number of the total precision (§8) & Kantorovich's inequality applied to fusion; sharp — attained at the equal-weight direction in $S$'s eigenbasis \\
In two dimensions the cue-line coordinate's half-swing splits as $h^{2} = \delta ^{2} + k^{2}$ (§8) & proved, given eigenvalues of $W$ real and inside $(0,1)$ — automatic in this model, and the range, not the realness, is what the overshoot criteria need \\
In two dimensions, some direction overshoots above the second cue \textbf{iff} $k^{2} > (1-\lambda _1)(1-\lambda _2) = \det(I-W)$, and some direction overshoots below the first cue \textbf{iff} $k^{2} > \lambda _1 \lambda _2 = \det W$ (§8) & both proved as necessary \textbf{and} sufficient conditions, by quantifier elimination over $0 < \lambda _1 \le  \lambda _2 < 1$ (negation unsatisfiable in both cases); the two thresholds are equal iff $\lambda _1 + \lambda _2 = 1$, and scoring the below-zero side with the above-one threshold misclassifies 8.6 per cent of 20,000 random cue pairs \\
Overshoot requires anisotropic fusion weights, and large overshoot requires an ill-conditioned total precision (§8) & proved for the first part (a matrix similar to $\lambda I$ is $\lambda I$, so equal eigenvalues force $k = 0$); the second is the exact attainability formula $k = \delta  \sin 2\theta  (\rho  - 1/\rho )/2$, checked numerically \\
$n(n+1)/2$ conflict directions identify the symmetric part $A$, and one fewer never does (§8) & proved by the polarization identity; identifies $A$ given the cue locations, which the design must anchor separately \\
The posterior along a null direction $v$ of $R$ equals the prior, in mean and variance and for every observation, \textbf{iff} $R \Sigma _s v = 0$ — equivalently, iff the unidentified contrast is a priori independent of the identified subspace (§8) & proved algebraically; necessary \textbf{and} sufficient, with the failure direction checked by construction \\
Absent that condition the posterior along $v$ is strictly tighter than the prior, bounded below by the prior conditional variance given the identified subspace (§8) & proved; the bound is sharp, attained in the zero-noise limit \\
The \emph{conditional} law of the null-space component given the identified component is unchanged by the data, with no condition on the prior (§8) & proved; this is what a flat likelihood buys, and all it buys \\
A deep source is identifiable from location data (§8) & conditional on the continuous-mixture assumption stated in §4/§8 and on full column rank of the noise-whitened $R$ \\
\bottomrule
\end{xltabular}}

\section{Discussion: what would falsify each result}

The models above are illustrative, so the useful question is not whether they fit but what would break them. Each result names its own failure condition.

\textbf{The identifiability floor (§4, §8).} Two results define this floor, and they carry different strengths; running them together would overstate both. An \textbf{exact} null direction of the referral operator — $\tilde{R} v = 0$ — is unrecoverable from location under the continuous-mixture source model of §4, at any measurement precision, because no observation bears on that contrast at all. Model E carries both cases. Its $R$ is 2$\times $3, so it has an exact null direction of its own, and that is what §9's sweep and Figure 15B draw. Its disc/facet \emph{pair} is the different and weaker case: those two columns are \textbf{near-collinear, not collinear} — the 2$\times $2 submatrix they form has minimum singular value 0.25 and condition number 9.3 (§8) — so that contrast is identifiable in principle, and the sweep re-run along it shows it stays unresolved across the whole noise-and-prior range rather than only at the illustrative parameters. What would falsify the second claim is a realistic provocation design that separates a same-segment pair from location data alone at an attainable precision. That would contradict Model E's parameterization — the numbers chosen for $R$ — and would leave the null-space theorem untouched, because the two claims are not the same claim.

\textbf{The information identity (§5).} Retaining another observation cannot reduce recoverable information about a \emph{fixed inferential target}. This is exact, so it cannot be falsified empirically; what can fail is its premise, and the premise is narrower than it first looks. A drifting generator does not break it: a trajectory is a perfectly good fixed target. What breaks it is a change of \textbf{estimand} — asking about the current state at each time rather than about the whole trajectory — and that is the clinically live case, an acute source centralizing over the observation window. Where the estimand changes, the identity no longer licenses the clinical inference drawn from it, and the quantity to track is a change point rather than an accumulating total. That is a testable claim about pain, and §5 states it as one.

\textbf{The bifurcation (§7).} Spatial extent is the order parameter and instability peaks at the transition. The falsifiable part is the \emph{falling limb}: past the transition, finite-window migration declines as the field saturates. The analysis does \textbf{not} establish the rising limb, and §7.1 says so — the interior maximum is real in the approximation and negligible in the process, a climb under five per cent against a decline over twenty-five per cent. A study that observed a large intermediate migration peak would not confirm the model; it would sit outside what the model licenses.

\textbf{The reporting model (§8).} Its parameters are recoverable only under two preconditions, both stated rather than assumed. First, the provocation must keep the mutual information above the noise floor. Second, because a shift in the anatomical bias is exactly cancelled by a shift in the cognitive prior within a single trial, the design needs trials whose report-weight contrasts span the whole report space — with two trials, $W_2 - W_1$ nonsingular, and a zero-expectation control trial is one sufficient and especially transparent choice, not the only one. That rank condition is necessary as well as sufficient, so what could fail is not the theorem but the premise that $W_j$ is known: §8.1 shows a scalar cue-line probe recovers only the symmetric part of $W$, so a design that keeps only the cue-line summary and then treats $W$ as known has assumed what it did not measure. A design with only one trial, with trials whose weight contrasts are singular, or with the full vector report discarded, cannot estimate the model, and a fit reported without that variation should not be believed.

\textbf{What cue-conflict data can identify (§8.1).} These are theorems, so like the information identity they cannot be falsified by an experiment; what can fail is a premise, and naming the right premise took some care. The results assume both cue precisions are positive definite and that the fusion is the linear precision-weighted blend of §8 — drop either and none of them applies. Two premises are easy to state wrongly. The two-dimensional decomposition needs the eigenvalues of $W$ to lie strictly inside $(0, 1)$, not merely to be real: $W = \operatorname{diag}(1.5, 2.0)$ has a real spectrum, overshoots, and satisfies neither of the two stated criteria. And the probe count identifies the symmetric part \emph{given the cue locations}, which the design must anchor separately; read without that condition it would appear to remove the very requirement §8 imposes. A study that estimated the reporting weight without anchoring the cue locations, and reported the symmetric part as identified, would not falsify these results — it would have assumed away their premise.

\textbf{What the models are not.} They are not a clinical decision tool. They are not fitted, and §10 marks which claims are numerical demonstrations under a chosen kernel, noise model and metric. Their value is that each one names a quantity a study could measure and a pattern that would count against it. The companion paper (Shavit, 2026) carries the empirical evidence and the design that would test the two-variable prediction the bifurcation implies.

\section{Statements and Declarations}

\textbf{Funding.} The author declares that no funds, grants, or other support were received during the preparation of this manuscript.

\textbf{Competing interests.} The author has no relevant financial or non-financial interests to disclose.

\textbf{Ethics approval.} Not required. This paper reports no study involving human participants or animals and analyses no individual-level human data. All quantitative results are simulations from the models described here, run on synthetic inputs.

\textbf{Consent to participate / consent to publish.} Not applicable, for the reason above.

\textbf{Author contributions.} A.Y.S. is the sole author. He specified every model — its structure, its assumptions and the claims it is asked to support — designed every analysis, and wrote and revised the manuscript. He did not write every line of the code: an AI assistant wrote parts of the simulation and figure code to his specification and under his direction, and he read, ran and verified all of it. The division of labour is disclosed in full immediately below.

\textbf{Use of AI tools.} AI tools assisted under the author's direction. Claude (Anthropic) wrote simulation and figure code to the author's specification and gave editorial support. Gemini (Google) and ChatGPT (OpenAI) were used for adversarial review and independent mathematical checking; ChatGPT's review of the mathematics prompted the claim-status audit in §10.

No claim rests on an AI's word. Every quantitative claim was checked against its primary source, and the verification trail is retained in the supplement: a per-reference registry records the class of evidence behind every source, and the reading ledger names every source whose full text remains unread. That trail also records what the checking caught — AI-supplied citations, since excluded; claims that misstated their source; and one figure the author had wrongly \emph{removed} on a bad inference.

The author is solely accountable for the content, including all errors. Consistent with ICMJE and COPE guidance, no AI system is an author or can hold accountability for the work.

\section{Data and Code Availability}

Every model, figure, and sensitivity sweep in this paper regenerates from the public code archive on the Open Science Framework (osf.io/zfhkq), persistent identifier https://doi.org/10.17605/OSF.IO/ZFHKQ. The study uses \textbf{no primary human-subjects data}: the models are illustrative and parameterized as described.

\section{References}

\par\hangindent=1.5em\noindent Amari, S. (1977). Dynamics of pattern formation in lateral-inhibition type neural fields. \emph{Biological Cybernetics, 27}(2), 77–87. https://doi.org/10.1007/BF00337259

\par\hangindent=1.5em\noindent Berthier, M., Starkstein, S., \& Leiguarda, R. (1988). Asymbolia for pain: A sensory-limbic disconnection syndrome. \emph{Annals of Neurology, 24}(1), 41–49. https://doi.org/10.1002/ana.410240109

\par\hangindent=1.5em\noindent Boström, K. J., de Lussanet, M. H. E., Weiss, T., Puta, C., \& Wagner, H. (2014). A computational model unifies apparently contradictory findings concerning phantom pain. \emph{Scientific Reports, 4}, 5298. https://doi.org/10.1038/srep05298

\par\hangindent=1.5em\noindent Botvinick, M., \& Cohen, J. (1998). Rubber hands "feel" touch that eyes see. \emph{Nature, 391}(6669), 756. https://doi.org/10.1038/35784

\par\hangindent=1.5em\noindent Colloca, L., \& Benedetti, F. (2005). Placebos and painkillers: Is mind as real as matter? \emph{Nature Reviews Neuroscience, 6}(7), 545–552. https://doi.org/10.1038/nrn1705

\par\hangindent=1.5em\noindent Coombes, S. (2005). Waves, bumps, and patterns in neural field theories. \emph{Biological Cybernetics, 93}(2), 91–108. https://doi.org/10.1007/s00422-005-0574-y

\par\hangindent=1.5em\noindent Cover, T. M., \& Thomas, J. A. (2006). \emph{Elements of information theory} (2nd ed.). Wiley.

\par\hangindent=1.5em\noindent Goldreich, D. (2007). A Bayesian perceptual model replicates the cutaneous rabbit and other tactile spatiotemporal illusions. \emph{PLoS ONE, 2}(3), e333. https://doi.org/10.1371/journal.pone.0000333

\par\hangindent=1.5em\noindent Helton, J. W., \& Howe, R. E. (1975). Traces of commutators of integral operators. \emph{Acta Mathematica, 135}, 271–305. https://doi.org/10.1007/BF02392022

\par\hangindent=1.5em\noindent Klit, H., Finnerup, N. B., \& Jensen, T. S. (2009). Central post-stroke pain: Clinical characteristics, pathophysiology, and management. \emph{The Lancet Neurology, 8}(9), 857–868. https://doi.org/10.1016/S1474-4422(09)70176-0

\par\hangindent=1.5em\noindent Körding, K. P., Beierholm, U., Ma, W. J., Quartz, S., Tenenbaum, J. B., \& Shams, L. (2007). Causal inference in multisensory perception. \emph{PLoS ONE, 2}(9), e943. https://doi.org/10.1371/journal.pone.0000943

\par\hangindent=1.5em\noindent Lee, K. K., Ferry, A. V., Anand, A., Strachan, F. E., Chapman, A. R., Kimenai, D. M., Meex, S. J. R., Berry, C., Findlay, I., Reid, A., Cruickshank, A., Gray, A., Collinson, P. O., Apple, F. S., McAllister, D. A., Maguire, D., Fox, K. A. A., Newby, D. E., Tuck, C., . . . Mills, N. L. (2019). Sex-specific thresholds of high-sensitivity troponin in patients with suspected acute coronary syndrome. \emph{Journal of the American College of Cardiology, 74}(16), 2032–2043. https://doi.org/10.1016/j.jacc.2019.07.082

\par\hangindent=1.5em\noindent Lewis, J. S., Kersten, P., McCabe, C. S., McPherson, K. M., \& Blake, D. R. (2007). Body perception disturbance: A contribution to pain in complex regional pain syndrome (CRPS). \emph{Pain, 133}(1–3), 111–119. https://doi.org/10.1016/j.pain.2007.03.013

\par\hangindent=1.5em\noindent Makin, T. R., Scholz, J., Filippini, N., Henderson Slater, D., Tracey, I., \& Johansen-Berg, H. (2013). Phantom pain is associated with preserved structure and function in the former hand area. \emph{Nature Communications, 4}, 1570. https://doi.org/10.1038/ncomms2571

\par\hangindent=1.5em\noindent Moseley, G. L. (2006). Graded motor imagery for pathologic pain: A randomized controlled trial. \emph{Neurology, 67}(12), 2129–2134. https://doi.org/10.1212/01.wnl.0000249112.56935.32

\par\hangindent=1.5em\noindent Rabin, J., Delon, J., \& Gousseau, Y. (2011). Transportation distances on the circle. \emph{Journal of Mathematical Imaging and Vision, 41}(1–2), 147–167. https://doi.org/10.1007/s10851-011-0284-0

\par\hangindent=1.5em\noindent Rainville, P., Duncan, G. H., Price, D. D., Carrier, B., \& Bushnell, M. C. (1997). Pain affect encoded in human anterior cingulate but not somatosensory cortex. \emph{Science, 277}(5328), 968–971. https://doi.org/10.1126/science.277.5328.968

\par\hangindent=1.5em\noindent Ramachandran, V. S., \& Hirstein, W. (1998). The perception of phantom limbs: The D. O. Hebb lecture. \emph{Brain, 121}(9), 1603–1630. https://doi.org/10.1093/brain/121.9.1603

\par\hangindent=1.5em\noindent Ramachandran, V. S., Rogers-Ramachandran, D., \& Stewart, M. (1992). Perceptual correlates of massive cortical reorganization. \emph{Science, 258}(5085), 1159–1160. https://doi.org/10.1126/science.1439826

\par\hangindent=1.5em\noindent Shavit, A. Y. (2026). Three failures of pain location: Why its diagnostic utility is three quantities, not one. \emph{arXiv}. https://arxiv.org/abs/2607.26297

\section*{Appendix A: Glossary}

Plain definitions of terms that are standard in one field and unfamiliar in another. The companion paper (Shavit, 2026) carries a longer glossary aimed at its clinical readers; this one covers the terms this paper uses.

{\footnotesize\begin{xltabular}{\textwidth}{>{\raggedright\arraybackslash}X>{\raggedright\arraybackslash}X}
\toprule
Term & Plain meaning \\
\midrule
\endhead
Referral matrix & the operator mapping deep sources to felt surface locations — the same object as the lead-field matrix, named for the clinical phenomenon it produces \\
Lead-field matrix & the mapping from internal sources to surface measurements (borrowed from EEG source localization) \\
Rank-deficient / null space & a mapping that collapses distinct inputs to the same output; the null space = causes giving an identical observation \\
Identifiability floor & the source contrasts a location measurement leaves unresolved — a limit of the mapping, not of the measurement. Two strengths, kept apart throughout: an \emph{exact} null direction — collinear columns, or simply more sources than measurements, as in Model E's 2$\times $3 $R$ — is unrecoverable at any precision, while a \emph{near}-null direction (near-collinear columns, as in Model E's disc/facet pair) is recoverable in principle and unresolved at every precision a study could deliver \\
Reporting node & the stage at which a felt location becomes a spoken or drawn one; where pain is \emph{said} to be, as distinct from where it is felt \\
Loss function & the cost attached to each way of being wrong; it sits in the decision layer, not in the probability model \\
Likelihood ratio (LR+, LR$-$) & how much a test result multiplies the odds of disease; LR+ > 1 raises those odds and LR$-$ < 1 lowers them, and the \emph{distance} from 1 says how much — a value just above or just below 1 barely moves the diagnosis \\
Covariate shift & in machine learning, usually means the input distribution $p(X)$ changes while $p(Y \mid X)$ stays fixed \\
Controlled direct effect / causal mediation & the effect of one variable holding a mediator fixed at a specified level, under potential-outcomes assumptions — isolates a mechanism. Stronger than the standardized policy contrast this paper computes (§6) \\
Critical slowing & near a transition the system returns to rest more and more slowly, so recovery time rises; with a noise source driving it, the mode's variance and its lag autocorrelation rise too — a generic early-warning signature. The slowing needs only the linearized dynamics; the variance and the autocorrelation need the noise model as well \\
Convex hull / convex-hull & the set of points lying between the cues; a blend that stays inside it can pull the estimate toward a cue but never past one \\
Repulsion & a report pushed \emph{away} from a cue rather than toward it; the opposite of the pull that cue integration produces, and outside what a convex blend can generate \\
\bottomrule
\end{xltabular}}
\end{document}